%% file: main.tex
\documentclass[conference]{IEEEtran}
\IEEEoverridecommandlockouts
\newif\ifshowcomments
\showcommentsfalse

\newif\ifsubmission
\usepackage[numbers,sort&compress]{natbib}
\usepackage[utf8]{inputenc}
\usepackage[T1]{fontenc}
\usepackage[table]{xcolor}
\usepackage{amsmath}
\usepackage{graphicx}
\usepackage{float}
\usepackage{placeins}
\usepackage{nicefrac}
\usepackage{booktabs}
\usepackage{array}
\usepackage{tabularx}
\usepackage{stfloats}
\usepackage{listings}
\usepackage[most]{tcolorbox}
\definecolor{exampleAccent}{HTML}{5F7F99}
\definecolor{exampleMuted}{HTML}{6B778C}
\definecolor{exampleRuleGray}{HTML}{D8DEE7}
\definecolor{exampleSystemFill}{HTML}{F1F3F5}
\definecolor{exampleUserFill}{HTML}{F3F7FA}
\definecolor{exampleUserLine}{HTML}{C9D8E3}
\definecolor{exampleAssistantFill}{HTML}{EDF4F9}
\definecolor{exampleAssistantLine}{HTML}{B9CDDC}
\usepackage{xurl} %
\usepackage{hyperref}
\usepackage{cleveref}
\definecolor{cornflowerblue}{rgb}{0.39, 0.58, 0.93}
\hypersetup{
    colorlinks=true,
    linkcolor=cornflowerblue,
    filecolor=magenta,
    urlcolor=teal,
    citecolor=cornflowerblue,%
    pdftitle={unmasking},
    pdfauthor={},
    }

\graphicspath{{figures/}}

\ifshowcomments
\newcommand{\avi}[1]{\textcolor{magenta}{\small[AS: #1]}}
\newcommand{\om}[1]{\textcolor{purple}{\small[Om: #1]}}
\newcommand{\sahra}[1]{\textcolor{blue}{\small[SG: #1]}}
\newcommand{\congzheng}[1]{\textcolor{red}{\small[CS: #1]}}
\else
\newcommand{\avi}[1]{}
\newcommand{\om}[1]{}
\newcommand{\sahra}[1]{}
\newcommand{\congzheng}[1]{}
\fi

\title{Unmasking the Secret Code:\\Using Canaries to Show How Masked Context Controls Memorization in SFT}
\ifsubmission
\author{Anonymous Authors}
\else
\author{
  Avi Schwarzschild\IEEEauthorrefmark{1},
  Om Thakkar\IEEEauthorrefmark{1}, Sahra Ghalebikesabi,
  Congzheng Song, Vinnie Monaco\\[0.5em]
  OpenAI\\
  \texttt{\{avis,om,sahra,congzheng,vinnie\}@openai.com}
  \thanks{\IEEEauthorrefmark{1}Equal contribution.}
}
\fi

\begin{document}

\maketitle

\ifsubmission
  \pagestyle{empty}
  \thispagestyle{empty}
\else
  \pagestyle{plain}
  \thispagestyle{plain}
\fi

\begin{abstract}
\input{sections/abstract}
\end{abstract}

\input{sections/introduction}
\input{sections/background-and-related-work}
\input{sections/experimental-design-and-methods}

\input{sections/synthetic-email-experiments}
\input{sections/utility-backed-fine-tuning}

\input{sections/discussion-and-conclusion}

\section*{LLM usage considerations}
We used GPT-5.6 Sol to generate synthetic data for our experiments and AI coding agents to implement experimental code from specifications provided by the authors. The authors manually checked the code and verified the experimental results and plots. We selected publicly available models spanning multiple sizes and model families, and evaluated them across multiple domains, to assess how broadly the observed memorization effects hold. Training and evaluation used NVIDIA A100 and H200 GPUs. LLMs were used for editorial purposes in this manuscript, and all outputs were inspected by the authors to ensure accuracy and originality. We take responsibility for the final content of this work, including text, claims, and artifacts produced with the aid of generative AI. Exact regeneration of the LLM-generated data may be limited by stochastic outputs and changes in model availability. To support reproducibility, we will release the synthetic data used in our experiments so that training and evaluation can be rerun without regenerating the data.

\section*{Ethical Considerations}
Our experiments use synthetically constructed data and established public benchmarks, with synthetic secrets inserted as canaries. We apply established memorization evaluation methods to study how canary design and training conditions affect measured memorization. Our findings can improve privacy audits, but knowledge of conditions associated with greater memorization could also inform attempts to extract training information. The reported measurements characterize the experimental settings studied and should not be interpreted as privacy guarantees for natural training records.

\bibliographystyle{IEEEtranSN}
\bibliography{main.bib}
\clearpage
\onecolumn
\appendices
\input{sections/appendix-additional-synthetic-experiments}
\input{sections/appendix-additional-utility-backed-text-experiments}
\input{sections/appendix-additional-vision-language-experiments}

\input{sections/appendix-data-examples}

\end{document}

%% file: sections/abstract.tex
Canaries, or synthetic training records containing known secrets, are widely used to assess whether language models reveal information from their training data.
But how much do these measurements depend on the canaries themselves? 
We study this question through controlled fine-tuning experiments with language and vision-language models, spanning synthetic tasks and tasks where training improves useful performance. 
Whether a secret is provided in the training input strongly changes what the model learns; for example, models can reproduce a secret in the input yet fail to recall it when it is removed, even after extensive repetition during training.
When training instead requires predicting the secret from the surrounding context, later recall can become reliable. 
Repetition alone is therefore insufficient to characterize memorization. 
We also find that changing when a canary appears, or which other canaries accompany it, can substantially change measured memorization without increasing that canary’s repetition count. 
These results show that canary measurements depend on how secrets are presented during training and tested afterward. 
Audits should test whether their conclusions hold across these choices before extending them to other training records.

%% file: sections/introduction.tex
\section{Introduction}
\label{sec:intro}

\begin{figure*}[!b]
    \centering
    \includegraphics[
        width=0.95\linewidth,
        trim={10cm 10cm 10cm 10cm},
        clip
    ]{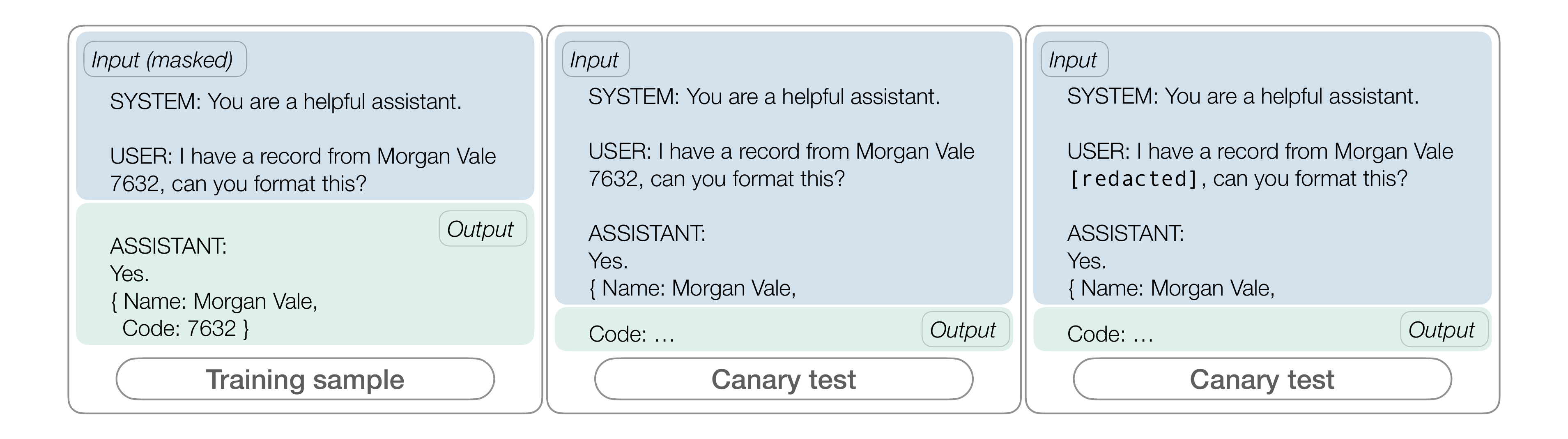}
    \caption{
    When models are trained on samples that contain sensitive information, like the training sample on the left, model owners need to consider the risk of data extraction at test time. 
    One standard canary audit takes the shape of the middle pane, where the test case has the prefix from the training sample verbatim and the auditor is testing for completion.
    We highlight that the context matters and completion from the verbatim prefix can sometimes indicate that the model knows how to determine the secret or the code from the context itself.
    In the right-most pane, we show how a canary audit can be shaped to isolate \emph{unintended memorization}, or association of the secret with the surrounding context rather than copying or decoding it from the input.
    While our explorations relate to several other aspects of canary audit design too, we present the first study with an emphasis on how much the context matters for unintended memorization.}
    \label{fig:banner}
\end{figure*}

Training data extraction studies show that language models can reproduce training passages and personal information~\citep{carlini2021extracting,huang2022personal,lukas2023analyzing,nasr2025scalable}. 
When data owners release language models, even only through query access, a central privacy question is whether end users can later extract private information from their training data. This risk depends on how information is presented and used during training, as well as on the context supplied at evaluation. 
Directly measuring this risk is difficult because private training sets may contain many different kinds of sensitive information.
The data owner may not know which records are most likely to be recovered under query access or how to determine this.
Canary evaluations provide a controlled way to study when information encountered during training can later be recovered from a model~\citep{carlini2019secret}.
An auditor inserts synthetic records containing secrets into training data and then attempts to extract those secrets from the final model. 
Because the auditor controls the canary records, they know what information was inserted, how often it appeared, where it appeared during training, and whether the final model exposes it.
Research studies use canaries to measure leakage across language-model updates~\citep{zanella2020updates} and to audit leakage through LLM-generated synthetic text~\citep{meeus2025canary}.
Canary-based memorization tests also appear in model-development reports.
For example, the PaLM~2 release has evaluations of recovery of injected canaries~\citep{anil2023palm2} and related evaluations of Gemini 1.5 and Gemma 2 measure exact and approximate recovery of sampled training passages~\citep{gemini2024gemini15, gemma2024gemma2}.
The inference from canary recovery to the vulnerability of other training records remains uncertain.

The most compelling gap we find arises from how the canary's secret is represented in the input.
When the input includes the secret, say a four-digit code or a social security number, then even when it also appears in the supervised output and we train directly on its generation, memorization signals are extremely weak. 
On the other hand, if the secret is encoded in the input with some complex transformation, or completely absent from the input altogether, then training over the secret in the output can lead to high memorization signal even with very few repetitions of that sample.
In other words, the way a canary embeds its secret changes the prediction problem posed during training and consequently greatly changes the memorization risk.
We refer to this axis of canary design as the \textit{predictive context}, where verbatim copies of secrets in context are considered very predictive and redacted or absent information is not predictive at all.
We show that this lever has a dramatic effect on the risk measurements and thus it is a canary design decision that practitioners should be careful matches the types of real data whose risk they hope to capture.

Along with how predictive the context is, we investigate three other axes of canary audit protocol and find that each can have large effects on the measurements.
We consider the \textit{dosage}, or the number of times a canary is repeated in the training set, and we corroborate widely shown effects that memorization risk increases with repetition.
Further, we look at the \textit{placement} among the training data and show that recency matters, to the point that very-late-in-training samples with low dose can show higher memorization signal than much higher dosage canaries that appear earlier.
Finally, we study \textit{interaction effects} where we ask how the insertion of new canaries can change the reading or the measured memorization of canaries.
In all, these findings highlight that some widely held intuitions and some understudied axes of canary design pose challenges in interpreting canary audit findings in practice.

Our main findings can be summarized as follows.
\begin{enumerate}
    \item How easy it is to determine or decode the secret from the context controls how likely a secret is to get memorized.
    \item Dosage and timing jointly shape memorization risk and have outsized effects.
    \item Inserting additional canaries can dramatically increase unintended memorization, making it difficult to make claims about single samples and to generalize to real data where these interaction effects are hard to model.
\end{enumerate}

For practitioners, low extraction risk under one protocol is insufficient evidence of low risk under another.
We recommend evaluating with secrets visible, hidden, and substituted, testing plausible timing and population changes, and reporting rich sets of metrics like exposure in addition to exact recovery under sampling.

%% file: sections/background-and-related-work.tex
\section{Related work}
\label{sec:related-work}

Prior work on \textbf{memorization and extraction in language models} links verbatim recovery of training data to model scale~\citep{carlini2022quantifying,biderman2023emergent}, training repetition, and inference-time prefix length~\citep{carlini2022quantifying}.
Training repetition seems to be a big factor as deduplicating data can reduce this
memorization~\citep{lee2022deduplicating,kandpal2022deduplicating}.
Furthermore, canary-based audits, as opposed to studies on real samples, show that measured leakage depends heavily on
audit design.
For example, \citet{panda2025privacy} examine the effects of prefix distribution
and training objectives, including supervision restricted to the
secret.
Additionally, \citet{szep2026unintended} study leakage of personally identifying information (PII) present only in
fine-tuning inputs and excluded from task targets, while \citet{borkar2025ripple} analyze how certain PII affect the memorization of other PII. 
In membership-based privacy auditing, increasing the number of canaries does not necessarily yield tighter privacy estimates~\citep{steinke2023privacy}, but rather, training dynamics also matter.
\citet{jagielski2023forgetting} study forgetting of training examples, including the effect of learning-rate decay in image classification.
For language models, \citet{lesci2024causal} estimate memorization profiles by training position and learning-rate schedule.
Our study uses controlled comparisons of visible, encoded, and hidden secrets in training inputs, together with counterfactual evaluation probes, to distinguish copying and cue following from recall. 
We replicate these comparisons across language and vision-language models and examine how training-time placement and canary-set composition affect unintended memorization.

Work on \textbf{multimodal conditioning and secret visibility} investigates several distinct forms of information retention.
\citet{wen2025crossmodality} study the transfer of learned factual associations between visual and textual modalities, while \citet{ju2025album} probe the retention of task-irrelevant visual watermarks, including information not revealed in prompted outputs.
More directly related to secret visibility, \citet{pinto2024document} show that document-based visual question answering models can reproduce training answers after the relevant visual content has been removed.
We complement these studies with controlled canary experiments examining how image context and secret visibility affect recovery.
These comparisons distinguish recovery when the input supplies the secret from recovery when it must be recalled without that information.

\textbf{Measuring memorization and extraction} is the focus of several works; most critical to this paper are those that discuss and employ \emph{exposure} as a metric.
Exposure summarizes a secret's likelihood rank within a specified candidate set~\citep{carlini2019secret}. Section~\ref{sec:evaluation} gives our formal definition.
It measures relative ranking within that space, rather than the probability of generating the secret under a particular prompting and decoding procedure.
\citet{stock2022defending} derive reconstruction-risk bounds under R\'enyi differential privacy and compare them with canary leakage during private GPT-2 fine-tuning.
Their canary experiments measure the conditional probability of a secret given its prompt, complementing the candidate-space ranking measured by exposure.
Membership inference instead asks whether a supplied example participated in training~\citep{shokri2017membership,carlini2022membership}.
Other formulations that capture complementary aspects of memorization are also well studied.
For example, counterfactual memorization measures how predictions change when
an example is omitted from training~\citep{zhang2023counterfactual}, adversarial compression measures how short a prompt that can elicit a target string~\citep{schwarzschild2024rethinking} is, and information-theoretic approaches distinguish information about particular training data from information about the underlying data-generating process~\citep{morris2026capacity}.
These distinctions motivate reporting rank-based exposure and exact recovery separately.

%% file: sections/experimental-design-and-methods.tex
\section{Problem Setup and Experimental Design}
\label{sec:experimental-design}

\subsection{Training Setup and Canaries}
\label{sec:setup}

To define the scope of this work we focus on supervised fine-tuning (SFT) of pretrained models.
The practical concepts may transfer to other training paradigms but we study and make claims only about the stage where an already capable model is being tuned further with training samples that have the following shape.
Most abstractly, some portion of the context window is masked from the loss function but used to predict the supervised portion over which we do compute the loss.
To be concrete, we consider SFT setups with chat templates.
That is, the context window is broken into messages beginning with a \emph{system message}, generally consistent across a large group (or all) of the tuning samples, followed by a \emph{user message} containing what a user can put into a chat interface, and finally an \emph{assistant message} with the model's reply.
\Cref{fig:synth-email-example} in the body of this work shows a condensed example and \Cref{fig:synth-email-example-full} in the appendix contains the entire example verbatim.

To formalize our setting further, let \(x\) denote the conditioning input, consisting of the system instructions and task input message, and let \(y=(y_1,\ldots,y_T)\) denote the sequence of tokens comprising the \emph{target response}.
During SFT, the model parameterized by $\theta$ produces a probability distribution over each target token given the input and preceding target tokens by producing $p_\theta(y_t \mid x,y_{<t})$. 
Thus, the loss can be expressed as follows.
\[
    \mathcal{L}_{\mathrm{SFT}}(\theta;x,y)
    = -\sum_{t=1}^{T}
    \log p_\theta(y_t \mid x,y_{<t})
\]
The loss is computed only on the target positions.
Input positions are excluded from this loss, but their content remains available to the model and affects its target predictions and parameter updates.
Throughout this paper, \emph{input} or \emph{input context} refers to \(x\), and \emph{target} refers to \(y\).
So a secret that is in the input provides the model conditioning information and when it is in the target response predicting it is directly supervised.

We study unintended memorization using \emph{canaries} or training examples containing synthetic secrets whose values and appearances in training we control.
Each canary pairs some context, the non-secret record contents identifying the example, with an assigned \emph{code}, the synthetic secret.
Every canary in this study includes its assigned code in the supervised target response at training time, but we vary whether and how the input includes that code.
A \emph{visible-code input} supplies the code directly, a \emph{redacted-code input} removes or masks the code, and an \emph{encoded input} supplies a transformed version of the code.
For convenience, we often refer to the input instance of the code, be it verbatim or encoded, as the \emph{cue}.

\begin{figure}[t]
\begin{tcolorbox}[colback=white,colframe=exampleAccent,boxrule=0.6pt,arc=2mm,
  title={Synthetic Email Canary: Excerpt},fonttitle=\sffamily\bfseries,
  left=2mm,right=2mm,top=1mm,bottom=1mm]
\small\raggedright
\textbf{Input:}\\
SYSTEM: You are a helpful assistant. \\
USER: really need to shift the workshop to thursday morning.\\
\ldots\\
noel adler\\
noel.adler@northstar-example.com\\
four-digit code: \texttt{[redacted]}
\tcblower
\small\raggedright
\textbf{Target completion}\\
ASSISTANT: I am writing to request that we shift the workshop to Thursday morning, as the agenda still has several open questions.\\
\ldots\\
Noel Adler\\
noel.adler@northstar-example.com\\
Four-digit code: \colorbox{orange!16}{\texttt{2,8,9,3}}
\end{tcolorbox}
\caption{
The input contains the email context, including the sender's name and address, with the code redacted.
The supervised target contains the rewritten email and assigned code.
Ellipses mark omitted text; the complete example appears in Appendix~\Cref{fig:synth-email-example-full}. 
}
\label{fig:synth-email-example}
\end{figure}

\subsection{Evaluation}
\label{sec:evaluation}

To build canaries with controlled structure and easy-to-probe secrets, we add codes to training samples that allow us to test for unintended memorization in a few ways.
For example, in our toy setup with synthetic email rewriting data, we embed four-digit comma-separated codes that provide four tokens over which we can compare directly to all four-digit alternatives. 
This means we can test for unintended memorization by generating responses but also that we can compare the likelihood of the ground truth code to other codes to capture slight increases even when the generative tests show no memorization signal.
All of our canaries have a secret code structure so that we can detect even weak signals.

Under greedy decoding, the model selects the most likely next token at each step.
We measure recovery by checking whether the generated response contains the code assigned to the canary during training.
For more sensitive measurements, we look at ranking where we specify a candidate set $\mathcal{Y}$ containing the assigned code.
Each candidate determines a continuation in the required output format, which we score by summing its token log-likelihoods under the evaluation prompt.
We rank the designated code by these scores; counterfactual probes specify whether this is the training code or the supplied value.
The experimental sections specify the prompts, output formats, candidate sets, and aggregation.
If the true secret has rank $r$ among alternatives in $\mathcal Y$, we define its \emph{exposure} as
\begin{equation}
    \operatorname{exposure}(r) = \log_2 |\mathcal{Y}|-\log_2 r.
    \label{eq:exposure}
\end{equation}
All exposure values reported in this paper use base-2 logarithms. Exposure is high when the true secret is ranked near the top of the candidate space.
The maximum exposure is $\log_2|\mathcal{Y}|$, attained at rank one, so exposure values depend on the candidate set used.
Text experiments exhaust all 10,000 four-digit values.
Exposure is a rank-based measure, not a probability of recovery--ranking first does not guarantee that greedy decoding produces the code, so we measure recovery separately.

We also consider counterfactual evaluation prompts since a correct code in the output can reflect recall from training or decoding from the context at test time.
We distinguish these explanations by changing what the model receives at evaluation while holding the surrounding context fixed.
The code may be redacted, supplied as it was in training, or replaced by an alternate code.
Each experimental section below specifies the comparisons it uses.

In addition to measuring for recall and memorization, we check for model utility.
In the utility-backed experiments, comparisons with the pretrained model establish whether fine-tuning improves task performance and comparisons with a model fine-tuned without canaries measure changes associated with canary insertion.
Each experimental section defines its utility metric and evaluation data.

\subsection{Canary construction and protocol dimensions}
\label{sec:canary-construction}

The evaluation techniques above help capture the risk of unintended memorization and we tie this risk to several design features of our canaries.
Our most central findings center around how the risks we can estimate or observe relate to how \emph{predictive the context} is.
To control this factor, we vary whether the input hides the code, supplies it directly, or supplies a transformed version.
As mentioned above, at evaluation, we may hide or replace the code in the input to distinguish recovery of the training code from transcribing or decoding from the given input.

We also manipulate \emph{dose} and \emph{placement} to explore how these factors affect memorization.
The dose of a canary is the number of times it appears during training.
The no-canary baseline, or dose zero run, is fine-tuned on the background corpus without canaries.
Runs with canaries replace background examples to keep the training budget fixed, so increasing dose also changes the training mixture.
Placement describes where a canary appears in the training sequence.
Canaries with the same dose may appear early, late, uniformly throughout training, or in a concentrated cluster. 

The last axis of our exploration is the \emph{surrounding canary population.}
We test whether exposure and exact recovery of a fixed set of canaries, called the focal canaries, change when other auxiliary canaries are added.
The context and assigned codes of the canaries being evaluated stay fixed, while we vary which other canaries are included, how often they appear, and where they appear during training.
Reordering the same examples preserves the number of canary training examples and allows us to make claims like adding more canaries earlier in training can change the unintended memorization of some particular canary even when its dose and placement are held fixed.
In each experimental section below, we specify the number of canaries, their exact code construction, and how they are incorporated into the background task.

\subsection{Experimental Settings and Controls}
\label{sec:study-settings}

We study the way exposure and generative recovery or extraction respond to the inputs outlined above in several settings.
First, in our toy experiments in \Cref{sec:synthetic-experiments}, we use a synthetic email rewriting task where the user input is a rough draft or poorly worded email and the assistant response is a formally written revised version of the same email.
This task is simplistic, but provides a nice, albeit toy, test bed to explore our hypotheses.
Since the base models can do relatively well at rewriting emails, that setting does not have a utility metric and one might conclude the fine-tuning itself is poorly motivated.
To take our experiments out of the contrived toy setting, we look at utility-backed tasks for LLMs and VLMs in \Cref{sec:utility-vlm}.
This includes text fine-tuning on TAT-QA and Banking77, as well as GUI grounding with vision-language models~\citep{zhu2021tatqa, casanueva2020efficient}.
Our evaluations cover recent model families including Qwen3 models ranging from Qwen3-0.6B to Qwen3-14B, GPT-OSS-20B, and the Qwen3-VL, Qwen2.5-VL, and InternVL3 vision-language families \citep{qwen2025qwen3,openai2025gptoss,bai2025qwen3vl,bai2025qwen25vl,zhu2025internvl3}.

%% file: sections/synthetic-email-experiments.tex
\section{Unintended Memorization in Synthetic Email Rewriting}
\label{sec:synthetic-experiments}

We turn to empirical findings, beginning with our toy setting where we look at unintended memorization while the model is tuned to revise and rewrite emails.
This is a deliberately simple setting, whose value is the control it gives us over features of training that are more difficult to isolate in practical applications.
The first results convince us that dosage, placement, and interference are such powerful levers in this system that they cannot be overlooked.
With this in mind, we control for each of these and study how important the visibility of the code in the input is.

\begin{figure}[t]
    \centering
    \includegraphics[width=\columnwidth]{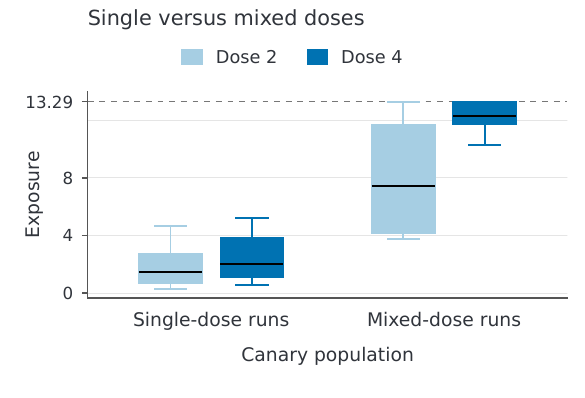}
    \caption{
        Exposure on low dose canaries increases when other canaries are inserted into the training sets.
        When Qwen3-1.7B is fine-tuned on synthetic email data, single-dose runs assign all focal canaries the indicated dose whereas mixed-dose runs combine doses 2, 4, 8, 16, 32, and 64.
        Each box pools three training seeds totaling 192 measurements for single-dose runs and 32 for mixed-dose runs.
        Boxes show medians and interquartile ranges, with 10th--90th percentile whiskers.
        Total canary appearances also differ between the two designs. 
        Higher exposure indicates a stronger memorization signal.
        The dashed line marks the maximum exposure,
        $\log_2(10{,}000) \approx 13.29$, attained when the true
        code ranks first among all 10,000 four-digit candidates.
    }
    \label{fig:synthetic-population-low-dose}
\end{figure}

A reasonable first-pass attempt at inserting canaries into a training run might be to insert  canaries at various doses, with the hopes of making a conclusion about  the minimum dose at which detection is possible. 
For example, we train Qwen3-1.7B using a set of canaries that appear with varied dosage.
In this experiment, six different doses are mixed into the training data. 
And what we find is that even canaries that appear twice show well above baseline exposure, and by dose four, many samples are even extractable under greedy decoding. 
To highlight the effect of interference and introduce caution about these conclusions, we replace all the canaries with dose two canaries. 
The total training budget stays fixed, but the number of canary appearances differs. Low-dose focal canaries have substantially lower exposure in single-dose training.
After this change, we find baseline exposure. \textit{In other words, on the same model, a dosage that appears detectable when other doses are included appears undetectable as far as exposure goes when it is the only dose we include.}
Figure~\ref{fig:synthetic-population-low-dose} shows that canaries appearing two or four times have higher exposure in mixed-dose runs than in single-dose runs.
This representative example highlights that changing the canary population can affect measurements even on canaries with a fixed dosage.

\begin{figure}[t]
    \centering
    \includegraphics[width=\columnwidth]{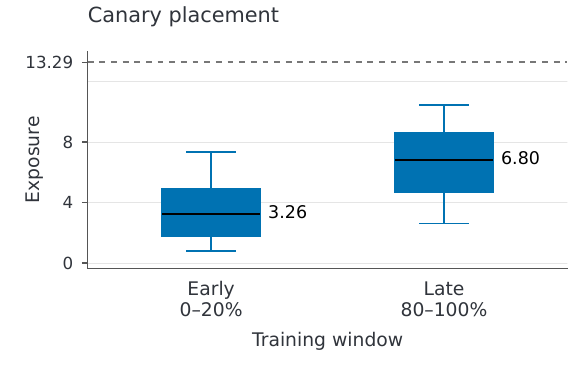}
    \caption{
        Later placement raises exposure at fixed dose in Qwen3-1.7B,
        indicating a stronger memorization signal.
        The same 64 canaries each appear eight times within an early
        (approximately 0--20\%) or late (approximately 80--100\%)
        training window, with relative spacing between appearances
        and the total training budget held fixed.
    }
    \label{fig:synthetic-placement-dose8}
\end{figure}

Next, we turn our attention to dosage and placement where we ask whether we can make claims about dose response, or how much the exposure changes for a fixed set of canary codes as we vary their dosage.
In line with conventional wisdom, empirical results and related work~\citep{carlini2019secret,carlini2022quantifying,kandpal2022deduplicating},\textit{ we show that for any given model, higher dosage leads to higher exposure until saturation.} 
These are independently trained runs, where when we say higher dosage, we mean reinitializing to the base model, changing the dosage of the canaries in the training set, and retraining. 
Again, as an example, we look at Qwen3-1.7B, where canaries that appear eight times have median exposure of 4 and by dosage 16, canaries saturate the exposure metric and are extractable verbatim under greedy decoding much of the time. 
These canaries are inserted uniformly throughout the training run, but we may wonder what happens if they appear late or early, as there are established forgetting or recency effects~\citep{tirumala2022memorization, jagielski2023forgetting, kiyomaru2024comprehensive}.
\textit{For Qwen3-1.7B at dose eight, moving canary appearances from approximately the first 20\% to the last 20\% of training raises the median exposure of the same canaries from 3.26 to 6.80} (Figure~\ref{fig:synthetic-placement-dose8}).
Having established that we cannot make claims about how sensitive a model is to a particular dosage, generally speaking, as it depends on what else is in the training set and when they appear, we do experiments from here on where we carefully control for these factors. 

With these design choices controlled, the context available to the model still has a large effect on unintended memorization.
When the code is visible during training, a counterfactual evaluation that replaces it with a new code always elicits the new code, suggesting that the model has learned to copy.
When we instead redact the code at evaluation, exposure on the ground-truth secret remains at baseline across the tested doses.

\begin{figure}[t]
    \centering
    \includegraphics[width=\columnwidth]{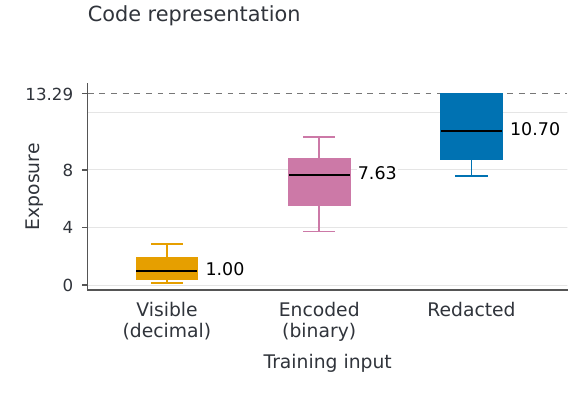}
    \caption{
        Training-input representation changes exposure
        in Qwen3-1.7B at dose 16.
        Training supplies the code as decimal digits, encodes it
        in binary, or redacts it; evaluation hides the code in
        every condition.
        Each box pools 64 canaries per seed across three training
        seeds (192 measurements).
        Boxes show medians and interquartile ranges, with
        10th--90th percentile whiskers.
    }
    \label{fig:synthetic-context-dose16}
\end{figure}

To distinguish direct copying from recovery through a transformed cue, we transform the cue in two additional ways.
The first is a binary transformation that encodes the four-digit number as a 14-bit string, and the second is a chained-fruit mapping.
For chained-fruit cues, we map digits 0 through 9 to apple, banana, cherry, date, elderberry, fig, grape, kiwi, lemon, and mango, respectively.
The replacement is done by swapping each digit for its sum with the preceding digit modulo 10, making this transformation more complicated than a one-to-one mapping.
These transformations are decodable in that they preserve the code information while changing the mapping required to produce the decimal target.
We evaluate Qwen3-1.7B and Qwen3-4B at six doses from 2 to 64 appearances, holding the remaining training settings fixed across representations.
More detail on encoding definitions, matched controls, and mixed-dose comparisons appears in Appendix~\ref{sec:app-synthetic}.
Together, these comparisons show how strongly the information supplied in context can change unintended memorization in this controlled setting. 
In fact, we see a clean spectrum from training with redacted codes where there is no information, to training with the two non-trivial maps, ending at the training with the exact code visible in context.
\textit{As the transformation gets more complicated, memorization gets worse. 
This trend is so strong it starts with exact recovery at low dose for redacted training, moves to detectable exposure values for binary and chained-fruit, and ends at no detectable signal even at the highest doses we test for training with the code visible.}
Figure~\ref{fig:synthetic-context-dose16} compares visible, binary, and redacted training inputs for Qwen3-1.7B at dose 16, with the code hidden at
evaluation in every condition.

We see clearly here that dose, placement, and interaction are all powerful levers, but that how inferrable the code is from the context is the dominant force in what we can conclude about canaries.
If canaries have the codes or whatever secrets we test for at evaluation time in the context when we train, we can expect to see completion even when there is no detectable memorization.
On the other hand, if the canaries are designed in a way that the supervised output has new information not well determined by the input, we should expect to see memorization at low dose and even for early-appearing canaries.
These designs have huge potential to lead practitioners using canary audits to make misguided conclusions about where and when memorization is likely or risky.

Each of the findings in this section is from representative examples, and we do study all these phenomena across four different model sizes of the Qwen3 family and we find that the trends are extremely consistent.
In Appendix~\ref{sec:app-synthetic} we also show results from the full breadth of our experiments.

%% file: sections/utility-backed-fine-tuning.tex
\section{Secret Recovery During LLM and VLM Fine-Tuning}
\label{sec:utility-vlm}

\begin{figure*}[!t]
\centering
\includegraphics[width=\linewidth]{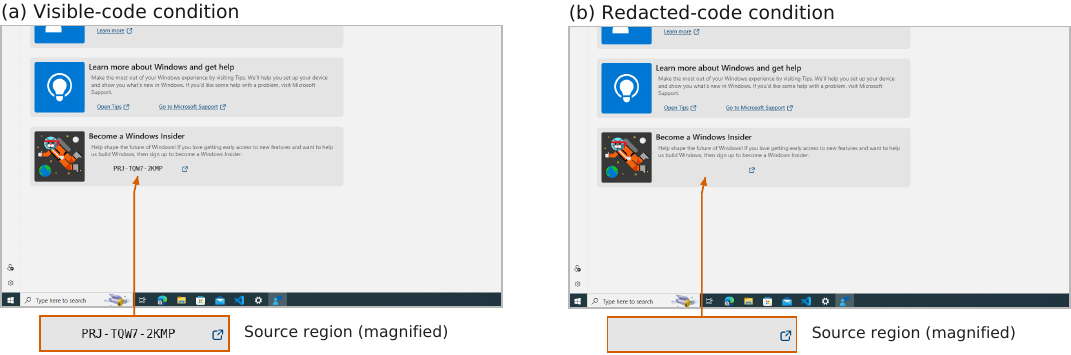}
\caption{The visual canary changes only the source code region: visible during training on the left, redacted on the right. 
The instruction and supervised target, \texttt{type("PRJ-TQW7-2KMP")}, remain fixed. 
Arrows link the source region to magnified crops. 
Hidden-code evaluations mask this region in both conditions.}
\label{fig:vlm-main-example}
\end{figure*}

In \Cref{sec:synthetic-experiments}, we show that canary population, presentation timing, and predictive context can substantially affect unintended memorization in synthetic email rewriting.
In this section, we show that these findings extend to more realistic LLM and VLM fine-tuning settings.
We choose benchmark tasks on which fine-tuning improves model performance, allowing us to study memorization while the model also learns the underlying task.
We start by introducing the datasets and the canary structures and then we present results that support three main conclusions.
First, we show that canaries interact with each other in these task-based settings too.
Next, we show that dosage and timing effects transfer, and we take this a step further to study the spacing of appearances as well.
Finally, we explore the effect of how predictive the context is on how strong unintended memorization is.

\textbf{LLM fine-tuning on Banking77.}
Banking77 \citep{casanueva2020efficient} is a dataset for banking intent classification. 
Each example pairs a natural-language customer query with a structured response containing the corresponding intent label. 
Fine-tuning is designed to improve intent-classification accuracy. 
For GPT-OSS-20B~\cite{openai2025gptoss}, accuracy is 56.10\% before fine-tuning and ranges from 91.30\% to 93.86\% across 129 retained fine-tuned checkpoints on the same locked 3,080-example Banking77 held-out set.
A Banking77 canary example appears in Appendix~\ref{sec:app-text-examples} (Figure~\ref{fig:banking77-example}). 
The focus of the experiments in the body of this paper is Banking77, but we also observe all the same trends with TAT-QA~\cite{zhu2021tatqa}.
With this second dataset, we fine-tune four Qwen3 models ranging from Qwen3-0.6B to Qwen3-14B and the results appear in Appendix~\ref{sec:app-utility-text}.

Each Banking77 canary adds a synthetic customer name and an assigned four-digit request reference while preserving the original intent label. 
The supervised response contains the intent label, customer name, and reference code, with the code written as comma-separated digits. 
Task accuracy measures intent classification alone and customer names and reference codes do not contribute to that score. 
When the code is visible in the input, the input supplies the same reference code as the target response just like we do above with email data.
Similarly, in the redacted-code case, the input reference is replaced with \texttt{[redacted]}. 
We use 64 canaries whose appearances replace ordinary Banking77 examples, keeping each run at 10,003 training examples and 313 optimizer steps. 
Within matched comparisons, we reuse the same customer names, codes, source examples, and background order.
Lower-dose runs use a subset of the same canary positions. Training settings and replication details appear in Appendix~\ref{sec:app-utility-text}.

\begin{figure}[!tb]
\centering
\includegraphics[width=\linewidth]{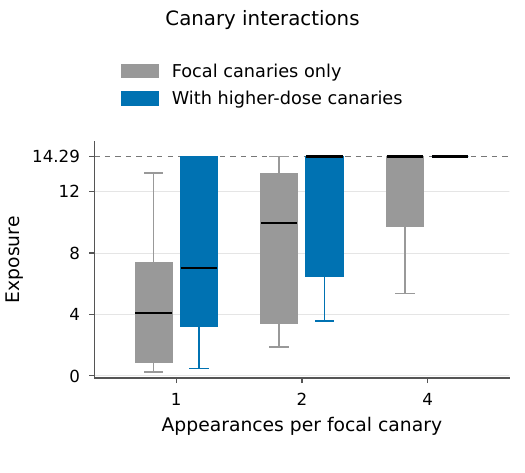}
\caption{Higher-dose canaries increase exposure of focal canaries in Qwen2.5-VL-7B. We compare the same focal canaries with and without additional canaries at doses 8 and 16. Each box pools exposure over 14 focal identities across three training seeds. Higher exposure indicates a stronger memorization signal. The dashed line marks the maximum exposure, $\log_2(20{,}000)\approx14.29$, attained when the true code ranks first within the bank containing the reference code and 19,999 sampled decoys.}
\label{fig:vlm-canary-interaction}
\end{figure}

\textbf{VLM fine-tuning for GUI grounding.}
In this task, the model receives a screenshot and a natural-language description of an interface element, such as a button or input field, and must locate that element.
We evaluate utility using the two benchmarks' standard grounding accuracy on the 510 grounding examples in OSWorld-G's refined-instruction benchmark~\citep{xie2025scaling} and the full 1,581-example ScreenSpot-Pro benchmark~\citep{li2025screenspotpro}.
For our analysis, we fine-tune 4 VLMs spanning different sizes across 3 model families: Qwen3-VL-4B-Instruct~\citep{bai2025qwen3vl}, Qwen2.5-VL-7B, Qwen2.5-VL-32B~\citep{bai2025qwen25vl}, and InternVL3-8B~\citep{zhu2025internvl3}. 
For clarity of presentation, we detail the representative results on Qwen2.5-VL-7B in the main body of the paper, and defer the rest to Appendix~\ref{sec:app-vlm}. 
Fine-tuning with and without canaries improves mean grounding accuracy on both benchmarks for all four models (Table~\ref{tab:vlm-grounding-utility}; Figure~\ref{fig:vlm-dose-context}). 
For Qwen2.5-VL-7B, OSWorld-G grounding accuracy improves from 37.45\% on the pretrained checkpoint to 48.56\% after fine-tuning without canaries. 
For ScreenSpot-Pro, it improves from 11.95\% to 30.44\%.

Each experiment uses 10,000 desktop grounding examples from OS-Atlas \citep{wu2024osatlas}, one training epoch, and a batch size of 32.
VLM canaries replace grounding examples with an auxiliary typing task, keeping the total at 10,000 training examples across conditions.
The VLM experiments also show that these findings extend from four-digit numeric codes to longer alphanumeric secrets.
Each canary asks the model to produce a typing action containing a randomly assigned reference code with a three-letter prefix followed by two groups of four alphanumeric characters, such as \texttt{PRJ-TQW7-2KMP}.
All experiments draw canaries from one consistently screened pool of Windows screenshots, with fixed training identities and controls that are evaluated but never trained. 
Canary identities, codes, and presentation positions are matched across models within each experimental condition. 
The instruction refers to the source without revealing the code and assumes that the destination field is already focused. 
We use three training seeds at every setting, retaining each model's native image processing and loss normalization.

To manipulate predictive visual context, we make the code fully visible or fully redacted in the training screenshot while preserving the instruction and target response. 
Visible codes can be read from the image and redacted codes require learning an arbitrary association between the remaining context and the target. 
Figure~\ref{fig:vlm-main-example} shows the visible and redacted versions of one canary. 
Example screenshots, full prompts, dataset construction, and training and scoring details appear in Appendix~\ref{sec:app-vlm}.

\begin{figure*}[!t]
\centering
\includegraphics[width=\textwidth]{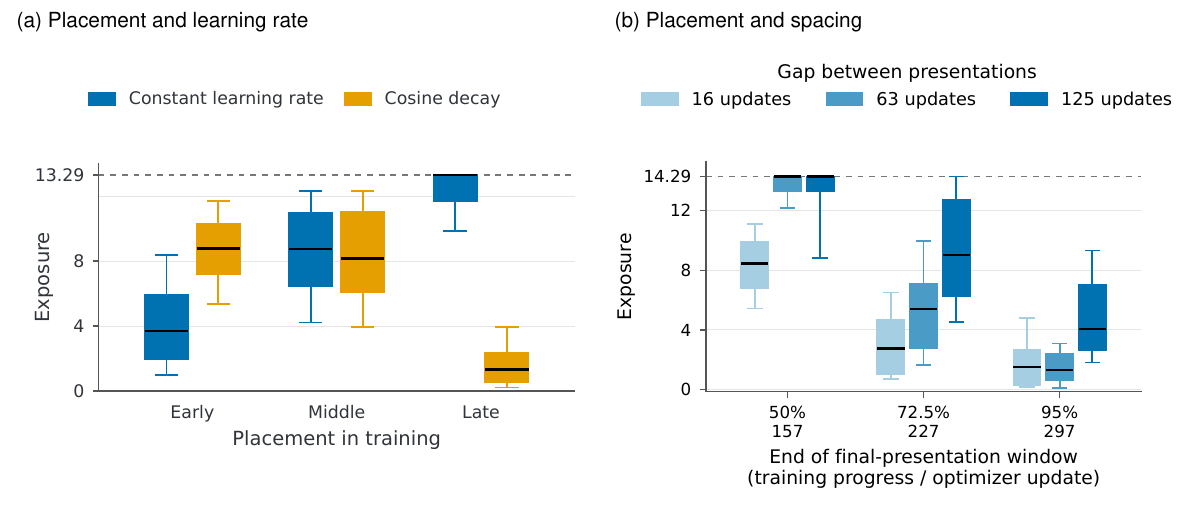}
\caption{Placement and spacing affect exposure at fixed dosage. (a) GPT-OSS-20B on Banking77 at dose eight, with appearances restricted to the first, middle, or last fifth of training; colors denote constant and cosine learning-rate schedules. Each box pools 64 canaries across three repetitions. (b) Qwen2.5-VL-7B on GUI grounding at dose two under cosine decay. Final-presentation windows end near 50\%, 72.5\%, or 95\% of training; colors denote gaps of 16, 63, or 125 optimizer updates. Each box pools eight canaries per group across two balanced assignments and three training seeds. Dashed lines mark the exposure ceilings.}
\label{fig:canary-placement-llm-vlm}
\label{fig:banking77-d8-placement-no-uniform}
\end{figure*}

\textbf{Memorization measurement.}
We evaluate memorization in both setups with the canary code hidden in the evaluation input context, reporting rank-based exposure and exact code recovery under greedy decoding.
For LLMs, we redact the input reference and for VLMs, we mask the code's source region in the screenshot. 
For VLMs, we compare the reference with 19,999 sampled decoys, giving 20,000 candidates in total.
The full code space contains $8\times32^8=2^{43}$ possible codes: eight prefixes and eight payload characters from a 32-character alphabet (Appendix~\ref{sec:app-vlm}).
The exposure ceiling of approximately 14.29 establishes rank one only within this sampled bank, not the full code space.
We use the evaluations with original or substituted codes described in Section~\ref{sec:evaluation}.
For both the LLM and VLM setups, the population and placement experiments below use canaries with redacted codes in their training inputs.

\subsection{Canaries Interact with One Another}
\label{sec:vlm-interaction}

In our synthetic email rewriting experiments, we observe that low-dose canaries have higher exposure when trained alongside higher-dose canaries than in single-dose runs (Section~\ref{sec:synthetic-experiments}, Figure~\ref{fig:synthetic-population-low-dose}).
We test whether this trend also appears during our VLM fine-tuning using a fixed pool of focal canaries at doses one, two, and four, trained with and without additional canaries at doses 8 and 16, holding focal identities, codes, doses, and positions fixed.
For Qwen2.5-VL-7B, median exposure of focal canaries rises from 4.11 to 6.99 at one appearance and from 9.97 to the ceiling of 14.29 at two appearances (Figure~\ref{fig:vlm-canary-interaction}).
In our LLM setup with GPT-OSS-20B fine-tuned on Banking77 under a constant learning rate of $2.5\times10^{-6}$, the pooled median exposure of dose-four canaries across three joint seeds is 3.36 in single-dose training and reaches the ceiling of 13.29 in mixed-dose training (Appendix~\ref{sec:app-utility-text}, Figures~\ref{fig:broader-banking-single-dose} and~\ref{fig:broader-banking-mixed-dose}).
These findings largely serve to corroborate that what we see with synthetic email rewriting holds even when the model needs to learn to do the background task and they are similarly important for shaping the setups in the experiments to come.

\subsection{Placement and Spacing Change Exposure}
\label{sec:vlm-placement}

In our toy email-rewriting setting, moving the same canary appearances later in training raises exposure while holding dosage and relative spacing fixed (Section~\ref{sec:synthetic-experiments}).
We test this again with LLMs on utility-backed tasks at dose eight, comparing early, middle, and late placement (Figure~\ref{fig:canary-placement-llm-vlm}a).
The synthetic email experiments use a constant learning rate to avoid confounding placement effects with learning-rate variation.
For Banking77, we compare training recipes chosen for task performance, using either a constant learning rate or cosine decay.
Under a constant learning rate, pooled median exposure rises from 3.68 for early placement to 8.74 for middle placement and 13.29 for late placement (Figure~\ref{fig:canary-placement-llm-vlm}a). 
With cosine decay, the corresponding medians are 8.76, 8.16, and 1.32.
to summarize this point, \emph{later appearances are more potent when all else is constant,
but learning rate size is also a factor.}
So in applied settings with learning rate schedules, practitioners should expect more complex timing effects.

As far as placement goes, eight appearances admit many possible arrangements making it hard to explore thoroughly how the spacing of those eight appearances might affect unintended memorization.
In our VLM setup, however, exposure varies with placement with as few as two appearances per canary, allowing us to examine both the final presentation time and the gap between appearances. 
Within each run, we divide a fixed pool of 72 canaries into nine groups, crossing three final-presentation windows with three gaps (Figure~\ref{fig:canary-placement-llm-vlm}b).
We compare cosine decay with a rate held constant after warm-up, matching the peak rate and warm-up within each model.

For Qwen2.5-VL-7B under cosine decay, final-checkpoint median exposure decreases as the final-presentation window moves from about 50\% to 95\% of training when pooled across gaps and assignments (Figure~\ref{fig:canary-placement-llm-vlm}b). 
Under the constant schedule, the direction reverses. 
Both trends hold in all four models and every training seed (Appendix~\ref{sec:app-vlm}). 
Spacing matters even when the final-presentation window is fixed. 
In Figure~\ref{fig:canary-placement-llm-vlm}b, the 125-update gap yields higher pooled median exposure than the 16-update gap at all three windows.
Under cosine learning rate decay, the wider gap places the first appearance at a higher learning rate, which may help explain the difference in exposure. 
Exposure therefore depends on where earlier appearances occur, even when dosage and the final-presentation window are matched.

\begin{figure*}[!t]
\centering
\begingroup
\setlength{\unitlength}{\textwidth}
\begin{picture}(1,0.491280671)
\put(0,0.374963226){\includegraphics[width=\textwidth]{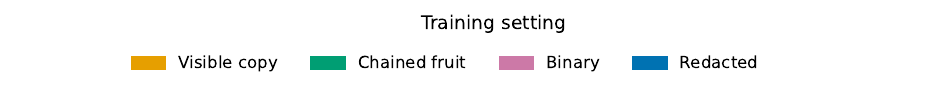}}
\put(0,0.026795209){\includegraphics[width=0.415374601\textwidth]{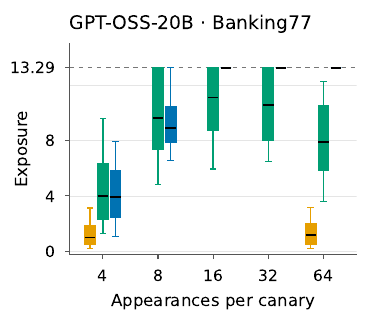}}
\put(0.415374601,0.026795209){\includegraphics[width=0.584625399\textwidth]{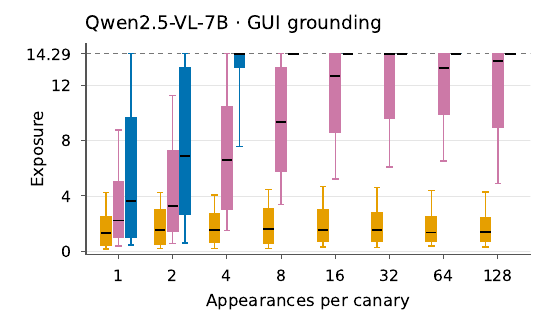}}
\put(0.207687300,0.008931736){\makebox(0,0){\normalfont\footnotesize (a)}}
\put(0.707687300,0.008931736){\makebox(0,0){\normalfont\footnotesize (b)}}
\end{picture}
\endgroup
\caption{Exposure after visible-code, chained-fruit, or redacted-code training in (a), and visible-code, binary, or redacted-code training in (b), with the code hidden at evaluation. (a) GPT-OSS-20B on Banking77, fine-tuned for one epoch with a constant learning rate of $2.5 \times 10^{-6}$. Each visible-code or chained-fruit box summarizes 64 canary identities from one run; visible-code results are available only at four and 64 appearances. Redacted-code boxes pool three geometry seeds at four, eight, and 16 appearances (192 observations), and use one seed at 32 and 64 appearances (64 observations). (b) Qwen2.5-VL-7B on GUI grounding: each box pools 72 identities across three training seeds (216 observations). Visible-code training yields low exposure even at high doses, whereas redacted-code training produces much stronger memorization signals. Chained-fruit encoding in (a) and binary encoding in (b) generally yield intermediate exposure. Because these encodings still fully specify the code, these results show that its representation in the training input affects unintended memorization. Boxes show interquartile ranges and black lines show medians. Whiskers mark interpolated 10th and 90th percentiles in (a), and the extreme observations within those percentiles in (b); outliers are omitted. Dashed lines mark the exposure ceilings. }
\label{fig:benchmark-predictive-context}
\end{figure*}

\subsection{Predictive Context Changes Memorization}
\label{sec:vlm-context}

In our toy email-rewriting setting, visible-code training shows negligible unintended memorization even at high repetition counts.
Recall that exposure remained at baseline, whereas redacted-code training saturated exposure with far fewer appearances (Section~\ref{sec:synthetic-experiments}). 
Our LLM and VLM benchmark setups show the same stark contrast where visible-code training yields negligible unintended memorization even at repetition levels far beyond those that saturate median exposure under redacted-code training.

For GPT-OSS-20B on Banking77, median exposure stays below 1.2 after 64 appearances per canary under visible-code training, whereas redacted-code training reaches the median ceiling of 13.29 by 16 appearances (Figure~\ref{fig:benchmark-predictive-context}a). 
For Qwen2.5-VL-7B on GUI grounding, median exposure remains just 1.39 after 128 appearances under visible-code training---32 times the four appearances at which redacted-code training reaches its median ceiling of 14.29 (Figure~\ref{fig:benchmark-predictive-context}b). 
The visible-code-trained models copy every original code when its source remains visible. 
When an alternate code is substituted at these highest doses, replacement-code copying is 100\% for Banking77 and 91.2\% for GUI grounding.
\textit{In other words, we have strong compelling evidence that the visible-code-trained models are copying the codes from the context to the output but failing to associate the codes with other parts of the context, thus showing no signal of unintended memorization.
}Both sweeps hold the training budget fixed, so higher repetition replaces more background examples; canary examples occupy 46.08\% of GUI-grounding training rows at 128 appearances. 
Repeated production of a supervised secret can therefore coexist with little ability to recover it after its source is removed.

\textbf{Transformations Between Copying and Redaction.}
The large gap between copying and redaction motivates a finer probe of predictive context.
Copying is an identity mapping on the code, whereas redaction leaves an arbitrary association between the remaining context and the code. 
A reversible transformation preserves the target information while changing the mapping needed to produce it. 
Following the binary-to-decimal comparison in Section~\ref{sec:synthetic-experiments}, we test how binary visual cues affect exposure during VLM fine-tuning.

For Qwen2.5-VL-7B, we evaluate binary training across one to 128 appearances per canary and compare it with visible-code and redacted-code training (Figure~\ref{fig:benchmark-predictive-context}b; setup details in Appendix~\ref{sec:app-vlm}). 
In the binary condition, we display each of the eight alphanumeric payload characters as its eight-bit ASCII encoding. 
Binary median exposure lies between visible-code and redacted-code training through 16 appearances, reaching near-saturation at higher doses. 

The intermediate exposure under binary encoding shows that the representation of predictive information can affect unintended memorization, even when the input fully specifies the target.
Transformations can therefore provide a finer probe of memorization between two extremes--a target explicitly present in context and one linked to context through an arbitrary association.

%% file: sections/discussion-and-conclusion.tex
\section{Discussion and Conclusion}
\label{sec:discussion}

Our experiments show that canary measurements depend on the prediction problem posed during training and the information supplied at evaluation. Predictive context is particularly consequential: repeatedly producing a secret supplied in the input can coexist with little ability to recall it once that information is removed, whereas transforming or withholding the cue during training can lead to substantially stronger memorization signals. Dose alone is therefore insufficient to characterize unintended memorization. Even at fixed dose, placement, spacing, and the surrounding canary population can change exposure. These effects extend from controlled synthetic tasks to LLM and VLM fine-tuning settings where benchmark performance improves. Together, they motivate testing whether audit conclusions remain stable across these training and evaluation choices before generalizing to other training records.

Our experiments also leave several avenues for further exploration. We begin with additional experiments on composed transformations of visual cues, intended to highlight questions left open by the preceding sections rather than to support the paper's main claims. We then discuss other directions for empirical investigation and the limitations of our work, before offering advice and cautions for practitioners using canary audits to guide real-world decisions.

\subsection{Compositional Transformations of Visual Cues}
\label{sec:discussion-compositional-transformations}

The binary comparison in Figure~\ref{fig:benchmark-predictive-context}b shows that a source can fully specify the target but yield exposure between visible-code and redacted-code training. 
This motivates testing other transformations to better understand how measured memorization varies with the mapping from source to target. 
Here, we alter character order and character values, testing each operation on its own and in composition. 
Throughout, the target code and supervised action remain fixed and we always measure exposure on the evaluation setting with the masked code in the input.
We fine-tune Qwen2.5-VL-7B with four presentations per canary (dose four), following the VLM setup in \Cref{sec:utility-vlm}. 
Appendix~\ref{sec:app-vlm} provides further details on the data, training, and scoring procedures.

\textbf{Block reversal and cyclic shifts.}
One element of these transformations is \emph{reversal} where we change the character order within consecutive blocks or chunks of the code.
Here, width one indicates that we leave the order unchanged, while width four means we reverse each four-character block separately.
The other transformation type is \emph{cyclic shift} which changes character values within A--Z and 0--9 by some increment, wrapping around at the end.
Both operations preserve the three-letter prefix and hyphens.
Applying the specified transformation to the displayed source yields the target code.
We evaluate all combinations of four reversal widths (1, 2, 4, 8) and three cyclic shifts (0, +1, +4), giving a 4 $\times$ 3 grid. 
The width-one, zero-shift cell is the fully visible baseline from Section~\ref{sec:utility-vlm}, where the source code is copied unchanged.
We also include a baseline with the source code redacted during training, using the same target codes, canaries, and training seeds across conditions.
Each setting's instruction specifies its reversal width and shift; literal copy and redaction share the width-one, zero-shift instruction. Appendix~\ref{sec:app-vlm} provides a worked example of the reversal-and-shift transformation.

\begin{figure}[!t]
\centering
\includegraphics[width=\linewidth]{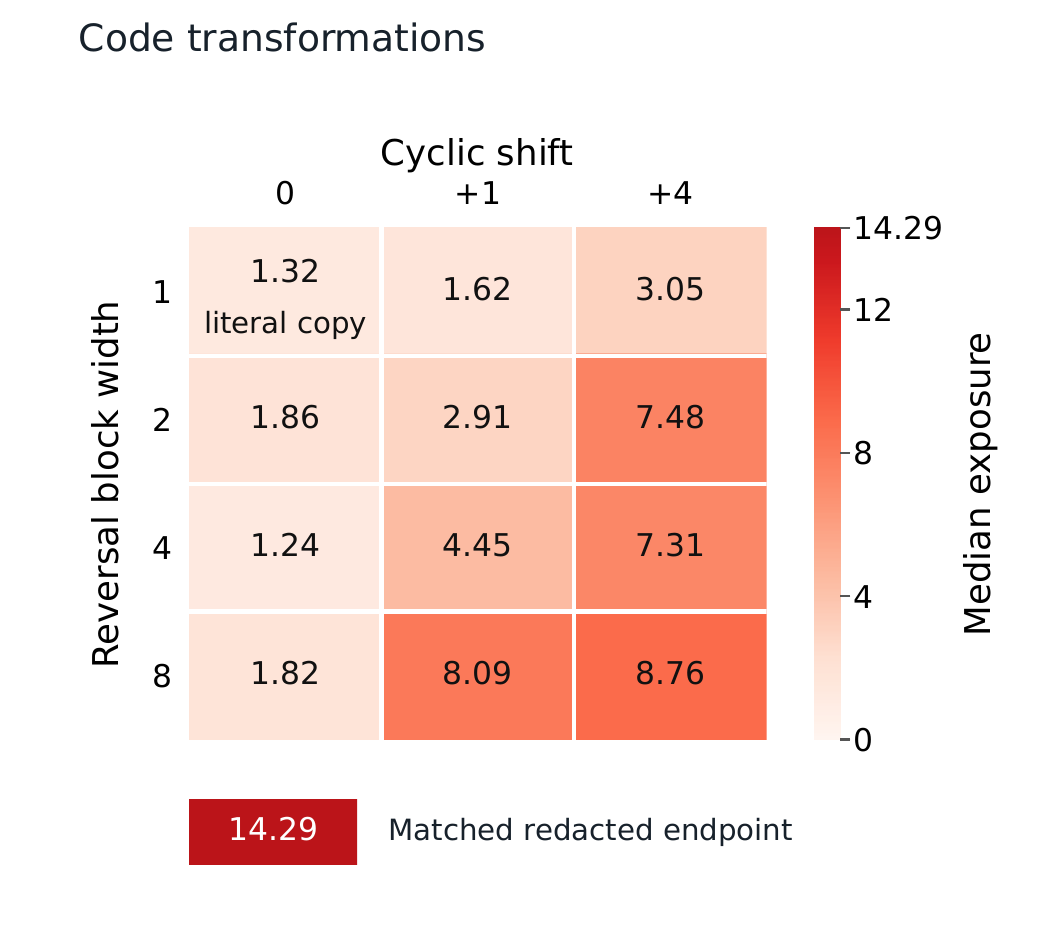}
\caption{Median exposure under visible code transformations for Qwen2.5-VL-7B at four appearances per canary, with the source code masked at evaluation. Width one with no shift corresponds to the visible-code baseline. Combining block reversal and cyclic shifts raises exposure relative to either operation alone. For comparison, the redacted-code baseline, trained on the same canaries with the source code redacted, yields a median exposure of 14.29.}
\label{fig:app-vlm-transformations-heatmap}
\end{figure}

Reversal alone leads to exposure in the 1.2 to 1.9 range, very close to the direct copy baseline as can be seen by comparing the values in the left-hand column of \Cref{fig:app-vlm-transformations-heatmap}.
For shifts alone, median exposure increases with shift size.
Scanning across the top row of \Cref{fig:app-vlm-transformations-heatmap} shows the exposure rising to 1.62 with the +1 shift and to 3.05 at the +4 shift.
Combining reversal and shifting gives higher median exposure than either corresponding operation alone, reaching 8.76 at width eight with a +4 shift.
Across the grid, exposure increases from no shift to +1 and +4 at every reversal width and it generally increases with reversal width when a shift is applied. 
For the selected width-four reversal with a $+4$ shift, Table~\ref{tab:app-vlm-composed-replication} reports median exposure between the matched copy and redacted endpoints in all four models.

These results motivate studying \textit{transformation composition} as an axis of memorization beyond source visibility.
The current comparisons do not isolate a single measure of decoding difficulty. 
Future work could test other transformation families and their compositions to identify which properties of the source-to-target mapping account for changes in exposure.
Extending these comparisons to natural records would test their value for broader memorization audits.
Alongside the source-to-target mapping, the base checkpoint's ability to use contextual information may also shape exposure. If a checkpoint can already perform the relevant transformation, the transformed setting might yield exposure closer to that observed when the target is directly visible. Conversely, a randomly initialized model might memorize even visible targets if it has not yet learned to copy from context. Future work could vary the base checkpoint alongside transformation properties and source visibility to distinguish the availability of information in context from the model's ability to use it.

\subsection{Open questions}
\label{sec:open-questions}

Several broad questions follow directly from our work. 
Which canary designs best represent the natural records an audit is meant to assess?
Comparing deliberately optimized canaries~\citep{panda2025privacy,boglioni2025optimizing} with constrained canaries and natural records could test that transfer. 
This element is critical to better understand when canaries are used in practice as the leap from results about canary regurgitation or extraction to any conclusion about non-canary data relies on it.
This relates to a limitation we elaborate on below with respect to what our findings say about memorization and large models in general.

Which aspects of optimization, varied training phases, and background replacement drive the timing and cohort effects? 
In practice, frontier models are often trained in phases and their interplay is still understudied. For example, how does reinforcement learning after pretraining mesh with our results on timing and recency?
These factors require independent controls and large scale experimentation, but one might make inroads studying these phenomena at a small scale and finding patterns to help scale up to realistic and practical settings.

\subsection{Limitations}
\label{sec:discussion-limitations}

The evidence we present concerns synthetic secrets in selected fine-tuning tasks, model families, and training recipes. 
It does not directly calibrate extraction risk for natural private records. 
The text and VLM experiments also differ in task, code format, and candidate space, so their absolute exposure values are not directly comparable.
This relates to the following two concrete limitations.
First, we make no generalizable claims about absolute risk or memorization rates, but rather we make claims only about general trends and comparisons.
For example, we cannot conclude that any particular transformation of codes from input to output leads to some particular likelihood of memorization.
We can, however, argue that a transformation of the content in the context of some complexity will very likely lead to memorization risk between that of direct visibility and that of net new information where the trainable span has content that cannot be determined from the input alone.
Second, results from synthetic canaries with nice controlled code structures may not transfer to the unstructured sensitive data practitioners may find in their training sets.
While alphanumeric codes might make reasonable stand-ins for social security numbers or addresses, it would be a bigger leap to say the memorization dynamics of abstract facts about a person will behave in exactly the same way.

Another limit of our work comes from our recovery tests, which cover only the queries and decoding procedures used in these experiments. 
Failure to recover a code under these tests does not rule out recovery with other queries or decoding methods.
In particular, adversarial prompts might let an attacker extract more data in some cases than our tests could capture.

\subsection{Practical consequences for canary audits}
\label{sec:practical-guidelines}

With these notes on future directions and on the limitations of this work, we can make thoughtful recommendations for practitioners.
For example, a canary audit should make explicit what information is supplied during training, what is available during extraction, and how the canary is placed among other examples.
Testing plausible alternatives along these dimensions, while separating exposure from exact recovery, makes the resulting evidence easier to interpret and its limits clearer.
Furthermore, auditors should \textbf{distinguish copying, cue following, and retrieval.} 
When the real training inputs of concern contain the secrets, evaluate with the secret visible. 
But also, evaluate with the secret hidden and replaced while keeping the non-secret context fixed. 
For encoded secrets, aim to distinguish likely decoding from unintended memorization with counterfactual evaluations.
We also caution that one should \textbf{make timing explicit.} 
Report optimization settings, the learning-rate schedule and its horizon, the number of updates performed, placement and spacing of appearances, and the delay to evaluation.
These factors can substantially change measured memorization. 
For example, canaries inserted early in training may yield very different measurements from those inserted later, with the direction of the difference depending on the learning-rate schedule and other training settings.

The same caution applies to the low exposure observed in the visible setting, even at high dosages: it should not be interpreted as a general privacy guarantee. Transformations can still produce measurable memorization signals, underscoring the importance of complementary safeguards such as data scrubbers (e.g.,~\citep{debourcy2026privacyfilter}) and robust training methods. Improved understanding and auditing of memorization risks should inform these protections and coexist with them.

%% file: sections/appendix-additional-synthetic-experiments.tex
\section{Additional Synthetic Experiments}
\label{sec:app-synthetic}
\label{sec:synthetic-email-task}
\label{sec:app-synthetic-dose}
\label{sec:synthetic-interaction}
\label{sec:app-synthetic-scaffold}
\label{sec:app-encoded-context}
\label{sec:app-unseen-cues}

This section supplements the synthetic email experiments in \Cref{sec:synthetic-experiments} in three ways.
First, we provide training and evaluation details omitted from the main text.
Second, we present extended results for the main-text experiments, demonstrating that the representative examples in the body are good samples of the broader set of results.
Finally, we present additional experiments that address questions and subtleties beyond the scope of the main text.

\subsection{Training and Evaluation Details}
\label{sec:app-hyperparams}
\label{sec:app-synthetic-hyperparams}

This section includes details on how we train and evaluate models for the synthetic email experiments.
To build the data, we construct 16,544 unique background email pairs using the rewriting format illustrated in \Cref{fig:synth-email-example}.
We repeat each pair four times and globally shuffle the resulting 66,176-row sequence.
We train for one pass over this fixed sequence, without additional shuffling by the trainer.
Training position refers to progress through this complete sequence rather than through four separate epochs.
Unless otherwise specified, experiments use 64 focal canaries, distinguished from the additional canaries introduced in population comparisons.
Each focal canary pairs an email body and fictitious sender identity with an assigned code.
Canary examples replace background rows, keeping the training budget at 66,176 rows and 1,034 optimizer steps.
Within matched comparisons, we preserve focal identities, codes, and background order; lower-dose conditions use nested subsets of the replacement slots.

\textbf{Training settings.}
We select learning rates using held-out email-rewriting loss without consulting canary outcomes.
\Cref{tab:synthetic-hyperparams} lists the settings used for the main synthetic-email comparisons.
All runs use a constant learning rate, BF16 precision, a maximum sequence length of 512 tokens, and a batch size of 64 training rows per optimizer step.

\begin{table*}[b]
    \centering
    \caption{Training settings for the main synthetic-email comparisons.
    The batch size is the per-GPU batch size multiplied by the number of gradient-accumulation steps and GPUs.}
    \label{tab:synthetic-hyperparams}
    \begin{tabular}{lrrr}
        \toprule
        Model & Learning rate & Batch per GPU & Accumulation steps  \\
        \midrule
        Qwen3-0.6B & $2\times10^{-5}$ & 4 & 16  \\
        Qwen3-1.7B & $8\times10^{-5}$ & 16 & 4  \\
        Qwen3-4B & $8\times10^{-5}$ & 8 & 8 \\
        Qwen3-14B & $1\times10^{-5}$ & 4 & 4  \\
        \bottomrule
    \end{tabular}
\end{table*}

\subsection{Cue Construction Evaluation}

\textbf{Code representations.}
We compare redacted, visible-decimal, binary, and chained-fruit input representations, while keeping the target code in the same comma-separated decimal format.
Binary cues encode the integer represented by the four decimal digits as a zero-padded 14-bit string.
For chained-fruit cues, let the decimal digits be $d_0,d_1,d_2,d_3$.
We transform them as follows and map indices 0 through 9 to apple, banana, cherry, date, elderberry, fig, grape, kiwi, lemon, and mango, respectively.
\begin{equation}
    (d_0,\ (d_1+d_0)\bmod 10,\ (d_2+d_1)\bmod 10,\ (d_3+d_2)\bmod 10)
\end{equation}
The cue is the resulting space-separated sequence of four words.
The transformation is invertible, but its rule is not supplied in the evaluation prompt.
All targets retain the same comma-separated decimal format.

\textbf{Evaluation.}
Exposure ranks the code being evaluated among all 10,000 decimal candidates after the fixed assistant prefix.
For substituted-cue evaluations, we distinguish the sample's assigned code from the replacement code supplied in the input.
Greedy recovery uses a deterministic continuation of at most 12 tokens.
Substituted familiar cues come from the next focal canary in a cyclic ordering, so they test recombination with another sample rather than generalization to an unseen value.

\subsection{Expanded Results for Experiments in the Main Body}

\begin{figure}[t]
    \centering
    \includegraphics[width=0.7\textwidth]{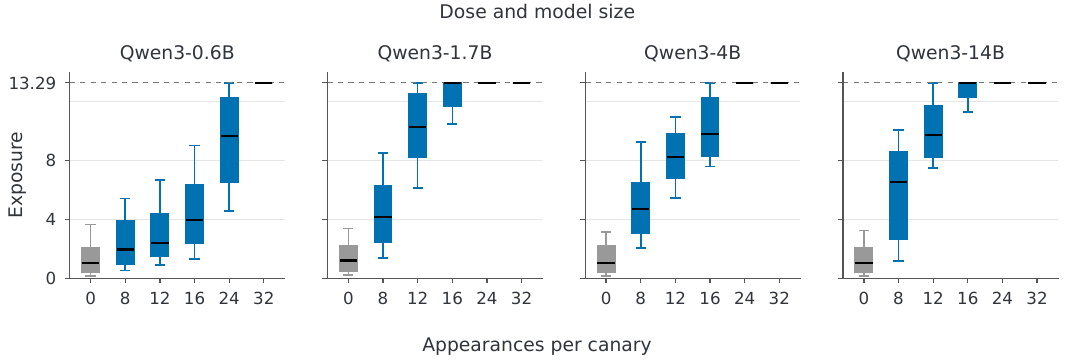}
    \caption{
        Synthetic-email dose response across four Qwen3 model sizes.
        Each box pools 64 canaries from each of three training runs (192 measurements): the center line marks the median, boxes span the interquartile range, and whiskers mark the 10th and 90th percentiles.
        Dose 0 denotes the no-canary baseline; the dashed line marks rank one among 10,000 candidates.
        }
    \label{fig:synthetic-dose}
\end{figure}

\textbf{Exposure Rises with Dose.}
\Cref{fig:synthetic-dose} supplements the representative Qwen3-1.7B results in \Cref{sec:synthetic-experiments} with dose-response results across four Qwen3 model sizes.
For each model, we train with the same 64 focal canaries at doses of 8, 12, 16, 24, and 32, alongside a no-canary baseline.
Across all four models, pooled median exposure rises with dose before reaching the ranking ceiling.
The Qwen3-1.7B example in the main text therefore captures the common direction of the effect.

\begin{figure}[ht]
    \centering
    \includegraphics[width=0.7\textwidth]{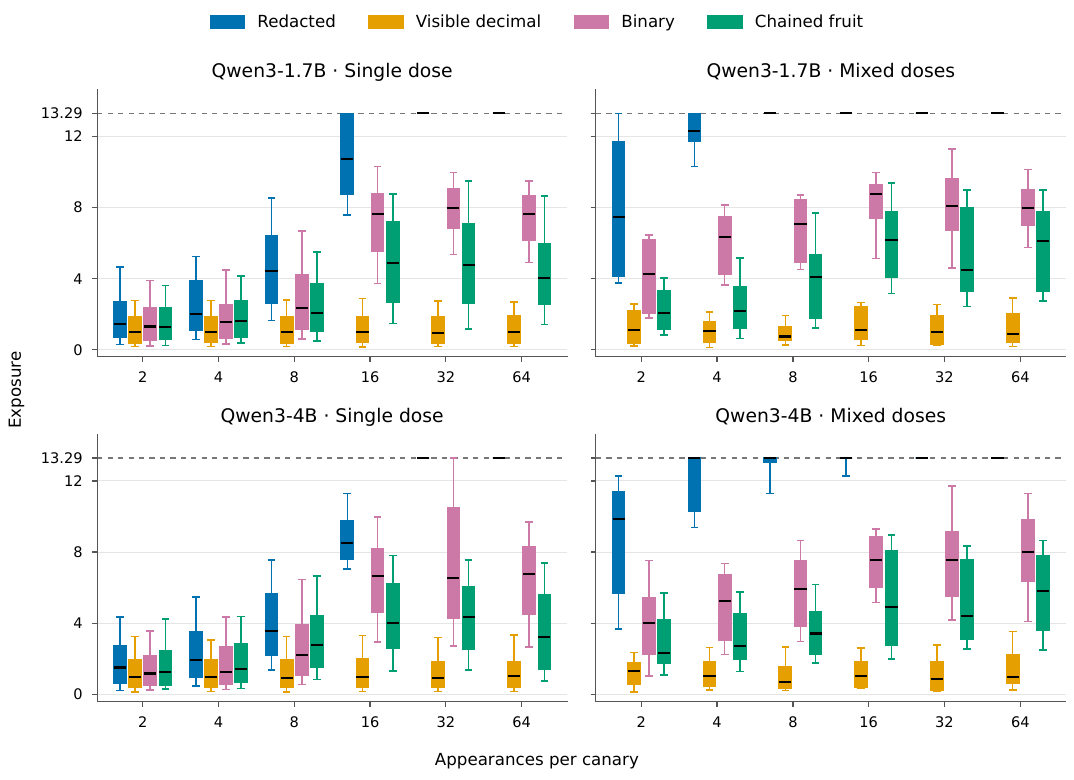}
    \caption{
    Training-input representation changes exposure across dosage in single-dose (left) and mixed-dose (right) runs.
    Rows show Qwen3-1.7B and Qwen3-4B.
    In single-dose runs, all 64 focal canaries receive the same dose; in mixed-dose runs, all six doses coexist within each training run.
    Every evaluation redacts the code; colors indicate the training representation.
    Each single-dose box pools 64 canary measurements from each of three training seeds (192 total); each mixed-dose box pools 32 canary measurements across three training seeds.
    Boxes show the pooled median, interquartile range, and 10th--90th percentile whiskers, describing canary distributions rather than confidence intervals.
    Within each seed, the six mixed-dose groups share a checkpoint and are not independent training replicates.
    The dashed line marks the exposure ceiling of 13.29.
    }
    \label{fig:representation-comparison}
    \label{fig:representation-mixed-dose-response}
\end{figure}

\textbf{Input Representation Across Models and Doses.}
\Cref{fig:representation-comparison} extends the main-text comparison to all four input representations across six doses for Qwen3-1.7B and Qwen3-4B, in both single-dose and mixed-dose runs.

The representation comparison uses 64 focal canaries and 16 additional support identities, each support identity appearing four times with its code redacted in the input.
We evaluate Qwen3-1.7B and Qwen3-4B with three training seeds at doses of 2, 4, 8, 16, 32, and 64 appearances.
Single-dose runs assign the same dose to all focal canaries, with appearances spread uniformly through training.
Within each seed, matched representation comparisons preserve focal anchors, assigned codes, background order, replacement positions, and training settings.
All four representation arms share the same system instruction and support population, including the matched redacted arm.
This differs from the original redacted-code experiments, whose system instruction explicitly described redaction and which included no added support population.
Across matched dose comparisons, lower doses use nested subsets of the same focal replacement positions.

Mixed runs assign 10 or 11 of the 64 focal canaries to each dose within a seed, yielding 32 canary observations per dose across the three seeds.
The six dose groups within a mixed run share model weights and are not independent training replicates.
Changing from single-dose to mixed-dose training changes neighboring canary doses and the number of replaced background rows, even when an individual canary's dose and positions are held fixed.

With the code hidden at evaluation, visible-decimal training leaves pooled median exposure near baseline across both models and both single-dose and mixed-dose runs.
At doses 16, 32, and 64, redacted training produces the highest pooled median exposure, with binary and chained-fruit training intermediate between redacted and visible-decimal training.
The main-text representation example therefore reflects a pattern shared across these conditions, although exposure after encoded-cue training does not increase monotonically with dose.

Exposure can vary across checkpoints, and so can which encoding produces more exposure (Figures~\ref{fig:broader-email-single-exposure} and~\ref{fig:broader-email-mixed-exposure}). At doses 16, 32, and 64 in both single-dose and mixed-dose runs, Qwen3-14B has lower binary-trained exposure than Qwen3-1.7B and Qwen3-4B. At these doses, binary training yields higher exposure than chained-fruit training for Qwen3-1.7B and Qwen3-4B, but lower exposure for Qwen3-14B. For canary audits, this motivates comparing encodings on the model being audited instead of carrying over the choice from another model.

The low-dose population effect also appears in both models: dose-2 and dose-4 redacted canaries have substantially higher pooled median exposure in mixed-dose runs than in single-dose runs.
For both binary and chained-fruit training in mixed-dose runs, median exposure at dose 64 exceeds that at dose 2 in every model and seed.

A separate substitution diagnostic checks whether visible-code training favors the supplied value.
For Qwen3-1.7B at dose 16, substituting a different decimal code after visible-code training gives high exposure to the supplied value while leaving the original training assignment near baseline.
Hiding the input code also leaves the original assignment near baseline, consistent with reliance on the supplied cue.

\subsubsection{Broader Representation and Dose Results}
\label{sec:broader-context}

The broader Email grids below and the placement and auxiliary-population distributions in Figure~\ref{fig:broader-email-timing} aggregate three seeds, with data, canary assignments, and training RNG varying jointly for canary conditions and training RNG varying over fixed data for clean controls.
Exposure boxes pool canary observations across seeds and show medians, interquartile ranges, and 10th/90th-percentile whiskers ending at the extreme observed values within those percentiles.
Metric curves show means of seed-level rates as specified in the captions.
For these Email results, exposure is computed over 10,000 candidate codes with the secret cue hidden, and greedy recovery is reported separately from supplied-cue controls.
Training budgets remain fixed within each dose comparison.

\begin{figure}[ht]
\centering
\includegraphics[width=0.7\textwidth]{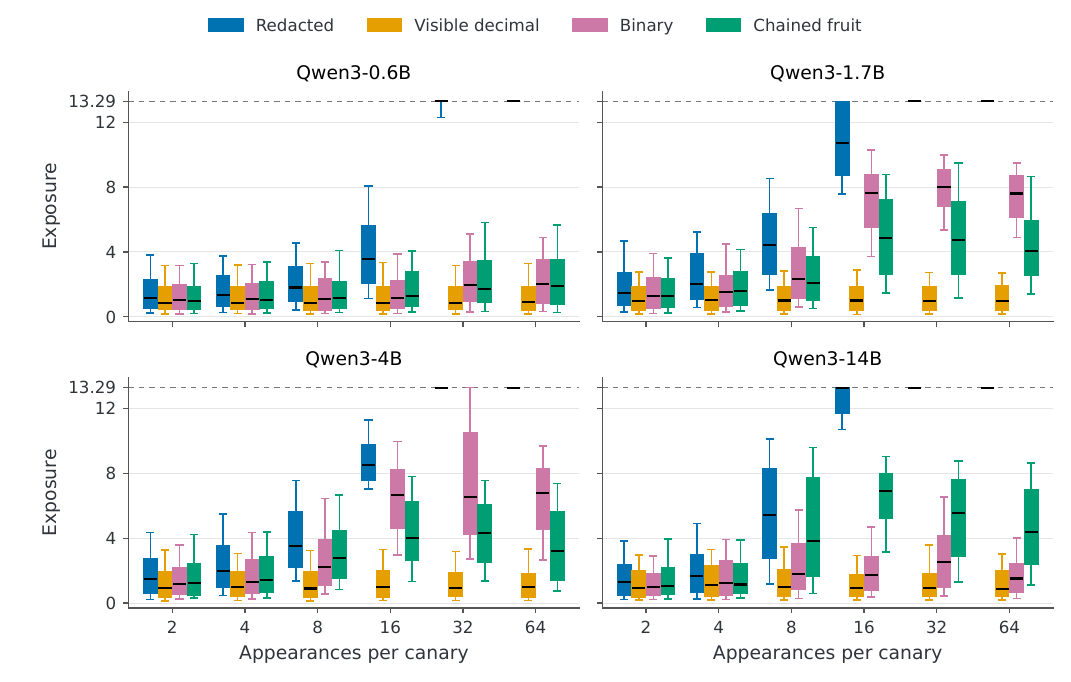}
\caption{Email single-dose exposure across four model sizes and four input representations.
Each box pools 64 canaries from each of three joint seeds, giving 192 observations per condition.}
\label{fig:broader-email-single-exposure}
\end{figure}

\begin{figure}[ht]
\centering
\includegraphics[width=0.7\textwidth]{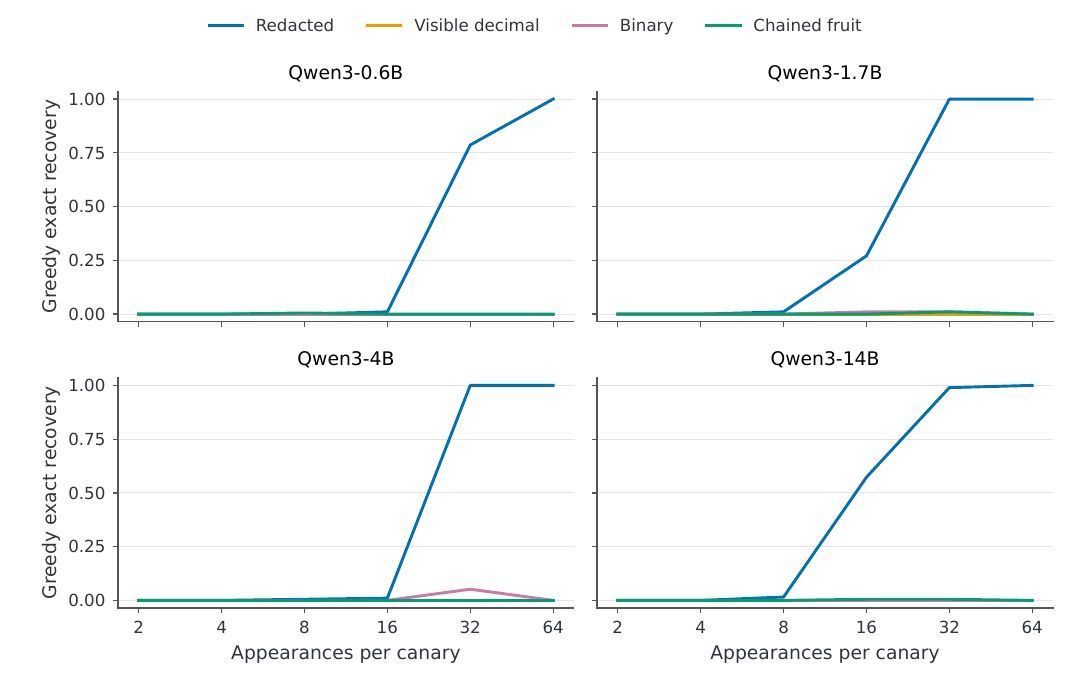}
\caption{Email single-dose greedy exact recovery.
Curves show the unweighted mean of three seed rates, with 64 canaries per seed.}
\label{fig:broader-email-single-recovery}
\end{figure}

\begin{figure}[ht]
\centering
\includegraphics[width=0.7\textwidth]{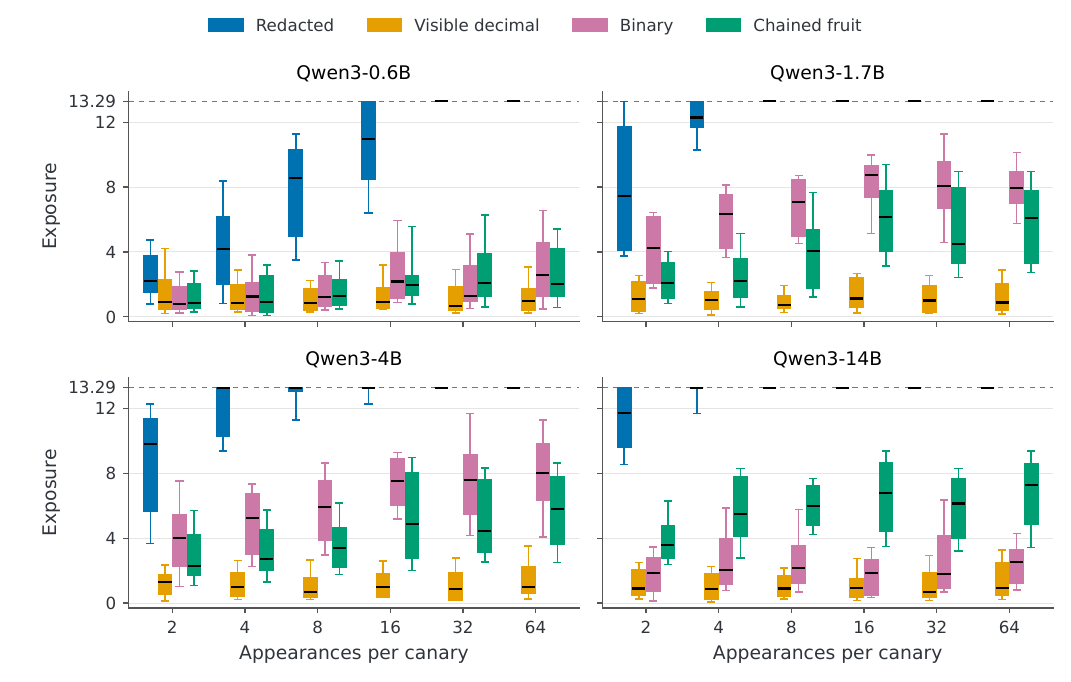}
\caption{Email mixed-dose exposure across four model sizes and four input representations.
Each box pools 32 canaries per dose across three joint seeds, with 10 or 11 per seed.
The six dose groups within a seed share one checkpoint.}
\label{fig:broader-email-mixed-exposure}
\end{figure}

\begin{figure}[!htbp]
\centering
\includegraphics[width=\textwidth]{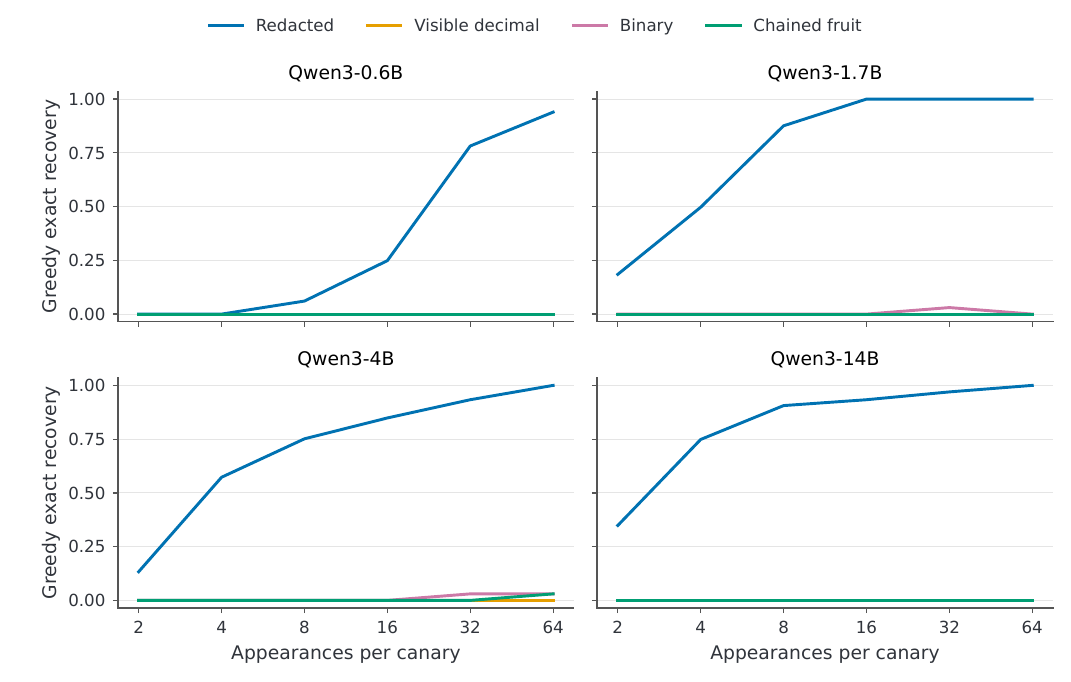}
\caption{Email mixed-dose greedy exact recovery.
Curves show the unweighted mean of three seed rates, with 10 or 11 canaries per dose and seed and 32 across seeds.}
\label{fig:broader-email-mixed-recovery}
\end{figure}

Figures~\ref{fig:broader-email-single-exposure} and~\ref{fig:broader-email-single-recovery} show single-dose results, while Figures~\ref{fig:broader-email-mixed-exposure} and~\ref{fig:broader-email-mixed-recovery} show mixed-dose results.
At single dose 64, every redacted checkpoint recovers all 64 secrets, while every visible-decimal, binary, and chained-fruit checkpoint recovers none.
The redacted single-dose curves are nondecreasing across doses, while encoded-input patterns vary across model recipes.

\subsection{Additional Experiments on Synthetic Email Data}
\label{sec:broader-email-timing}

\textbf{Does the Background Ordering Affect Learning?}
Our synthetic email experiments combine four copies of the background corpus into one globally shuffled training sequence.
A natural concern is whether this ordering changes how the model learns the background task compared with conventional epochs.
To check this, we train Qwen3-1.7B without canaries using four passes over the 16,544 background examples, independently shuffling the examples within each pass.
We hold the starting weights, training examples, optimizer settings, and training seed fixed, with both runs processing 66,176 rows in 1,034 optimizer updates at a constant learning rate of $8\times10^{-5}$.
Batches and optimizer state remain continuous across pass boundaries, so the comparison isolates the ordering of the examples.

\begin{figure}[t]
    \centering
    \includegraphics[width=0.5\textwidth]{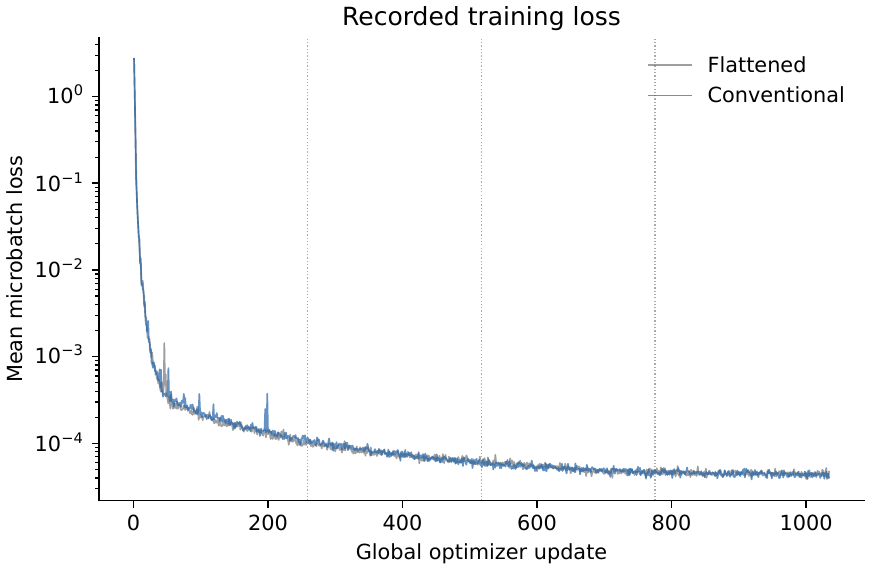}
    \caption{
Training loss for Qwen3-1.7B without canaries under two
background orderings. ``Flattened'' denotes the globally
shuffled four-copy sequence used in our experiments;
``Conventional'' denotes four independently shuffled
corpus passes. The plot shows recorded mean microbatch
loss at each optimizer update on a logarithmic loss axis.
Dotted vertical lines mark the conventional corpus-pass
boundaries. Each curve represents one training run, with
initialization and training settings matched across orderings.
}
    \label{fig:synthetic-email-epoching}
\end{figure}

The training-loss curves closely overlap throughout the run (\Cref{fig:synthetic-email-epoching}).
On the same held-out set of 1,024 email examples, the final assistant-token negative log-likelihood is 0.003580 for our globally shuffled sequence and 0.003073 for the separately shuffled passes.
Both orders therefore yield similar training behavior and low held-out loss, although conventional ordering gives the lower held-out loss in this comparison.
This control uses one training seed and no canaries; it addresses learning of the background task rather than whether the ordering affects canary memorization.

\begin{figure}[t]
    \centering
    \includegraphics[width=0.5\textwidth]{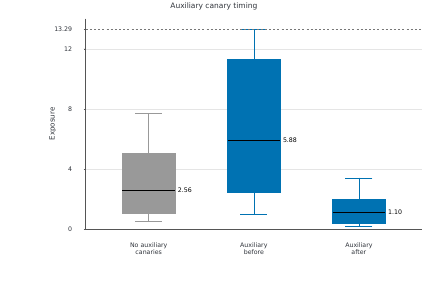}
    \caption{
    Focal-canary exposure under different auxiliary-cohort timings in Qwen3-4B.
    Both cohorts contain 64 dose-8 canaries.
    Focal final appearances remain fixed near the 95th percentile of training, while auxiliary final appearances fall near the 90th or 99th percentile; a reference condition omits the auxiliary cohort.
    Boxes summarize pooled focal-canary measurements: center lines mark medians, boxes span the interquartile range, and whiskers mark the 10th and 90th percentiles.
    The dashed line marks rank one among 10,000 candidates.
    }
    \label{fig:synthetic-interaction}
\end{figure}

\textbf{Auxiliary-cohort Order Changes Focal Exposure.}
The single-dose versus mixed-dose comparison changes both neighboring canary doses and the total number of canary appearances.
Here, we ask whether changing the timing of a fixed auxiliary cohort can alter exposure of focal canaries whose dose and placement remain fixed.
With Qwen3-4B, we train on 64 dose-8 focal canaries whose final appearances lie in a narrow band around the 95th percentile of training.
Two conditions introduce the same auxiliary cohort of 64 dose-8 canaries, with final appearances near either the 90th or 99th percentile.
These conditions preserve the focal canaries and their positions, auxiliary examples, removed background examples, remaining background order, and training recipe.
Only the placement of the auxiliary cohort changes.
We also train a reference model containing the focal canaries without the auxiliary cohort.

Across 192 paired canary measurements from three training seeds, moving the auxiliary cohort from after to before the focal cohort raises exposure for 86\% of focal canaries, with a median paired increase of 4.77 (\Cref{fig:synthetic-interaction}).
The reference without auxiliary canaries has intermediate median exposure.
Thus, relative cohort timing affects exposure even when the canary population and total number of appearances remain fixed.
The comparison does not identify a pairwise interaction mechanism or distinguish facilitation by earlier examples from interference by later ones.

\textbf{A Single-Seed Control with One-Shot Auxiliary Examples.}
One possible explanation for the increase from an earlier auxiliary cohort is that those examples familiarize the model with the canary format.
If exposure to that format alone is sufficient, distinct canaries appearing once each might produce a similar increase.
To test this possibility, we replace the 64 auxiliary canaries appearing eight times each with 512 distinct canaries appearing once each.
This single-seed diagnostic preserves the 512 auxiliary training slots, the focal canaries and their slots, the removed background examples, and the remaining background order.

Median focal exposure is 1.16 with these one-shot examples, compared with 1.91 without auxiliaries and 7.21 with the repeated auxiliary cohort before the focal cohort.
The median paired change relative to the no-auxiliary reference is $-1.17$, and only 25.0\% of focal canaries increase in exposure.
This single-seed control does not rule out format learning or distinguish the effects of repeating identities, codes, or templates.

\textbf{Placement and Auxiliary-Population Distributions Across Seeds.}

\begin{figure}[!htbp]
\centering
\includegraphics[width=\textwidth]{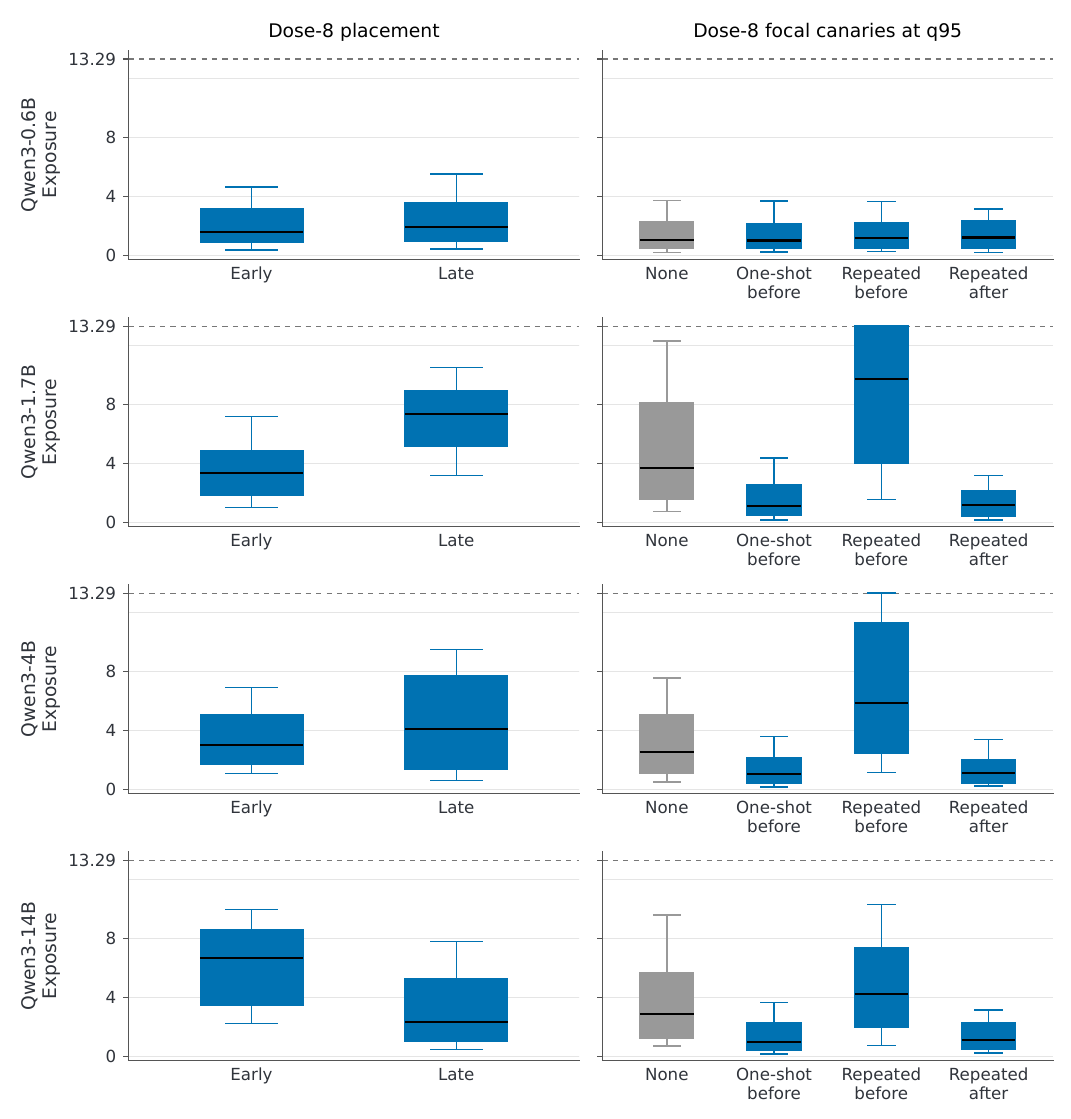}
\caption{Email placement and auxiliary-population exposure for redacted canaries at dose eight.
Each box pools 64 focal canaries from each of three joint seeds.
Repeated populations contain 64 identities at dose eight, while the one-shot condition uses 512 identities once.}
\label{fig:broader-email-timing}
\end{figure}

Figure~\ref{fig:broader-email-timing} provides the placement and auxiliary-population distributions.
The timing and population plots use their separate redacted-only protocols.

\clearpage

\begin{figure}[!htbp]
    \centering
    \includegraphics[width=\textwidth]{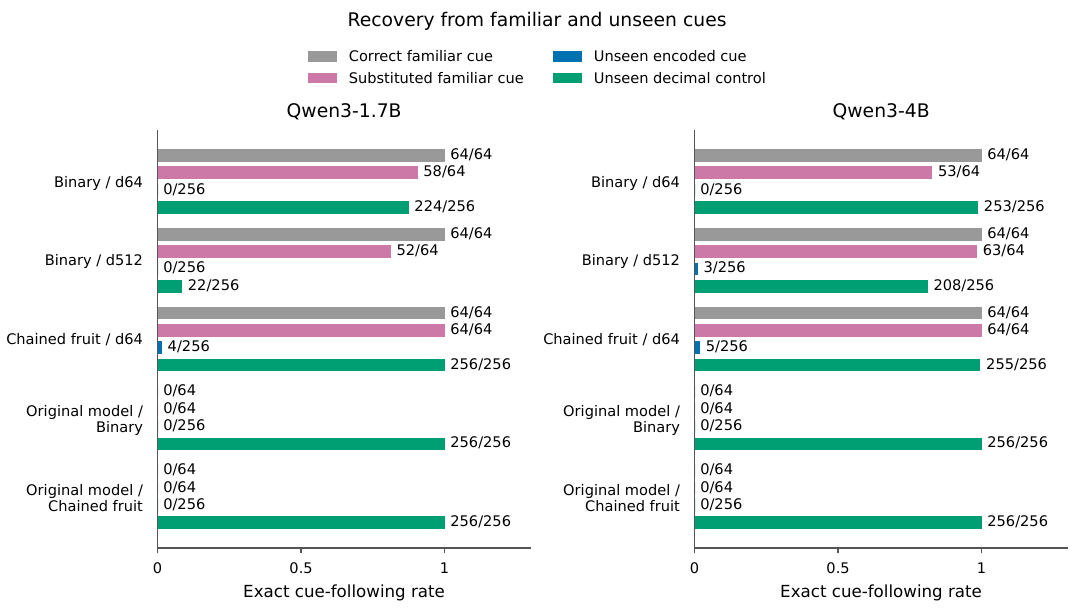}
    \caption{Recovery from familiar encoded cues transfers poorly to unseen values under the fixed evaluation prompt.
    Correct and substituted training-set cues each use 64 probes; unseen encoded cues and readable-decimal controls each use the same 256 held-out values.
    Fine-tuned checkpoints use one training seed per condition; original-model controls are also shown.
    For original models, ``familiar'' denotes membership in the subsequent fine-tuning set, which those controls have not received.
    Held-out status is verified against the selected fine-tuning streams, not pretraining.
    Decimal controls help distinguish poor encoded-value recovery from broader failures to reproduce new values in the required output format.}
    \label{fig:unseen-cue-generalization}
\end{figure}

\textbf{Generalization to Unseen Encoded Cues.}
The representation experiments leave open what the model learns from an encoded cue.
Recovering a familiar decimal code could reflect learning the conversion rule, remembering particular cue--code associations, or both.
We therefore test whether recovery extends to encoded values that did not occur in the selected fine-tuning data.
By also presenting the same values directly as decimal digits, we check whether failures reflect difficulty interpreting encoded cues or a broader difficulty reproducing new codes in the required format.
We compare Qwen3-1.7B and Qwen3-4B before fine-tuning with models trained on binary cues at doses 64 and 512 and chained-fruit cues at dose 64 (\Cref{fig:unseen-cue-generalization}).
For each of the 64 focal canaries, we test four new codes, giving 256 distinct values used across all models and conditions.
We exclude the 64 focal and 16 support secrets and verify that none of the new codes appears in the selected fine-tuning data as decimal digits, comma-separated decimal digits, binary strings, or chained-fruit sequences.
For each canary, we keep the surrounding prompt and generation settings fixed and change only the code cue.

If a model fails to recover a new code, we want to know whether it struggles with the encoding or with producing a new value in the required format.
We therefore also supply each of the same 256 codes directly as comma-separated decimal digits in place of the encoded cue.
Before fine-tuning, both models recover all 256 codes when given the decimal digits, but none when given either encoding under the same prompt.
The models can therefore copy these new values into the required output format, even though they do not recover them from the encoded cues.

\textbf{Recovery from Familiar Cues Transfers Poorly to New Values.}
After fine-tuning at dose 64, both models recover all 64 training codes when given their encoded cues.
They also usually follow a familiar cue when we pair it with a different canary: both recover the supplied code in all 64 chained-fruit probes, while Qwen3-1.7B and Qwen3-4B succeed in 58 and 53 of the 64 binary probes, respectively.
Yet this success transfers poorly to new values.
Neither model recovers any of the 256 unseen binary codes, and they recover only four and five unseen chained-fruit codes, respectively.

The same models can still reproduce most or all of these new values when we supply them directly as decimal digits.
The binary-trained models recover 224 of 256 values for Qwen3-1.7B and 253 for Qwen3-4B; the fruit-trained models recover 256 and 255, respectively.
The large gap between familiar and unseen encoded values therefore cannot be explained simply by an inability to produce new codes in the required output format.

\textbf{Higher-Dose Binary Diagnostic.}
The poor recovery of new binary codes at dose 64 might reflect insufficient training on encoded examples.
We therefore examine binary-trained checkpoints at dose 512, where each training code appears eight times as often.
Qwen3-1.7B still recovers none of the 256 new binary codes, while Qwen3-4B recovers only three.
However, when we supply the same values directly as decimal digits, the models recover only 22 and 208 of 256, respectively.
For Qwen3-1.7B in particular, poor recovery from binary cues therefore accompanies a broader difficulty reproducing new values in the required format.

More repetition also does not guarantee recovery when the encoded cue is removed.
In separate exploratory binary runs at doses 128, 256, and 512, neither model recovers any of its 64 training codes or ranks any of them first when the input cue is redacted.
Yet every checkpoint recovers all 64 codes when given the corresponding binary cues.
These are single-seed diagnostics, and increasing dose also changes the training mixture: at dose 512, focal canaries occupy 32,768 of the 66,176 training rows.

\begin{figure}[!htbp]
    \centering
    \includegraphics[width=\textwidth]{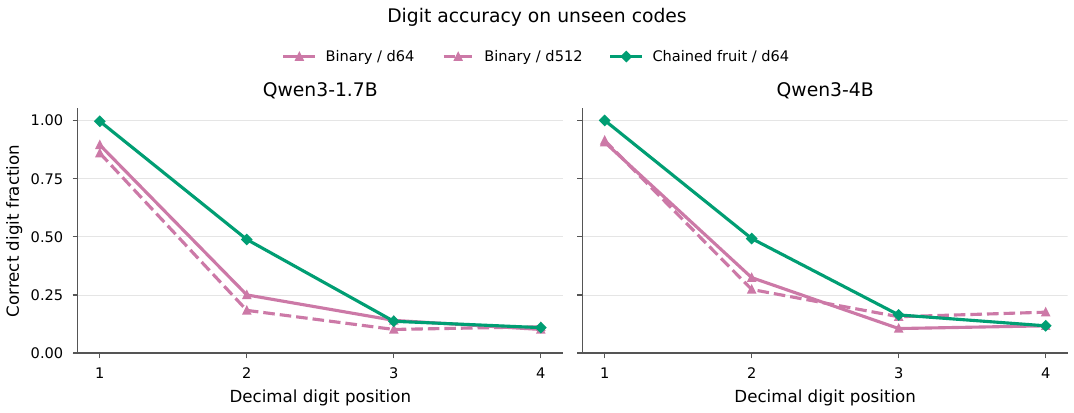}
    \caption{Exploratory digit-level accuracy on 256 unseen values for selected single-seed checkpoints.
    Position 1 is the leading decimal digit. Invalid outputs count as incorrect.
    Strong leading-digit performance coexists with poor complete-code recovery.
    Lines connect digit positions within a checkpoint; they do not represent training trajectories.}
    \label{fig:unseen-cue-digits}
\end{figure}

\textbf{Partial Generalization.}
Failure to recover an entire code can hide success on part of it.
After observing poor whole-code recovery, we examine accuracy at each digit position (\Cref{fig:unseen-cue-digits}).
Across the selected trained checkpoints, the leading digit is correct for 220--234 of the 256 new binary values and 255--256 of the chained-fruit values.
By contrast, the final digit is correct for only 26--45 values.
Invalid outputs count as incorrect at every position.
The models therefore show partial generalization to new values, even though they rarely recover a complete code.
This does not identify how they perform the conversion as learned sub-patterns or similarities to training codes could also explain the result.

We also examine whether the correct code ranks highly even when exact recovery is poor.
For this analysis, we use a fixed subset of 64 unseen probes, one per focal canary, drawn from the 256 generation probes.
At dose 64, median exposure with binary cues is 4.28 for Qwen3-1.7B and 4.05 for Qwen3-4B; with chained-fruit cues, the corresponding values are 6.64 and 5.67.
Chained-fruit cues thus yield stronger leading-digit accuracy and higher median exposure in these checkpoints, but these single-seed comparisons do not establish a general ordering of transformation difficulty.

\clearpage

%% file: sections/appendix-additional-utility-backed-text-experiments.tex
\section{Additional Utility-Backed Text Experiments}
\label{sec:app-utility-text}\suppressfloats[t]

The main text focuses on Banking77 as our representative text fine-tuning task.
Here, we extend the study to financial question answering on TAT-QA and present fuller results for Banking77.
In both settings, we examine how dosage, placement, predictive context, and the surrounding canary population affect exposure while checking that fine-tuning still improves performance on the intended task.
We also report results separately for each repetition, so readers can see how consistently the findings hold across runs.

\begin{table}[htbp]
\centering
\small
\caption{Protocol overview. All canary appearances replace background rows within a fixed training budget. Counts describe the principal dose sweeps; population and placement experiments specify their own allocations. A pooled observation is not necessarily a distinct identity or an independent training replicate.}
\label{tab:protocol-overview}
\begin{tabularx}{\textwidth}{@{}lXrrX@{}}
\toprule
Task & Canary target & Training rows & Identities/run & Dose-sweep controls\\
\midrule
Synthetic email & Rewrite plus signature code & 66,176 & 64 & Nested slots\\
TAT-QA & Answer, scale, and reference & 26,418 & 32 & Nested slots \\
Banking77 & Intent, customer, and reference & 10,003 & 64 & Nested slots \\
GUI grounding & Auxiliary typing action & 10,000 & 36 & Two disjoint panels reshuffled at each dose. \\
\bottomrule
\end{tabularx}
\end{table}

\begin{table}[htbp]
\centering
\caption{Text-task accuracy (\%) before and after fine-tuning without canaries. TAT-QA uses strict answer-and-scale accuracy; Banking77 uses intent accuracy. The Banking77 row uses a constant learning rate.}
\label{tab:text-utility}
\label{tab:tatqa-utility}
\begin{tabular}{@{}llrr@{}}
\toprule
Task & Model / schedule & Base & Fine-tuned \\
\midrule
TAT-QA & Qwen3-0.6B & 3.2 & 49.4 \\
 & Qwen3-1.7B & 8.2 & 53.8 \\
 & Qwen3-4B & 15.0 & 63.6 \\
 & Qwen3-14B & 21.8 & 70.4 \\
Banking77 & GPT-OSS-20B / const. & 56.1 & 93.1 \\
 
\bottomrule
\end{tabular}
\end{table}

\subsection{Qwen3 on TAT-QA}
\label{sec:tatqa}

To test whether our findings extend beyond intent classification, we fine-tune four Qwen3 models on TAT-QA.
Each example contains a financial table, supporting passages, and a question, and the model returns its answer and scale in structured JSON.
The scale field indicates whether an answer is expressed in thousands, millions, billions, or percent.
For canary examples, we add a synthetic four-digit request reference while preserving the original answer and scale.
This reference serves as the canary secret, allowing us to measure its recovery separately from whether the model answers the question correctly (\Cref{fig:tatqa-example}).

We start from 13,209 ordinary TAT-QA training examples, include each example twice, and shuffle the resulting 26,418 examples into a single training sequence.
Canary examples replace ordinary examples, so adding canaries or increasing their dose does not increase the training budget.
All conditions therefore use the same number of training examples and optimizer steps.
We fine-tune all model parameters using AdamW with BF16 precision and loss computed only on assistant responses.
We use constant learning rates of $2 \times 10^{-5}$, $10^{-5}$, $5 \times 10^{-6}$, and $10^{-6}$ for Qwen3-0.6B, Qwen3-1.7B, Qwen3-4B, and Qwen3-14B, respectively, with no warm-up.
The effective batch size is 32 examples and the maximum sequence length is 4,096 tokens.
Each run processes the fixed training sequence for 826 optimizer steps without additional shuffling by the trainer.

We use 32 canaries, each assigned a four-digit request reference.
The codes are balanced so that each digit appears three or four times in every position.
In visible-code training, the input supplies the assigned reference and in redacted-code training, it hides the reference.
The target response contains the reference in both cases, alongside the original answer and scale.

To compare these conditions, we reuse the same canary identities, prompts, assigned codes, and background order.
When varying dose, we construct lower-dose runs using subsets of the same canary insertion positions.
We measure exposure with the input reference hidden, ranking the assigned code against all 10,000 possible comma-separated four-digit codes.

Before turning to canary exposure, we check that fine-tuning improves performance on the underlying question-answering task.
We use strict accuracy: a prediction is correct only if its answer matches the reference under the dataset's normalization rules and its scale is correct.
Recovery of the added request reference is scored separately and does not contribute to task accuracy.
Fine-tuning without canaries improves accuracy for all four Qwen3 model sizes; for example, Qwen3-14B rises from 21.8\% to 70.4\% (\Cref{tab:tatqa-utility}).
These gains let us study memorization in a setting where fine-tuning substantially improves the model's ability to perform the intended task.

\begin{figure}[t]
    \centering
    \includegraphics[width=\textwidth]{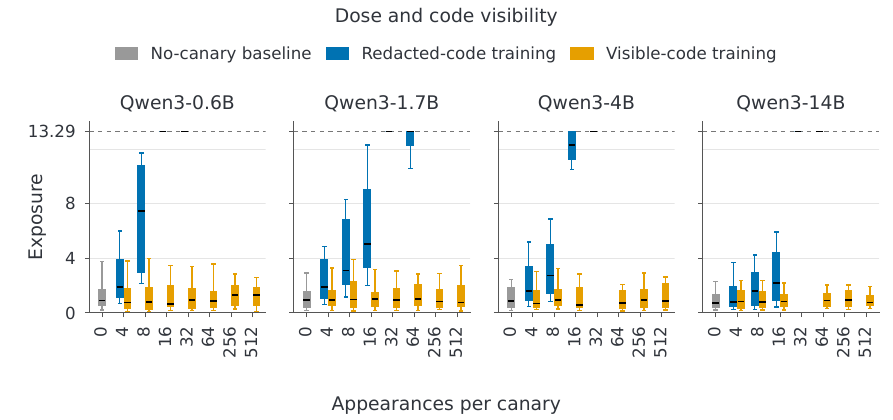}
    \caption{
    Dosage changes exposure across Qwen3 model sizes while TAT-QA accuracy remains above the base-model accuracy in \Cref{tab:tatqa-utility}.
    Exposure is evaluated with the request reference redacted, for models trained with redacted or visible references.
    Each box summarizes 32 canaries from one run: the center shows the median, the box the interquartile range, and the whiskers the 10th--90th percentiles.
    Gray marks show the no-canary baseline.
    Across the evaluated doses and training variants, held-out task accuracy spans 9.0 percentage points for Qwen3-0.6B and 2.2--4.8 points for the larger models.
    }
    \label{fig:tatqa-context-ladder}
\end{figure}

\begin{figure}[!htbp]
    \centering
    \includegraphics[width=0.75\linewidth]{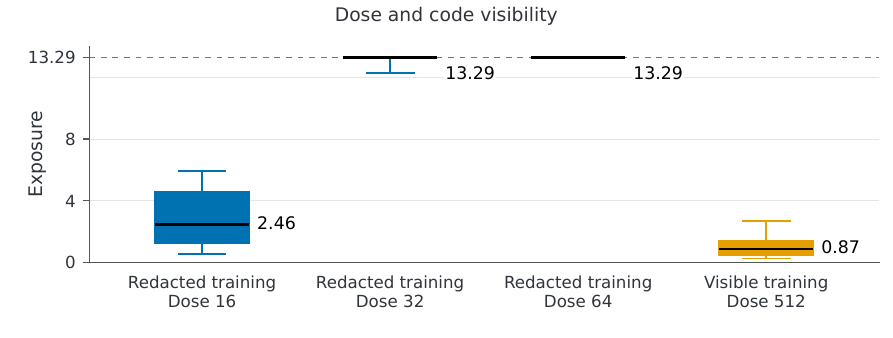}
    \caption{
    Qwen3-14B reaches high exposure with redacted-code training by dose 32, while visible-code training leaves exposure low at dose 512.
    All evaluations redact the request reference.
    Each box pools 96 canary measurements across three runs; centers show medians, boxes show interquartile ranges, and whiskers show the 10th--90th percentiles.
    The dose-32 median reaches the ceiling; at dose 64, every evaluated code ranks first.
    }
    \label{fig:tatqa-selective-replication}
\end{figure}

\textbf{Dosage.}
Increasing the number of canary appearances raises exposure on TAT-QA, as in the synthetic email experiments.
With the request reference hidden during training, the dose needed to reach high exposure varies across the four Qwen3 model sizes (\Cref{fig:tatqa-context-ladder}).
Across three Qwen3-14B runs with 32 canaries each, pooled median exposure rises from 2.46 at dose 16 to the ceiling of 13.29 at dose 32 (\Cref{fig:tatqa-selective-replication}).
At dose 64, the assigned reference ranks first among all 10,000 candidates for every canary in all three runs.
The sharp increase from dose 16 to dose 32 occurs in each repetition, with results reported separately in \Cref{tab:tatqa-replication}.

\textbf{Predictive context.}
We next ask whether frequent repetition still produces high exposure when the training input supplies the request reference.
We train Qwen3-14B with visible references at 512 appearances per canary.
Across three runs with 32 canaries each, the model copies every supplied reference, including when we substitute a different reference at evaluation.
This substitution checks whether the model follows the visible input even when it differs from the reference assigned during training.
When we hide the reference at evaluation, however, pooled median exposure is only 0.87 (\Cref{fig:tatqa-selective-replication}).
Thus, even after 512 appearances, successful copying of a visible reference coexists with low exposure when that reference is hidden.

\begin{figure}[!htbp]
    \centering
    \includegraphics[width=\textwidth]{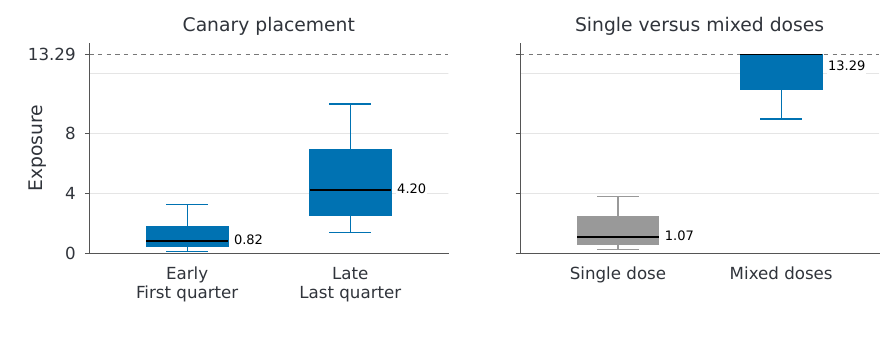}
    \caption{
    Later placement and mixed-dose training raise TAT-QA canary exposure with Qwen3-14B.
    Left: the same dose-16 canaries are placed in the first or last quarter of training.
    Right: dose-4 canaries are measured in single-dose and mixed-dose training.
    Each box pools 96 canary measurements across three repetitions of the comparison; centers show medians, boxes show interquartile ranges, and whiskers show the 10th--90th percentiles.
    The mixed-dose comparison also changes the total number of canary appearances.
    }
    \label{fig:tatqa-followups}
\end{figure}

\textbf{Placement.}
Later placement raises exposure on TAT-QA as it does in the synthetic email setting.
We hold dosage at 16 appearances and compare placement in the first and last quarters of training using Qwen3-14B.
Each run contains 16 early canaries and 16 late canaries; a second run exchanges the assignments so that every canary is measured in both positions.
Late placement raises exposure for 87 of 96 canaries, with a median paired increase of 3.12 (\Cref{fig:tatqa-followups}, left).
Because early and late canaries share each training run, this comparison measures placement effects in the presence of other canaries.

\textbf{Mixed-dose training.}
Dose-4 canaries have higher exposure when trained alongside higher-dose canaries.
We divide the 32 canaries into four groups of eight with doses of 4, 16, 64, and 256 appearances, and rotate the assignments across four runs so that every canary is measured at each dose.
We compare the dose-4 measurements with a run in which all 32 canaries appear four times.
Median dose-4 exposure rises from 1.07 in single-dose training to 13.29 in mixed-dose training (\Cref{fig:tatqa-followups}, right).
This comparison changes both the surrounding canaries' doses and the total number of canary appearances: mixed-dose runs contain 2,720 canary examples, compared with 128 in single-dose training.
It therefore does not isolate an interaction between canaries or provide an independent dose-response curve.

TAT-QA accuracy remains between 68.2\% and 71.4\% across the dose and context comparisons, compared with 21.8\% for the base Qwen3-14B model.
The placement and mixed-dose comparisons also retain high task accuracy (\Cref{sec:app-tatqa-replication}).

\FloatBarrier\subsection{GPT-OSS-20B on Banking77}
\label{sec:banking77}
We use the GPT-OSS-20B/Banking77 canary construction and fixed-budget protocol described in Section~\ref{sec:utility-vlm}.
The dosage, placement, and mixed-dose comparisons below use a constant learning rate of \(2.5 \times 10^{-6}\), selected using task accuracy without inspecting canary outcomes.
The canary identities, codes, and insertion positions vary across three repetitions while the optimizer seed is fixed.
Broader results with data, canary assignments, and training RNG varying jointly are reported in \Cref{sec:broader-banking-coverage}.

\textbf{Task accuracy.}
Banking77 accuracy is the fraction of predictions with the correct intent label; customer names and request references do not contribute to this score.
Fine-tuning on ordinary Banking77 examples raises held-out accuracy from 56.1\% to 93.1\% with the constant learning rate.
The constant-rate canary runs also retain higher accuracy than the base model.

\begin{figure}[!htbp]
\centering
\includegraphics[width=\linewidth]{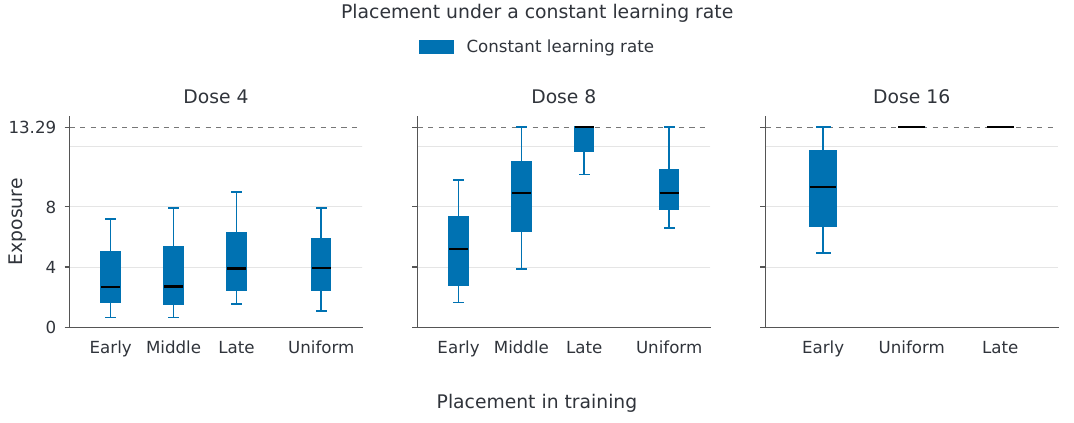}
\caption{Banking77 placement comparisons at doses 4, 8, and 16 under a constant learning rate of \(2.5 \times 10^{-6}\).
At dose 4, pooled median exposure remains below 4 across placements; at dose 8, pooled median exposure increases from early to middle to late placement.
At dose 16, uniform and late placement reach the median exposure ceiling.
Early, middle, and late windows span the first, middle, and last fifths of training; uniform placement spreads appearances throughout training.
Each box pools 192 canary measurements across three repetitions with varying canary identities, codes, and insertion positions and a fixed optimizer seed.
Centers show medians, boxes show interquartile ranges, and whiskers show the interpolated 10th and 90th percentiles.}
\label{fig:banking77-dose16-placement}
\label{fig:banking77-low-dose-placement-full}
\end{figure}

\textbf{Dosage.}
Pooled median exposure rises from 3.95 at dose 4 to 8.90 at dose 8, and reaches the ceiling of 13.29 at dose 16 under uniform placement (\Cref{fig:banking77-low-dose-placement-full}, uniform-placement boxes).
The pooled no-canary baseline is 0.98.
This sharp increase repeats the dosage effect seen in the earlier tasks, although the dose at which exposure reaches the ceiling differs.
Results for each repetition appear in \Cref{sec:app-banking77-replication}.

\textbf{Placement.}
At dose 16, restricting appearances to the first fifth of training lowers pooled median exposure to 9.29, while uniform placement and placement in the last fifth reach the median exposure ceiling of 13.29 (\Cref{fig:banking77-dose16-placement}, right).
The ceiling limits what this comparison reveals about differences between uniform and late placement.
We also compare early, middle, and late placement at doses 4 and 8, restricting appearances to the first, middle, or last fifth of training.
At dose 4, pooled median exposure remains below 4 at every placement (\Cref{fig:banking77-low-dose-placement-full}).
At dose 8, pooled median exposure increases from 5.17 for early placement to 8.93 for middle placement and 13.29 for late placement.

\textbf{Predictive context.}
Visible-code training again produces copying.
In a control using one set of 64 canaries, models trained with the constant learning rate at doses 4 or 64 copy both the original and substituted visible codes for every canary.
When the code is redacted at evaluation, median exposure is 0.99 at dose 4 and 1.18 at dose 64.
Thus, successful copying of a visible reference again coexists with low exposure when the reference is hidden.

\begin{figure}[!htbp]
\centering
\includegraphics[width=\linewidth]{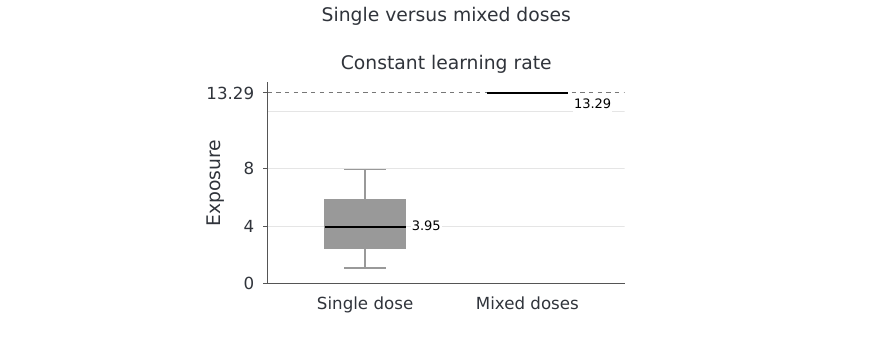}
\caption{Mixed-dose training raises dose-4 Banking77 canary exposure under a constant learning rate of \(2.5 \times 10^{-6}\).
Single-dose runs give all 64 canaries four appearances; mixed-dose runs assign 16 canaries to each of doses 4, 16, 32, and 64.
Each box pools 192 dose-4 measurements across three repetitions with varying canary identities, codes, and insertion positions and a fixed optimizer seed.
Centers show medians, boxes show interquartile ranges, and whiskers show the interpolated 10th and 90th percentiles.
The comparison also changes the total number of canary appearances.}
\label{fig:banking77-mixed-interaction}
\end{figure}

\textbf{Mixed-dose training.}
Dose-4 canaries have higher exposure when trained alongside higher-dose canaries.
We divide the 64 canaries into four groups of 16 with doses of 4, 16, 32, and 64 appearances, and rotate the assignments across four runs so that every canary is measured at each dose.
Pooled median dose-4 exposure rises from 3.95 in single-dose training to the ceiling of 13.29 in mixed-dose training (\Cref{fig:banking77-mixed-interaction}).
As in TAT-QA, this comparison changes both the surrounding canaries' doses and the total number of canary appearances: mixed-dose runs contain 1,856 canary examples, compared with 256 in the dose-4 reference.
The result shows that exposure depends on the training population, but does not isolate an interaction between canaries.

\FloatBarrier\subsection{TAT-QA replication}
\label{sec:app-tatqa-replication}

The Qwen3-14B comparisons in \Cref{sec:tatqa} are repeated three times with 32 canaries per repetition.
\Cref{tab:tatqa-replication} reports the exposure measurements separately for each repetition.
The dosage threshold, low exposure after visible-code training, late-placement increase, and mixed-dose increase recur in all three repetitions.
The figures pool the individual canary measurements; their boxes describe variation across canaries, not uncertainty intervals over training runs.

\begin{table*}[!htbp]
\centering
\caption{TAT-QA exposure with Qwen3-14B in each repetition.}
\label{tab:tatqa-replication}
\begin{tabular}{lrrr}
\toprule
Comparison & Repetition 1 & Repetition 2 & Repetition 3 \\
\midrule
Redacted code, dose 16 & 2.18 & 2.55 & 2.69 \\
Redacted code, dose 32 & 13.29 & 13.29 & 13.29 \\
Redacted code, dose 64 & 13.29 & 13.29 & 13.29 \\
Visible code, dose 512 & 0.76 & 1.07 & 0.81 \\
\midrule
Early placement, dose 16 & 0.66 & 1.10 & 0.98 \\
Late placement, dose 16 & 4.33 & 4.07 & 4.29 \\
Median paired late-placement gain & 3.49 & 2.86 & 3.08 \\
\midrule
Fixed-dose training, dose 4 & 0.81 & 1.18 & 1.21 \\
Mixed-dose training, dose 4 & 11.79 & 13.29 & 12.29 \\
\bottomrule
\end{tabular}
\end{table*}

All exposure evaluations redact the request reference.
Entries are medians over 32 canaries, except for the paired placement gain, which is the median of the 32 within-canary differences.

For each repetition of the placement or mixed-dose comparison, we average task accuracy over the two placement runs or four mixed-dose runs, respectively.
These averages range from 69.8\% to 70.8\%.

\subsubsection{Broader TAT-QA Results Across Models and Seeds}
\label{sec:broader-tatqa-grid}
\label{sec:broader-tatqa-timing}
\label{sec:broader-tatqa-lr}
\label{sec:broader-tatqa-audit}
These expanded TAT-QA results complement the single-run context-ladder results in \Cref{fig:tatqa-context-ladder} and the selected three-run Qwen3-14B comparisons above.
Each plotted condition in the expanded TAT-QA figures below aggregates three seeds, with data, canary assignments, and training RNG varying jointly for canary conditions and training RNG varying over fixed data for clean controls.
Exposure boxes pool canary observations across seeds and show medians, interquartile ranges, and 10th/90th-percentile whiskers; the whiskers end at the extreme observed values within those percentiles.
Metric curves show means of seed-level rates or checkpoint statistics as specified in the captions.
Exposure uses 10,000 candidate codes with the secret cue hidden, and greedy recovery is reported separately from supplied-cue controls.
Training budgets remain fixed when dose changes, and the timing and population plots retain their separate redacted-only protocols.
The expanded TAT-QA grid retains the model-specific constant learning-rate recipes used in the paper.

\begin{figure}[!htbp]
\centering
\includegraphics[width=\linewidth]{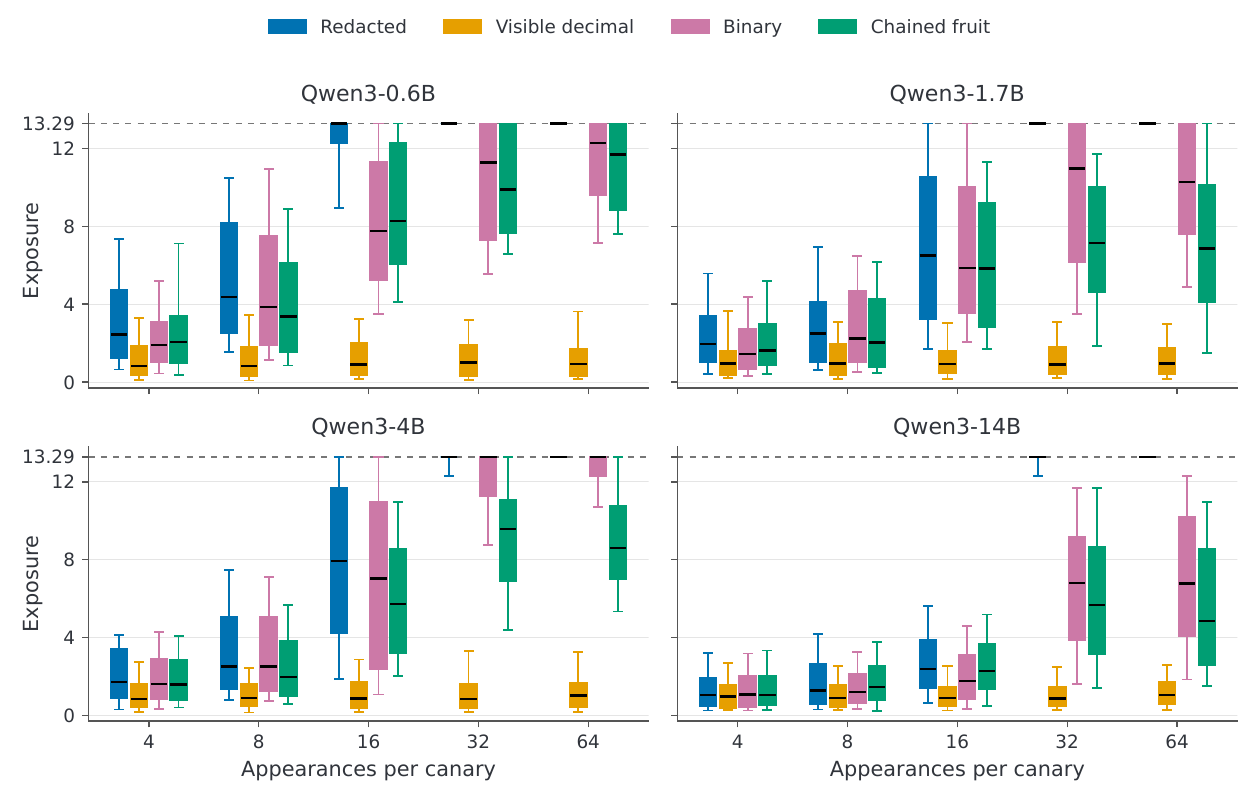}
\caption{TAT-QA single-dose exposure across four input representations.
Each box pools 32 canaries from each of three joint seeds, giving 96 observations per condition.}
\label{fig:broader-tatqa-single-exposure}
\end{figure}

\begin{figure}[!htbp]
\centering
\includegraphics[width=\linewidth]{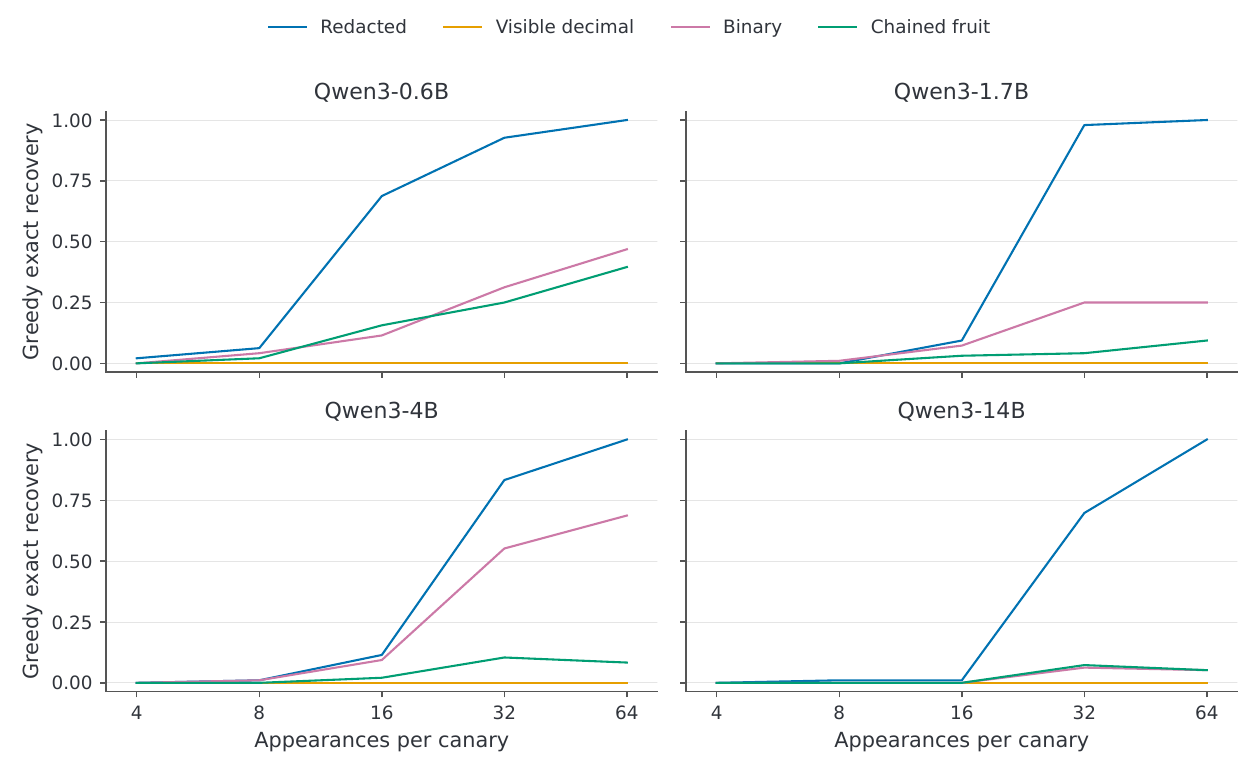}
\caption{TAT-QA single-dose greedy recovery.
Each curve is the mean of three equal-denominator seed rates, with 32 canaries per seed.}
\label{fig:broader-tatqa-single-recovery}
\end{figure}

\textbf{Representation and dose.}
Figures~\ref{fig:broader-tatqa-single-exposure} and~\ref{fig:broader-tatqa-single-recovery} summarize the primary representation and dose grid.
At dose 64 in this grid, every redacted checkpoint recovers all 32 secrets, while every visible-decimal checkpoint recovers none when the cue is hidden.

\begin{figure}[!htbp]
\centering
\includegraphics[width=\linewidth]{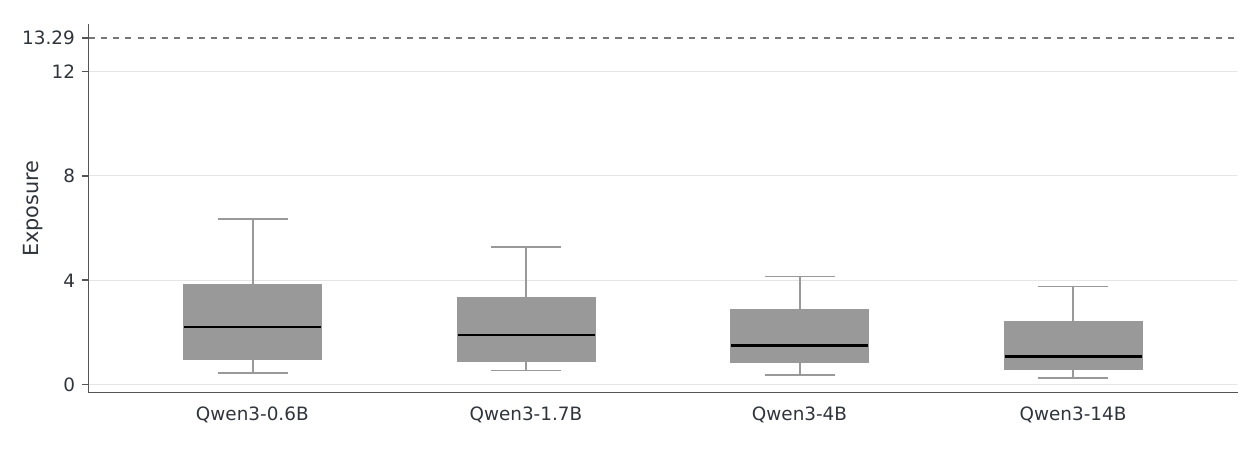}
\caption{TAT-QA fixed-dose-four exposure under the redacted-only instruction.
Each box pools 32 canaries from each of three joint seeds.}
\label{fig:broader-tatqa-fixed-dose-four}
\end{figure}

\begin{figure}[!htbp]
\centering
\includegraphics[width=\linewidth]{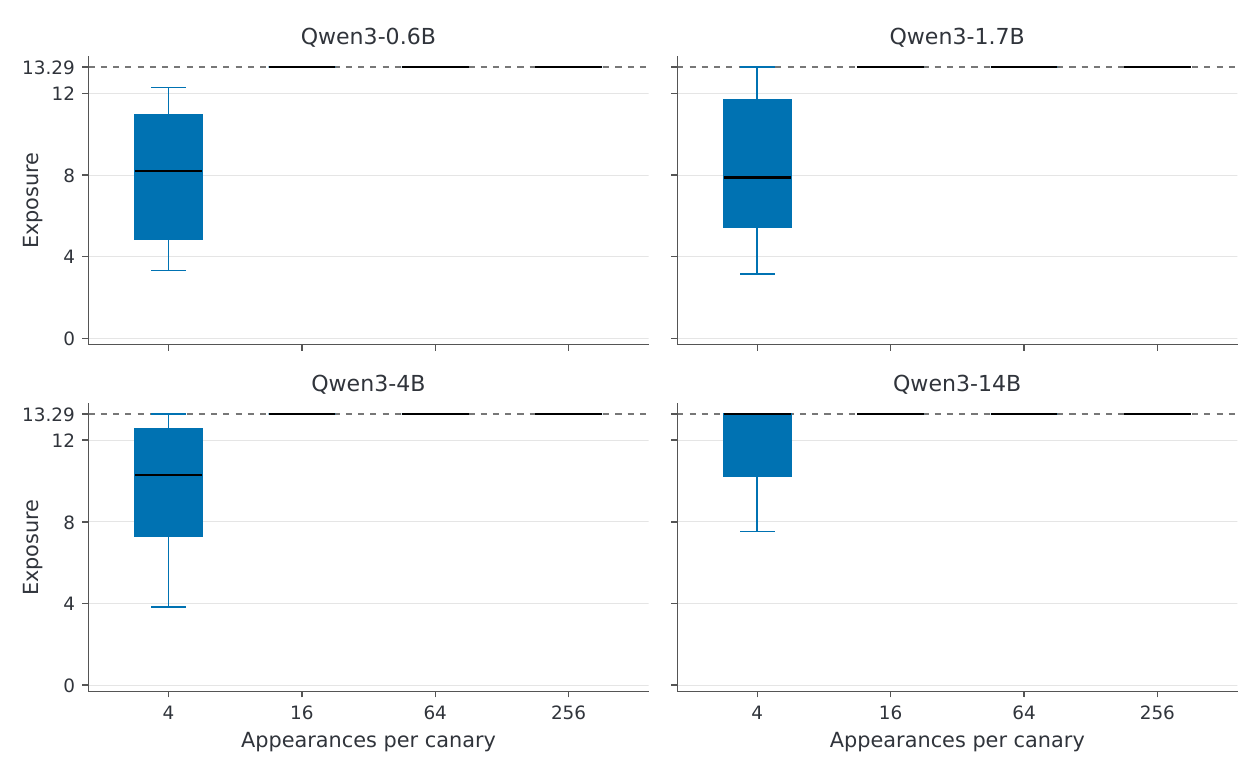}
\caption{TAT-QA mixed-dose exposure under the redacted-only instruction.
Each box pools 96 observations per dose, combining four rotations within each of three joint seeds.}
\label{fig:broader-tatqa-mixed}
\end{figure}

\begin{figure}[!htbp]
\centering
\includegraphics[width=\linewidth]{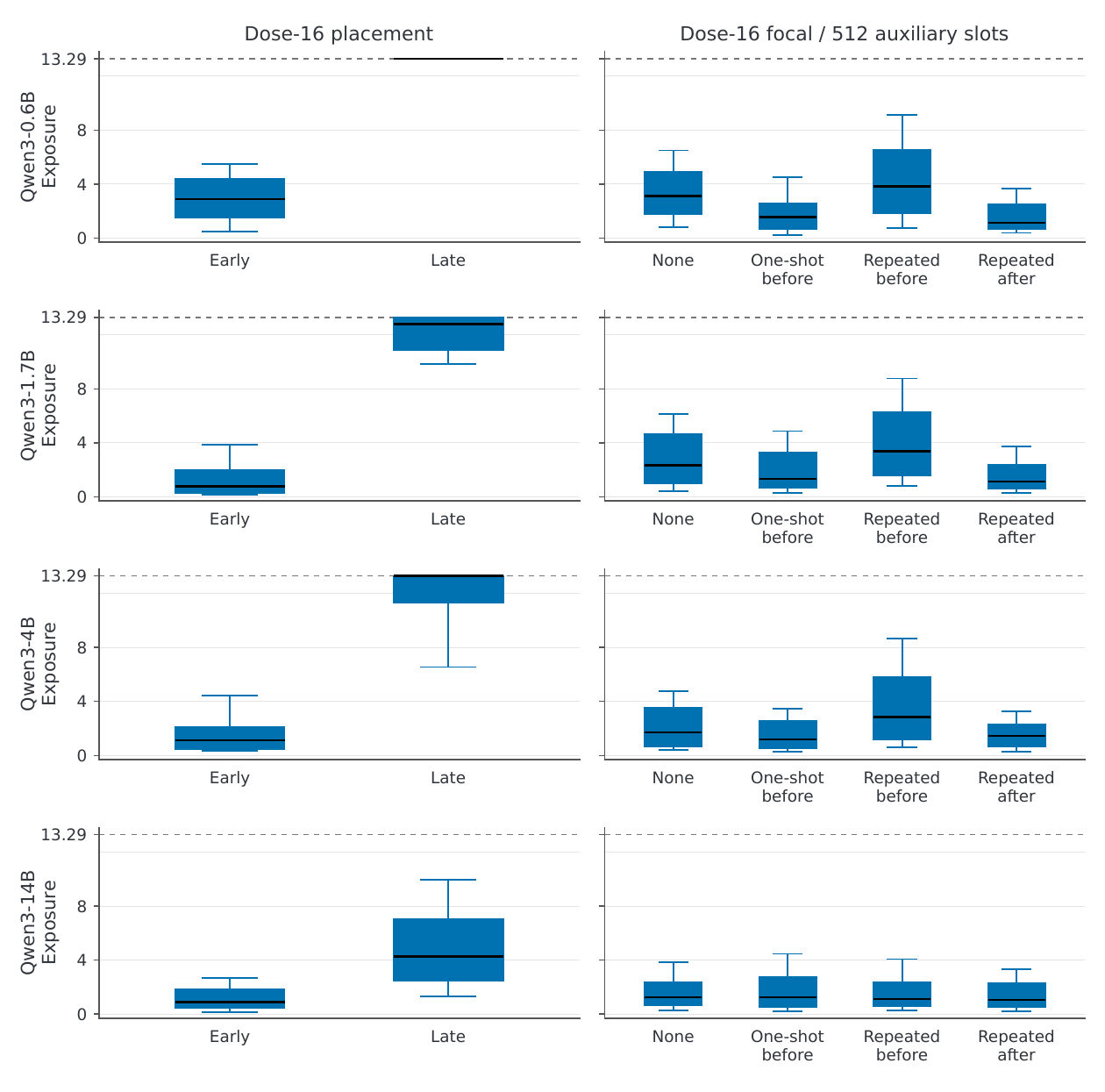}
\caption{TAT-QA dose-16 placement and auxiliary-population conditions.
Each box pools 32 focal observations per seed across three joint seeds, with complementary placement assignments within each seed.
The one-shot condition uses 512 identities once, while repeated conditions use 32 identities 16 times.}
\label{fig:broader-tatqa-timing}
\end{figure}

\textbf{Redacted-only dose, placement, and population controls.}
Figures~\ref{fig:broader-tatqa-fixed-dose-four} and~\ref{fig:broader-tatqa-mixed} show the redacted-only dose protocols separately, and Figure~\ref{fig:broader-tatqa-timing} shows placement and auxiliary-population results.

\clearpage

\begin{figure}[!htbp]
\centering
\includegraphics[width=\linewidth]{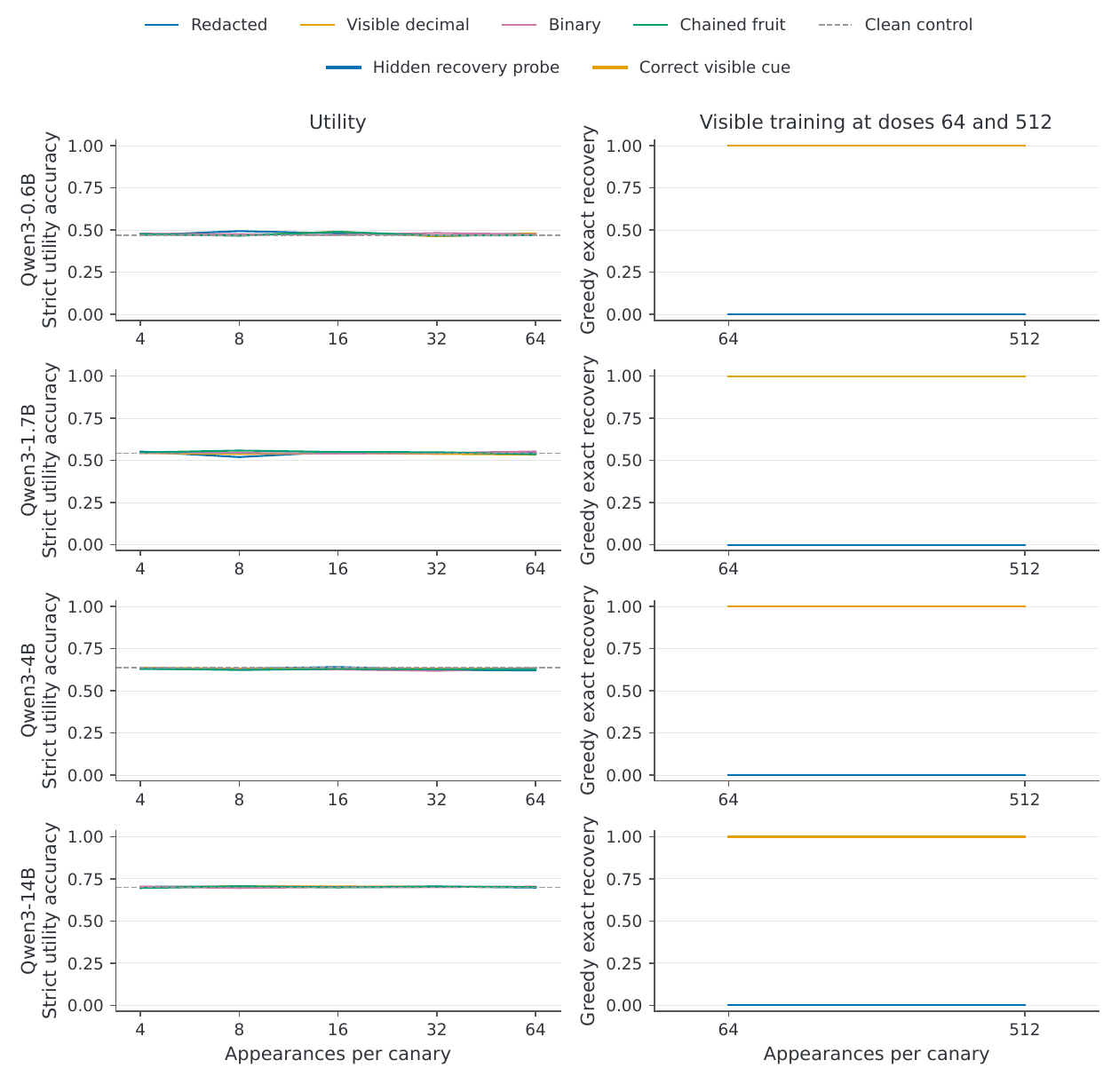}
\caption{TAT-QA utility and visible-decimal high-dose recovery.
Curves show means across three seeds, with utility evaluated on the same 500 questions and recovery on 32 canaries per seed.
Dashed lines mark clean-control means.}
\label{fig:broader-tatqa-utility-high-dose}
\end{figure}

\begin{figure}[!htbp]
\centering
\includegraphics[width=\linewidth]{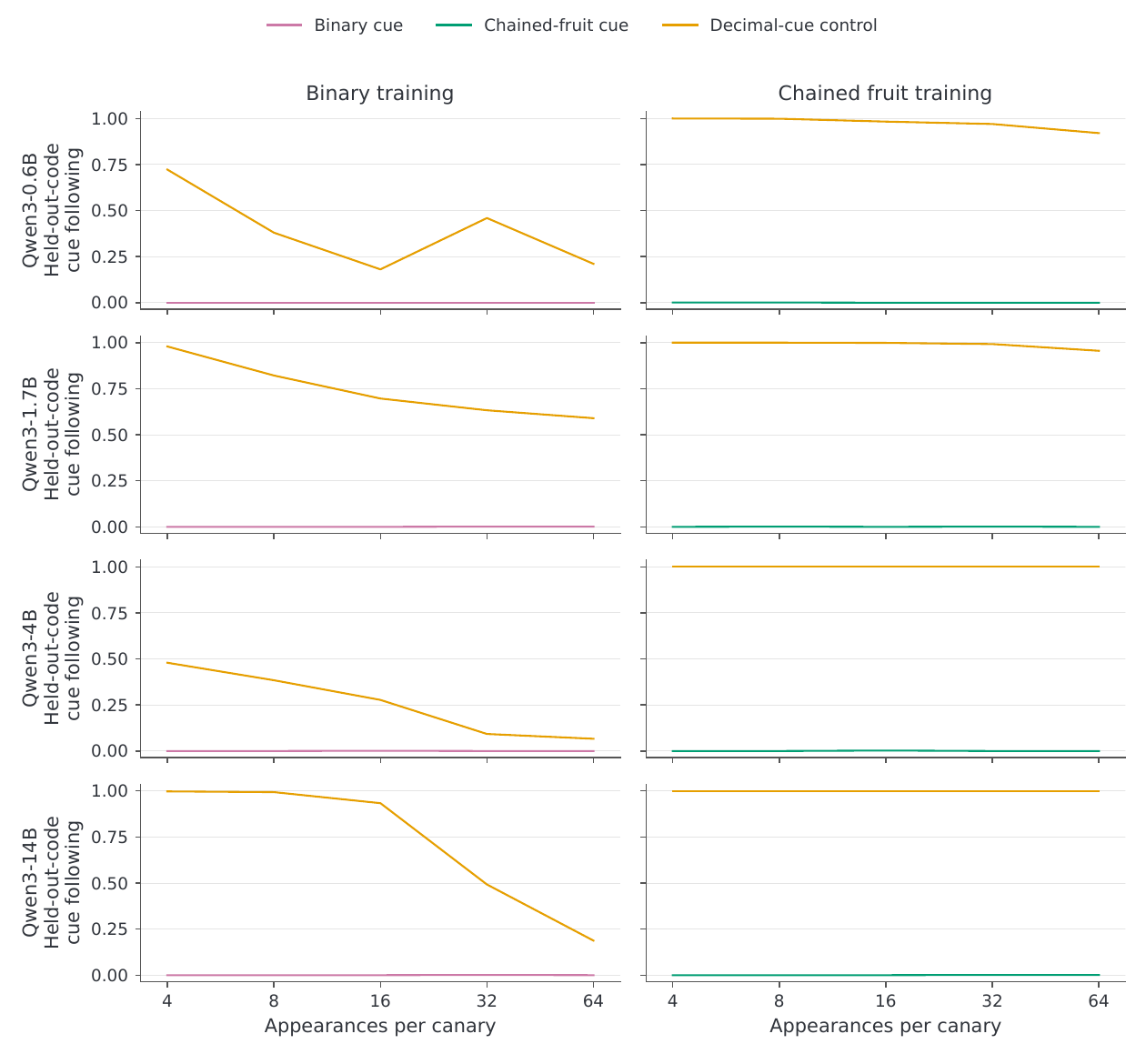}
\caption{TAT-QA cue following for held-out codes on known identities.
Curves show mean rates across three joint seeds, with 256 probes per cue format per checkpoint.
Held out refers to canary assignments, not global absence of the numeric values from training.}
\label{fig:broader-tatqa-unseen-codes}
\end{figure}

\textbf{Utility and supplied-cue controls.}
In a canary-free check on the fixed 500-question held-out set, Qwen3-14B at \(10^{-6}\) achieved 70.4\% strict answer-and-scale accuracy.
Figures~\ref{fig:broader-tatqa-utility-high-dose} and~\ref{fig:broader-tatqa-unseen-codes} provide the utility, high-dose, and supplied-cue controls.

\clearpage

\clearpage
\clearpage

\subsection{Banking77 replication}
\label{sec:app-banking77-replication}

The dosage, placement, and mixed-dose comparisons in \Cref{sec:banking77} are repeated three times with 64 canaries per repetition.
The canary identities, codes, and insertion positions vary across repetitions, while the optimizer seed is fixed.
\Cref{tab:banking77-replication} reports median exposure separately for each repetition under the constant learning rate.
The figures pool individual canary measurements across repetitions; their boxes describe variation across canaries, not uncertainty intervals over training runs.

\begin{table*}[!htbp]
\centering
\small
\caption{Banking77 median exposure under a constant learning rate of \(2.5 \times 10^{-6}\) in each repetition. Each entry is the median over 64 canaries.}
\label{tab:banking77-replication}
\begin{tabular}{lrrr}
\toprule
Comparison & Rep. 1 & Rep. 2 & Rep. 3 \\
\midrule
Fixed dose 4 & 2.86 & 4.89 & 3.60 \\
Fixed dose 8 & 9.70 & 9.04 & 8.53 \\
Fixed dose 16 & 13.29 & 13.29 & 13.29 \\
Early placement, dose 16 & 10.59 & 9.04 & 8.24 \\
Early placement, dose 8 & 6.24 & 2.86 & 6.55 \\
Middle placement, dose 8 & 9.12 & 7.44 & 9.90 \\
Late placement, dose 8 & 12.29 & 13.29 & 13.29 \\
Mixed doses, dose-4 canaries & 13.29 & 13.29 & 13.29 \\
\bottomrule
\end{tabular}
\end{table*}

Median exposure reaches the ceiling of 13.29 at dose 16 with uniform or late placement in every repetition.
At dose 8, median exposure increases from early to middle to late placement in every repetition.
In mixed-dose training, median exposure reaches the ceiling at all four assigned doses (4, 16, 32, and 64) in every repetition.
This agreement does not remove the confounding between the surrounding canaries and the total number of canary appearances in the mixed-dose comparison.
In the visible-code control described above, neither constant-rate checkpoint greedily recovers a trained code when it is redacted at evaluation.

\subsubsection{Broader Banking77 Results Across Joint Seeds}
\label{sec:broader-banking-coverage}
\label{sec:broader-banking-composition}
\label{sec:broader-banking-placement}
\label{sec:broader-banking-utility}
These broader Banking77 results complement the constant-rate repetitions above.
For canary conditions in this grid, data, canary assignments, and training RNG vary jointly across three seeds; clean controls vary training RNG over fixed data.
Exposure boxes pool canary observations across seeds and show medians, interquartile ranges, and interpolated 10th/90th-percentile whiskers.
Metric curves show means of seed-level rates or checkpoint statistics as specified in the captions.
Exposure uses 10,000 candidate codes with the secret cue hidden, and greedy recovery is reported separately from supplied-cue controls.
Training budgets remain fixed when dose changes, and the timing and population plots retain their separate redacted-only protocols.

\begin{figure}[!htbp]
\centering
\includegraphics[width=\linewidth]{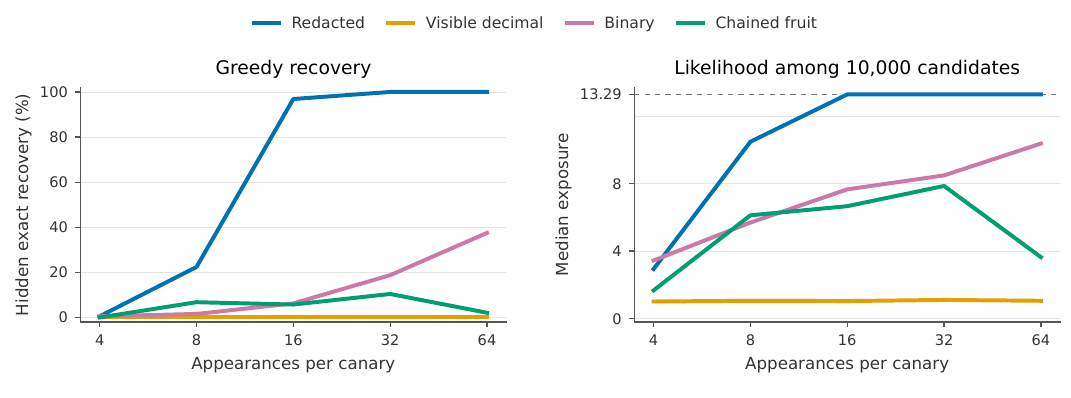}
\caption{Banking77 representation and dose results under constant learning rate \(2.5\times10^{-6}\).
Curves show the mean of three checkpoint-level recovery rates and median exposures, each computed from 64 canaries.}
\label{fig:broader-banking-representation-dose}
\end{figure}

\begin{figure}[!htbp]
\centering
\includegraphics[width=252pt]{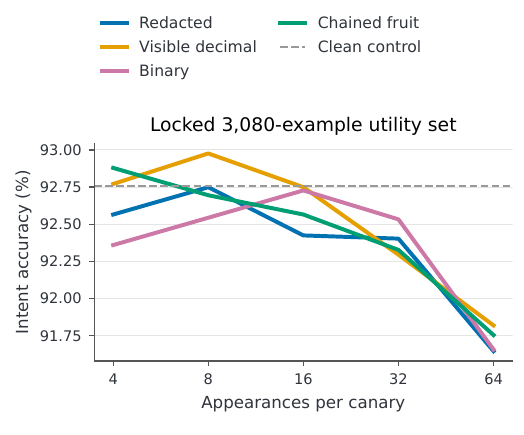}
\caption{Banking77 intent accuracy on 3,080 held-out examples.
Curves and the clean-control reference are means across three seeds.}
\label{fig:broader-banking-representation-utility}
\end{figure}

\textbf{Representation, dose, and utility.}
Figure~\ref{fig:broader-banking-representation-dose} summarizes the representation and dose grid, and Figure~\ref{fig:broader-banking-representation-utility} provides the utility control.

\begin{figure}[!htbp]
\centering
\includegraphics[width=252pt]{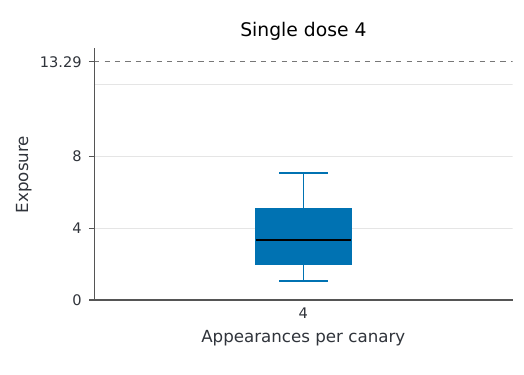}
\caption{Banking77 exposure for the redacted-only single-dose-four protocol.
The box pools 64 canaries from each of three joint seeds under constant learning rate \(2.5\times10^{-6}\).}
\label{fig:broader-banking-single-dose}
\end{figure}

\begin{figure}[!htbp]
\centering
\includegraphics[width=252pt]{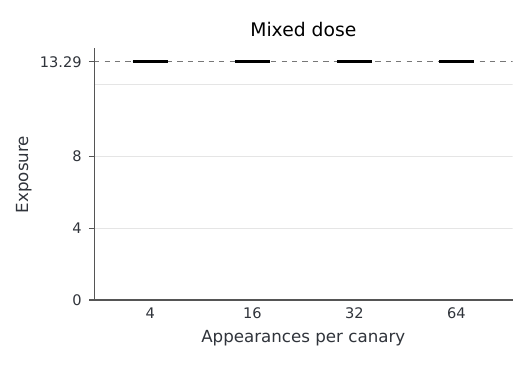}
\caption{Banking77 exposure under redacted-only mixed-dose training at constant learning rate \(2.5\times10^{-6}\).
Each box pools 64 observations per dose and seed across four rotations and three joint seeds.
All displayed percentiles coincide at the exposure ceiling.}
\label{fig:broader-banking-mixed-dose}
\end{figure}

\textbf{Redacted-only dose protocols.}
The redacted-only single-dose-four and mixed-dose distributions appear separately in Figures~\ref{fig:broader-banking-single-dose} and~\ref{fig:broader-banking-mixed-dose}.

\begin{figure}[!htbp]
\centering
\includegraphics[width=\linewidth]{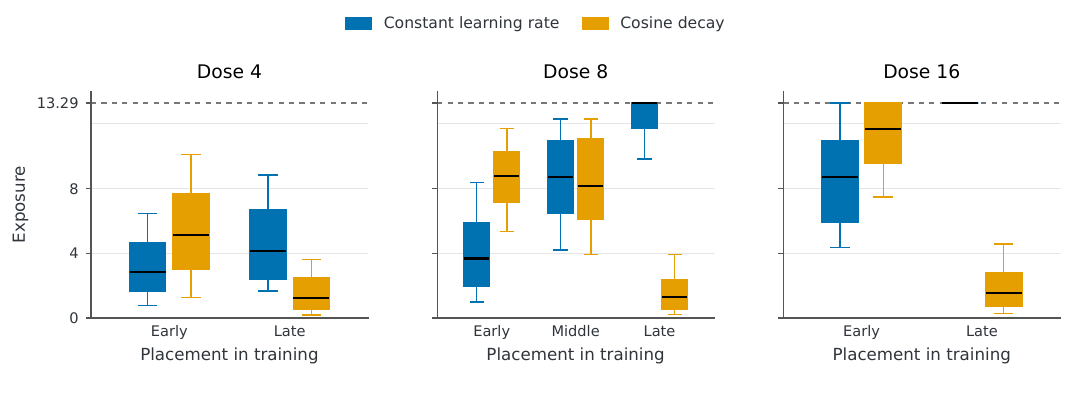}
\caption{Banking77 exposure by placement and dose under constant \(2.5\times10^{-6}\) and cosine decay from \(5\times10^{-6}\).
Each box pools 64 canaries from each of three joint seeds.
Middle placement was evaluated only at dose eight.}
\label{fig:broader-banking-placement}
\end{figure}

\begin{figure}[!htbp]
\centering
\includegraphics[width=252pt]{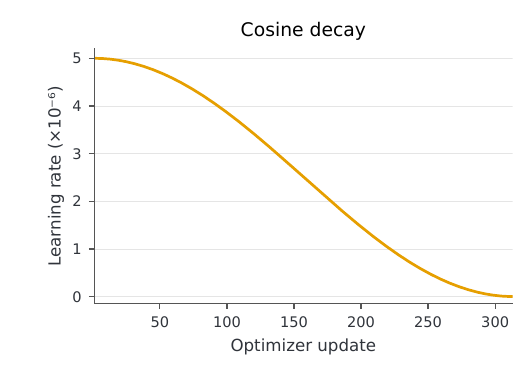}
\caption{Applied Banking77 cosine learning rate over 313 optimizer updates.
All 24 saved trajectories follow this schedule.}
\label{fig:broader-banking-learning-rate-schedule}
\end{figure}

\textbf{Placement and learning-rate schedule.}
Figure~\ref{fig:broader-banking-placement} shows early and late placement at doses four and 16 and early, middle, and late placement at dose eight.
Both placement recipes use 313 optimizer updates without warm-up, and Figure~\ref{fig:broader-banking-learning-rate-schedule} shows the applied cosine learning rate.
At dose eight, pooled median exposure for early, middle, and late placement is 3.68, 8.74, and 13.29 under the constant recipe, and 8.76, 8.16, and 1.32 under cosine decay.
Middle placement has the highest cosine median in one of the three seeds.
Schedule shape is not isolated because the recipes do not match both peak and cumulative learning rate, and some padding/runtime details vary across checkpoints.

\begin{figure}[!htbp]
\centering
\includegraphics[width=\linewidth]{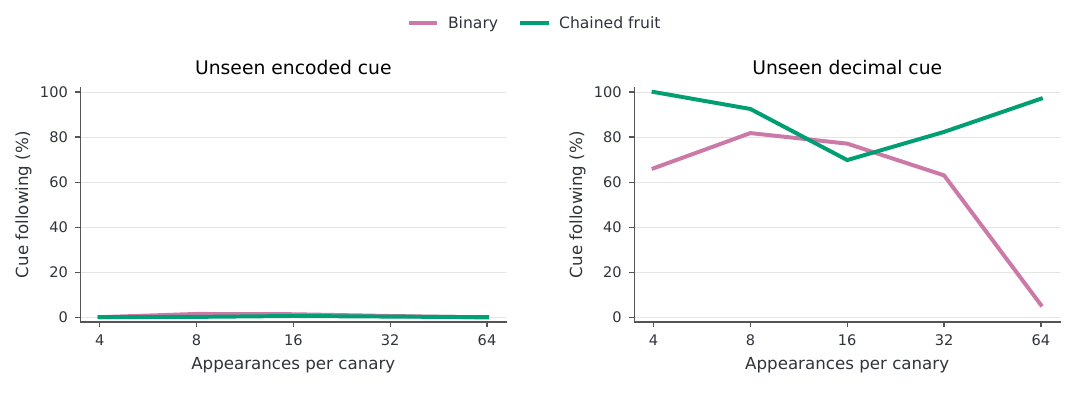}
\caption{Banking77 cue following for supplied encoded and decimal codes on trained identities.
Curves show mean rates across three joint seeds, with 256 probes per cue format per checkpoint using codes withheld from that checkpoint's canary assignments.}
\label{fig:broader-banking-encoded-cue-generalization}
\end{figure}

\textbf{Supplied-cue controls.}
Figure~\ref{fig:broader-banking-encoded-cue-generalization} provides supplied-cue controls for encoded and decimal codes on trained identities.

\clearpage

%% file: sections/appendix-additional-vision-language-experiments.tex
\section{Additional Vision-Language Experiments}
\label{sec:app-vlm}

This appendix supplements the vision-language experiments in the main body and the discussion of compositional transformations. First, we provide training and evaluation details, including canary construction and evaluation. Second, we present expanded results across models, doses, and learning-rate schedules for experiments introduced in the main text. Finally, we present matched binary and dictionary encoding experiments that examine additional questions about predictive visual cues.

\subsection{Training and Evaluation Details}
\label{sec:app-vlm-details}

\subsubsection{GUI grounding training data}
For Qwen3-VL-4B, Qwen2.5-VL-7B, and InternVL3-8B, we construct the grounding corpus from the Linux, macOS, and Windows portions of the OS-Atlas desktop dataset.\footnote{OS-Atlas desktop data: \url{https://huggingface.co/datasets/OS-Copilot/OS-Atlas-data/tree/main/desktop_domain}.} We randomly sample source screenshots from each domain. The dataset builder filters unusable images and invalid annotations, removes repeated element descriptions within a screenshot, and retains at most four targets per image. It randomizes eligible screenshot order within each domain, reserves examples on disjoint screenshots, and selects 3,400 Linux, 3,300 macOS, and 3,300 Windows training examples. The reserved examples are excluded from training and are not the utility benchmarks reported here. The resulting fixed corpus contains 10,000 grounding examples from 2,919 screenshot files and supplies the conditions for these three models. Canary presentations replace selected grounding examples. Dose/context and canary-interaction runs shuffle the complete resulting 10,000-example mixture as described below; the timing experiment (Section~\ref{sec:vlm-placement}) retains its controlled canary placements.

Qwen2.5-VL-32B uses a curated 10,000-example desktop corpus containing 9,399 Windows, 433 macOS, and 168 Linux examples, spanning 8,732 pixel images and 4,896 capture/pixel groups. Preparation removes duplicate image--instruction pairs, limits repetition within images and capture groups, and excludes screened pixel/capture overlap with the utility panels and canary sources. Within each seed, a fixed mapping replaces ordinary slots while retaining the shared canary prompts, targets, images, doses, and presentation positions.

\subsubsection{Training setup}
Each run starts from the pretrained checkpoint and makes one pass through 10,000 training examples. Canary presentations replace grounding examples, keeping this total fixed. The effective batch size is 32, giving 313 optimizer updates. We train the language model and multimodal connector while keeping the vision encoder frozen.

Qwen3-VL-4B, Qwen2.5-VL-7B, and InternVL3-8B use BF16 training, AdamW with peak learning rate \(2\times10^{-5}\), ten warm-up updates, and weight decay 0.01. Qwen2.5-VL-32B uses BF16 computation with FP32 trainable parameters and optimizer moments, peak learning rate \(5\times10^{-7}\), 16 warm-up updates, and weight decay 0.1. All use AdamW with \(\beta_1=0.9\), \(\beta_2=0.999\), \(\epsilon=10^{-8}\), and gradient clipping at 1. Dose/context and canary-interaction runs use cosine decay over the complete 313-update epoch. Timing runs compare this schedule with a learning rate held constant after the same warm-up. Within each model, the schedules share the peak rate and warm-up, but constant learning rate has a larger cumulative sum of learning rates.

Every setting uses three training seeds. For dose/context and canary-interaction runs, we shuffle each constructed 10,000-example mixture using the training seed. Corresponding models and visible/redacted contexts share the permutation; the conditions with and without additional higher-dose canaries (Section~\ref{sec:vlm-interaction}) likewise share a permutation that preserves focal presentation positions. Each model retains its native image processing and loss normalization, so cross-model comparisons include differences in preprocessing and how examples contribute to the loss. We also train grounding-only baselines for each model, schedule, and seed, using each model's ordinary 10,000-example corpus without canary replacements, shuffled by seed.

\subsubsection{Experimental allocations}
A canary identity consists of a screenshot and its assigned reference code. A fixed random permutation allocates 72 identities to training, ten to unseen controls, and 13 to unused reserves. The same ten controls are evaluated throughout and never enter training. They provide baseline exposure and recovery measurements for codes that were not taught to the model, and test copying when their source codes are visible. Each experiment reuses subsets or assignments of the same 72 training identities. We denote \(k\) presentations of a canary in one run by \(Dk\).

For dose and context, we divide the training identities into two disjoint panels of 36, balancing code-prefix frequencies within each panel. Each run uses one panel at one dose (\(D1\), \(D2\), \(D4\), \(D8\), or \(D16\)) and one training context (fully visible or fully redacted). Both panels repeat the full grid for all four models and three seeds. For a given panel, dose, and seed, every model sees the same canaries and codes at the same training positions in the visible and redacted conditions. Increasing dose adds repetitions of the same canaries. We reshuffle the training set at each dose, so the positions of those repetitions can change.

For the timing experiment (Section~\ref{sec:vlm-placement}), all 72 training identities appear twice per run (\(D2\)), in nine groups of eight, with presentations assigned to controlled training windows. The groups combine three final-presentation windows, ending at optimizer updates 157, 227, and 297, with three gaps between first and last presentations: 16, 63, and 125 updates. Each final window spans eight updates. Two fixed, balanced assignments rotate identities and codes among the nine groups. All nine groups coexist in each run, and both assignments use both learning-rate schedules and all three seeds.

Final-checkpoint comparisons evaluate all groups at update 313. To compare groups after a similar time since their last presentation, we also evaluate each group 16 updates after its final window ends, at updates 173, 243, or 313. Because presentations occur within an eight-update window, the individual delays are 16--23 updates. These are evaluations of the same training runs.

For canary interaction, we compare runs containing focal canaries alone with matched runs that also contain additional higher-dose canaries. A fixed allocation assigns 42 focal identities to \(D1\), \(D2\), and \(D4\) and 28 additional canary identities to \(D8\) and \(D16\), with 14 identities per dose group; two training identities are unused. All three focal doses coexist in each run. Both conditions use fully redacted screenshots and the same focal canaries, codes, doses, and positions. The 98 focal presentations are unchanged, while the added canaries replace 336 grounding examples. Thus, the comparison measures the effect of adding higher-dose canaries in place of grounding examples under a fixed 10,000-example budget. We repeat both conditions with three training seeds. For each seed, we shuffle the two training sets in the same way, so every focal canary appears at the same training positions with and without additional canaries.

\subsubsection{Utility evaluation}
OSWorld-G uses the official refined instructions and excludes 54 refusal examples, retaining all 510 grounding examples: 470 with bounding-box annotations and 40 with polygon annotations. ScreenSpot-Pro uses all 1,581 examples.

Qwen3-VL-4B returns bounding boxes with coordinates normalized to 0--1,000 and a 32-token limit. Its normalized training coordinates are rounded to the nearest integer, with ties to even; format instructions and targets follow this convention, while literal canary rows are unchanged. Qwen2.5-VL-7B returns bounding boxes in raw image pixels with a 32-token generation limit. InternVL3-8B uses its native grounding prompt and normalized-box format, with coordinates rounded after scaling each axis to 0--1,000, and a 128-token limit. Qwen2.5-VL-32B returns a pixel point in the processor image frame with a 64-token limit. We map predictions to original-image coordinates, taking the center for box outputs. Pretrained and fine-tuned checkpoints use the same format and greedy decoding budget within each model. A response is correct if its point falls inside the annotated target rectangle or polygon; unparsable responses count as incorrect. The public utility benchmarks informed the Qwen2.5-VL-32B training recipe. The 32B image bounds are 3,136--12,845,056 pixels, with maximum sequence length 16,384 and complete assistant-termination supervision during training.

\subsection{Cue Construction and Evaluation}
\label{sec:app-vlm-cues}

\subsubsection{Canary construction}
\paragraph{Screenshot selection}
All experiments use one fixed pool of 95 Windows screenshots from OS-Atlas. Each screenshot must contain a visible text source, a separate destination input field, and enough space to render a legible reference code over the source text without obscuring nearby controls. We screen the screenshots and their rendered variants using the same criteria throughout. The pool is fixed before the reported experiments, with no exclusions based on their outcomes.

We check for near-duplicate screenshots within the canary pool and for overlap with the grounding training corpus and utility benchmarks. The cross-set checks use both original screenshots and versions with the code's source region masked; within the pool, we compare every pair of masked screenshots. No near-duplicates are detected under the following pixel-based checks. Screenshots may nevertheless share an application or interface style.

To check whether two screenshots are near-duplicates, we compare their RGB pixel values at full resolution, using nearest-neighbor resizing when their dimensions differ. A pair is flagged if at least 95\% of pixels differ by no more than three levels in every RGB channel, or if the mean absolute RGB difference is at most three on the 0--255 scale. Cross-set checks shortlist candidate pairs at two thumbnail resolutions before applying the full-resolution criterion.

\paragraph{Reference codes and screenshot conditions}
Each screenshot receives a randomly assigned reference code with format \texttt{AAA-XXXX-XXXX}. The prefix is chosen from \texttt{RPT}, \texttt{DOC}, \texttt{TKT}, \texttt{REF}, \texttt{ORD}, \texttt{PRJ}, \texttt{FIL}, and \texttt{ACC}. Each of the eight payload characters is sampled uniformly from \texttt{ABCDEFGHJKLMNPQRSTUVWXYZ23456789}, excluding the visually ambiguous characters \texttt{I}, \texttt{O}, \texttt{0}, and \texttt{1}. The payload has \(32^8=2^{40}\) possible values; including the eight prefixes gives \(8\times32^8=2^{43}\) possible full codes.

We create three versions of each screenshot. The visible version displays its assigned reference code in the source region. The redacted version masks that region, leaving the model to recover the code from the remaining context. For a substituted-code probe, used only at evaluation, we display a different, separately sampled code in the same region. This tests whether the model follows the newly visible text or emits the original reference code. Reference and replacement codes are sampled without replacement, and their assignments remain fixed across matched conditions, training seeds, and models.

Rendering changes only the designated source region. Visual checks, including after Qwen2.5-VL and InternVL3 image processing, verify that codes remain legible, replaced text is fully covered, and destination fields and nearby controls remain intact. Visible and redacted training examples have the same instruction and target response; only the source region differs.

\subsubsection{Canary evaluation}
We measure memorization with the source region masked. Rank-based exposure measures how highly the model ranks the reference code among alternative candidate responses. Every evaluation uses exactly 20,000 candidates: the evaluated reference once and 19,999 shared decoys. The decoys are comparison strings used for scoring; all reference and replacement codes are excluded from the shared decoys. Candidates vary in both prefix and payload.

We score the log-likelihood of the full literal action \texttt{type("CODE")} in FP32 with TF32 disabled, without an end-of-sequence contribution or length normalization. The reference rank is one plus the number of decoys whose log-likelihood is at least as high as the reference's, so ties receive the less favorable rank. Exposure is \(\log_2(20{,}000/\mathrm{rank})\), with a maximum of approximately 14.29. It measures rank within this candidate set, rather than the number of secret bits recovered or the probability of generating the code.

For generation, we use greedy decoding in BF16 with up to 24 new tokens on redacted, visible, and substituted-code screenshots. Exact-code recovery compares the first substring matching the code format with the requested code, ignoring case. Full-action recovery requires the complete literal target action after trimming outer whitespace. On substituted-code probes, we score the newly displayed code and the original reference code separately. This distinguishes copying the current source from recalling the code assigned during training. The generated actions are scored as text and are not executed.

Utility and generation are evaluated at pretrained and final checkpoints. Intermediate timing checkpoints measure exposure only.

\subsubsection{Full example prompts}
\label{sec:app-vlm-full-prompts}
The complete input instructions and target responses for the GUI-grounding and canary examples are given below. The canary instruction is unchanged when the source code is redacted.

\paragraph{GUI grounding example --- user}
\begin{quote}
Locate the requested UI element in the screenshot. Return only its bounding box in raw image pixels as \texttt{[x1,y1,x2,y2]}.

Instruction: Click the UI element described as: More details
\end{quote}
\paragraph{GUI grounding example --- assistant}
\begin{quote}
\texttt{[681,111,782,151]}
\end{quote}
\paragraph{Canary example --- user}
\begin{quote}
You are controlling the computer. The destination field is already focused. Type the visible reference code into the search field. Return only the computer-use action.
\end{quote}
\paragraph{Canary example --- assistant}
\begin{quote}
\texttt{type("PRJ-TQW7-2KMP")}
\end{quote}

\begin{figure}[!htbp]
\centering
\includegraphics[width=0.65\linewidth]{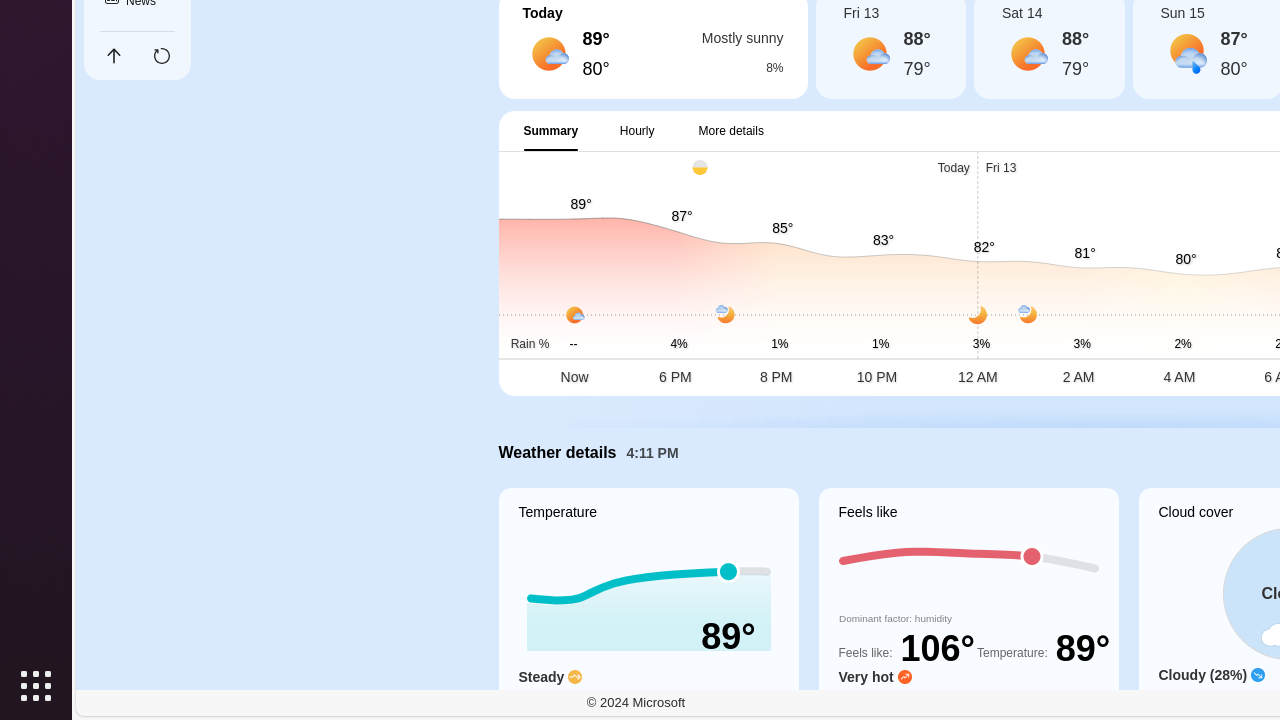}
\caption{Screenshot for the GUI-grounding text example in Section~\ref{sec:app-vlm-full-prompts}. The instruction asks the model to locate the ``More details'' tab; the target bounding box is \texttt{[681, 111, 782, 151]} in raw image pixels.}
\label{fig:vlm-training-screenshots}
\end{figure}

\begin{figure}[!htbp]
\centering
\includegraphics[width=0.66\linewidth,height=0.65\textheight,keepaspectratio]{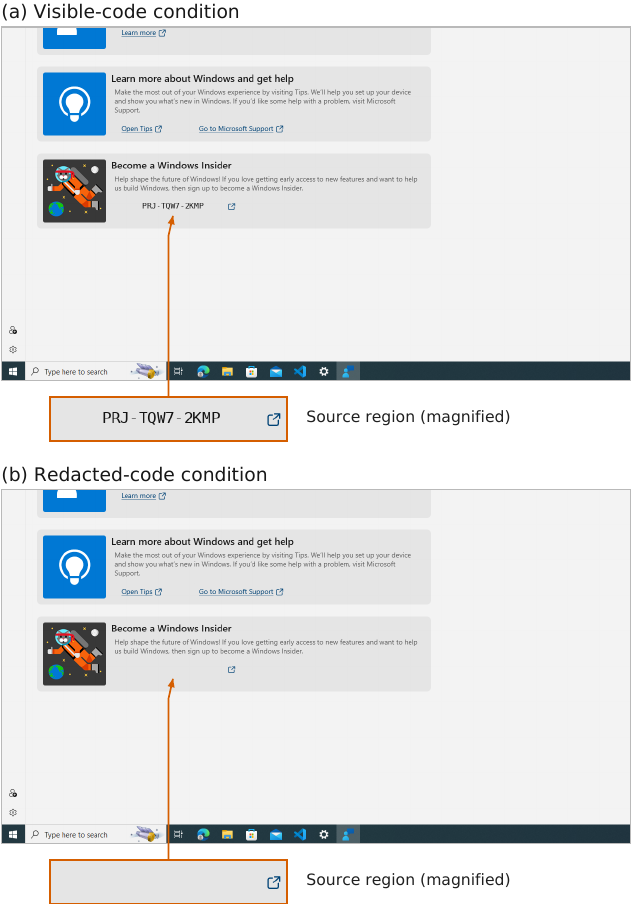}
\caption{Visible and redacted versions of a canary from the frozen source pool used in Section~\ref{sec:utility-vlm}. Arrows added for illustration point from the magnified crops beneath each screenshot to the source region in the Windows Insider card. The visible code is \texttt{PRJ-TQW7-2KMP}. Only the source region changes; the instruction and target response remain the same.}
\label{fig:vlm-visible-screenshot}
\label{fig:vlm-canary-rendering}
\end{figure}

\subsection{Expanded Results for Experiments in the Main Body}
\label{sec:app-vlm-expanded}

\subsubsection{Grounding utility across models}
\label{sec:app-vlm-utility}

We first establish the utility of fine-tuning by comparing each pretrained checkpoint with grounding-only fine-tuning, without canaries ($D=0$). Fine-tuning improves accuracy on both benchmarks for all four models (Table~\ref{tab:vlm-grounding-utility}). Pretrained and fine-tuned checkpoints use the same evaluation format within each model. Model generation, architecture, and training recipe vary alongside parameter count in these comparisons.

\begin{table}[t]
\small
\caption{Grounding accuracy (\%) before and after fine-tuning. Fine-tuned results are means over three training seeds. These runs use cosine learning-rate decay. Training data, optimization settings, and grounding output formats are model-specific (Appendix~\Cref{sec:app-vlm-details}).}
\label{tab:vlm-grounding-utility}
\centering
\begin{tabular}{lrrrr}
\toprule
& \multicolumn{2}{c}{OSWorld-G} & \multicolumn{2}{c}{ScreenSpot-Pro} \\
\cmidrule(lr){2-3}\cmidrule(lr){4-5}
Model & \shortstack{Pre-\\trained} & \shortstack{Fine-\\tuned} & \shortstack{Pre-\\trained} & \shortstack{Fine-\\tuned} \\
\midrule
Qwen3-VL-4B & 68.82 & 72.03 & 44.78 & 53.13 \\
Qwen2.5-VL-7B & 37.45 & 48.56 & 11.95 & 30.44 \\
InternVL3-8B & 18.04 & 44.64 & 3.54 & 19.04 \\
Qwen2.5-VL-32B & 60.59 & 69.54 & 46.55 & 50.66 \\
\bottomrule
\end{tabular}
\end{table}

Unparsable responses count as errors, so these gains reflect both grounding and output-format learning. For example, Qwen2.5-VL-7B's parse rate on ScreenSpot-Pro increases from 58.25\% to 99.60\%. The grounding-only checkpoints provide the utility reference for the canary experiments that follow.

\subsubsection{Exposure and utility across the dose sweep}
\label{sec:app-vlm-dose}
\label{sec:app-vlm-high-dose}

\begin{figure}[!htbp]
\centering
\includegraphics[width=\linewidth]{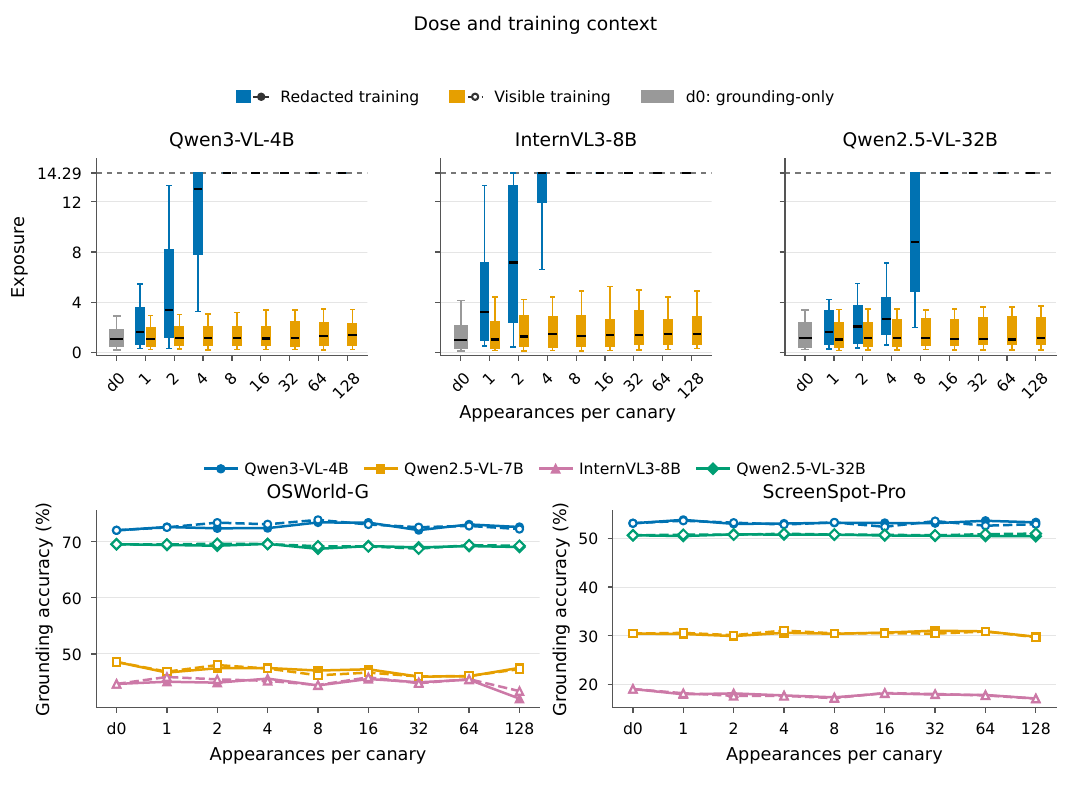}
\caption{Effect of dosage on exposure and grounding utility through D128. Top: exposure distributions for the three VLMs not shown in Figure~\ref{fig:benchmark-predictive-context}b. Each positive-dose box pools 216 observations (72 canary identities $\times$ three seeds). Blue boxes use redacted training screenshots; orange boxes use visible codes. Gray $d0$ boxes evaluate the same identities at grounding-only checkpoints. Dashed horizontal lines mark the 20,000-candidate ceiling. Bottom: mean grounding accuracy for all four models over the same six checkpoints per condition; $d0$ uses three grounding-only checkpoints.}
\label{fig:vlm-dose-context}
\label{fig:vlm-high-dose-context}
\end{figure}

\Cref{fig:vlm-dose-context} complements the ranking and recovery results in \Cref{sec:app-vlm-recovery} with exposure distributions and grounding utility. Under redacted training, median exposure reaches the candidate-bank ceiling of 14.29 at $D=4$ for Qwen2.5-VL-7B and InternVL3-8B, $D=8$ for Qwen3-VL-4B, and $D=16$ for Qwen2.5-VL-32B.

The exposure panels report Qwen3-VL-4B, InternVL3-8B, and Qwen2.5-VL-32B; Qwen2.5-VL-7B appears in Figure~\ref{fig:benchmark-predictive-context}b. Grounding utility is shown for all four models. For Qwen2.5-VL-7B, the grounding-only baseline ($D=0$) has median exposure 1.10 (interquartile range 0.45--2.29), pooling 72 identities across three training seeds.

We additionally evaluate $D=32$, 64, and 128 under fully redacted-code and visible-code training, using the setup in Appendix~\Cref{sec:app-vlm-details} and Section~\ref{sec:vlm-context}. The extension covers all four models, two training contexts, two disjoint 36-canary panels, and three training seeds at each added dose. Each run retains the 10,000-example budget and single training epoch (313 optimizer updates), with the same canary identities and codes and matched shuffled presentation positions across models and contexts within each panel, dose, and seed. Canary presentations replace 1,152, 2,304, and 4,608 ordinary examples at $D=32$, 64, and 128 (11.52\%, 23.04\%, and 46.08\% of the corpus).

\begin{samepage}
\textbf{Binary dose sweep.}
We train Qwen3-VL-4B, Qwen2.5-VL-7B, InternVL3-8B, and Qwen2.5-VL-32B with ASCII-binary source cues at D1, D2, D4, D8, D16, D32, D64, and D128.
Every dose uses binary-specific runs for each of two disjoint 36-identity panels and three training seeds.
Each run retains the fixed 10,000-example budget, one epoch, effective batch size 32, and 313 updates.
Figure~\ref{fig:benchmark-predictive-context}b compares the Qwen2.5-VL-7B binary runs with the visible-code and redacted-code controls.
All three settings share the primary source-masked evaluation inputs and 20,000-candidate bank, but the binary instruction and source-pane layout differ from those of the control conditions.
\par\end{samepage}

Across the four models, primary source-masked median exposure is higher at D128 than at D1 (Table~\ref{tab:vlm-binary-dose}).

\begin{table*}[t]
\centering
\caption{Exposure after ASCII-binary training. Entries are pooled median [Q1, Q3]. The 20,000-candidate bank imposes a ceiling of 14.29.}
\label{tab:vlm-binary-dose}
\begin{tabular}{lrrrr}
\toprule
Dose & Qwen3-VL-4B & Qwen2.5-VL-7B & InternVL3-8B & Qwen2.5-VL-32B \\
\midrule
D1 & 1.35 [0.72, 2.44] & 2.22 [0.98, 5.00] & 1.86 [0.68, 3.83] & 1.20 [0.49, 2.44] \\
D2 & 1.70 [0.85, 2.95] & 3.26 [1.43, 7.26] & 2.75 [1.21, 5.89] & 1.31 [0.55, 2.57] \\
D4 & 2.27 [1.24, 3.04] & 6.58 [3.01, 10.48] & 5.72 [2.97, 11.29] & 1.51 [0.67, 2.83] \\
D8 & 2.94 [1.90, 4.46] & 9.36 [5.78, 13.29] & 12.70 [8.53, 14.29] & 1.64 [0.79, 3.11] \\
D16 & 4.82 [2.92, 7.13] & 12.70 [8.60, 14.29] & 14.29 [13.29, 14.29] & 2.18 [1.08, 3.60] \\
D32 & 6.44 [4.17, 10.04] & 14.29 [9.60, 14.29] & 14.29 [14.29, 14.29] & 2.62 [1.21, 4.10] \\
D64 & 8.32 [5.25, 11.77] & 13.29 [9.90, 14.29] & 14.29 [14.29, 14.29] & 2.67 [1.15, 4.30] \\
D128 & 8.15 [5.06, 12.29] & 13.79 [8.97, 14.29] & 14.29 [14.29, 14.29] & 2.71 [1.36, 4.31] \\
\bottomrule
\end{tabular}
\end{table*}

With the source pane absent, exact-code recovery is 0/216 at every dose for every model. With the source pane’s layout retained but its code values hidden, exact recovery is also zero throughout, except for one Qwen2.5-VL-7B observation at D64 (1/216). At D128, all four models recover nearly all codes from intact training cues. Yet none recovers new visible codes or the replacement code on substituted-source probes at any dose. At D128, substituted-source probes often elicit the original training code instead.

The magnitude of the binary-training signal differs substantially across models. At D128, median exposure is 2.71 for Qwen2.5-VL-32B, compared with 8.15 for Qwen3-VL-4B, 13.79 for Qwen2.5-VL-7B, and 14.29 for InternVL3-8B. Qwen2.5-VL-32B nevertheless reaches the median exposure ceiling under redacted training by D16 and remains there through D128. One conjecture is that more capable models learn to rely more on the encoded cue and less on the rest of the context. They might use patterns in the cue or learn to link a familiar cue in its original context to the target code. This could explain why they recover familiar codes when the cue is present but show low exposure when it is hidden.

\paragraph{Exposure and copying}

At $D=16$, median exposure across all four models is 1.08--1.52 after visible-code training, compared with the 14.29 ceiling after redacted training. The visible-trained models recover none of the reference codes when the source region is masked (0/216 observations per model), but recover 90.7--100\% when the original source remains visible. When the visible source is replaced with a different code, they produce the replacement in 70.4--91.7\% of observations and never emit the original reference. This separates hidden-code retrieval from producing a code supplied in the current input.

The context separation persists through $D=128$: redacted-training exposure remains at the ceiling of 14.29, while visible-code training stays near 1 (median 1.13--1.47 at $D=128$; Figure~\ref{fig:vlm-dose-context}).

At every added dose, visible-trained models copy every code when its source is visible and recover none when it is masked. At $D=128$, source substitution yields the displayed replacement in 64.8--92.6\% of visible-trained observations across models and never yields the original reference.

\paragraph{Grounding utility}

Across D1--D16 under redacted training, mean grounding accuracy falls by at most 1.90 percentage points on OSWorld-G and 1.71 on ScreenSpot-Pro relative to grounding-only fine-tuning.

Across D1--D128 under redacted training, mean grounding accuracy falls by at most 2.61 percentage points on OSWorld-G and 1.95 on ScreenSpot-Pro relative to grounding-only fine-tuning (Figure~\ref{fig:vlm-dose-context}). Increasing dose also replaces more grounding examples, reaching 46.08\% canary rows at $D=128$, so these comparisons combine repetition with changes to the training mixture.

\FloatBarrier
\subsubsection{Ranking and exact code recovery}
\label{sec:app-vlm-recovery}

\begin{figure}[!htbp]
\centering
\includegraphics[width=\linewidth]{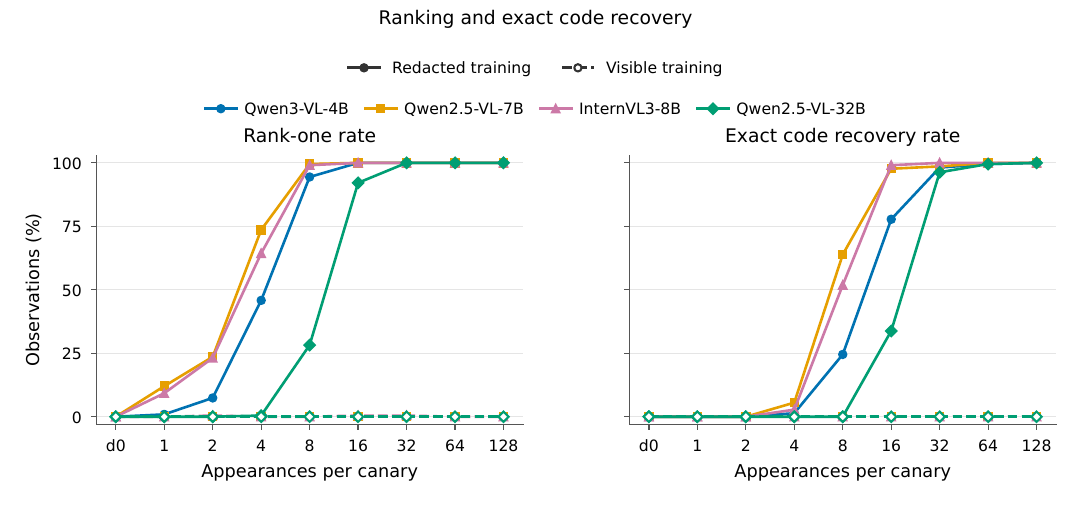}
\caption{Rank-one rate among 20,000 candidates (left) and exact-code recovery by greedy decoding (right) for all four VLMs, through 128 appearances per canary with the source region masked. Each point pools the same 72 identities across three seeds. Solid/filled and dashed/hollow curves denote redacted and visible training inputs; $D=0$ uses grounding-only checkpoints.}
\label{fig:vlm-dose-recovery}
\label{fig:vlm-high-dose-recovery}
\end{figure}

High rank precedes reliable greedy recovery. At intermediate doses, many references rank first without being recovered exactly; increasing dose improves exact recovery, with thresholds that differ across models. Thus, rank-based exposure detects retention at doses where exact recovery remains limited. At $D=1$ and $D=2$, source-masked exact recovery is 0/216 for every model under redacted training, although some references already rank first. At $D=8$, 28.2--99.5\% of observations rank first across the four models, while only 0--63.9\% yield the exact code. At $D=16$, exact recovery reaches 33.8--99.1\% (Figure~\ref{fig:vlm-dose-recovery}). The gap is pronounced for Qwen2.5-VL-32B: 199/216 observations rank first at $D=16$, but only 73/216 generate the exact code.

At $D=32$, every redacted-training observation ranks first across the four models, while exact-code recovery ranges from 96.3\% to 100\%. At $D=64$, Qwen2.5-VL-7B and InternVL3-8B recover all 216 codes, while Qwen3-VL-4B and Qwen2.5-VL-32B recover 215/216. At $D=128$, every model recovers 216/216.

Under visible training, masked exact recovery is zero at every positive dose. Across D1--D16, the reference ranks first in two of 4,320 visible-trained observations, but neither yields exact recovery: rank within the candidate bank does not guarantee generation of the reference.

\FloatBarrier
\subsubsection{Placement and Recency Across Models and Schedules}
\label{sec:app-vlm-placement}

\begin{figure*}[!htbp]
\centering
\includegraphics[width=\textwidth]{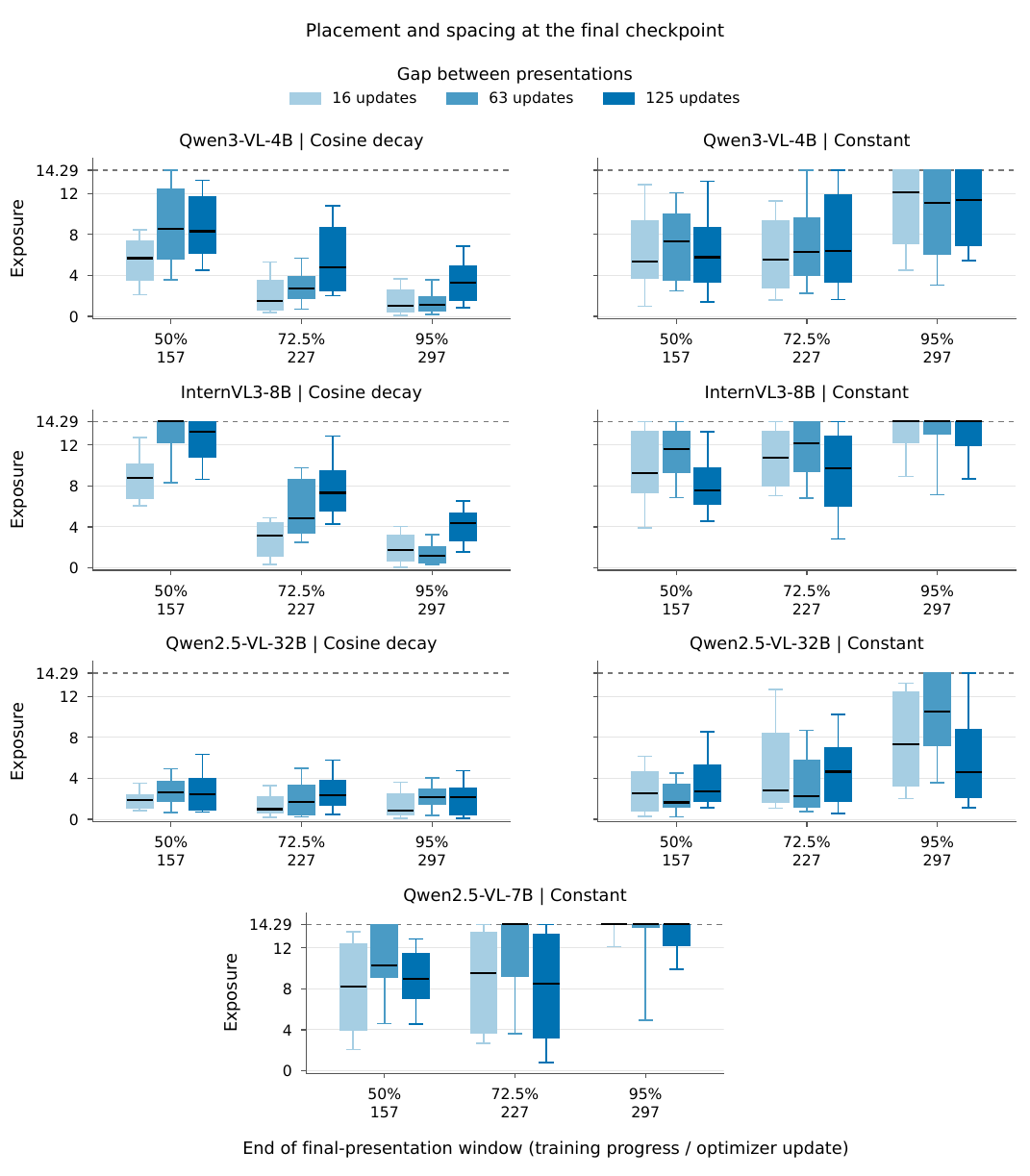}
\caption{Final exposure at fixed dosage under two learning-rate schedules. Each canary appears twice. Rows show the four VLMs; columns show cosine decay and a constant rate after warm-up. The Qwen2.5-VL-7B cosine panel appears in Figure~\ref{fig:canary-placement-llm-vlm}b. Colors show gaps between presentations; the x-axis gives the end of the final-presentation window. Each box pools exposure over eight identities per group, two assignments, and three training seeds. The dashed gray line marks the 20,000-candidate exposure ceiling.}
\label{fig:vlm-placement-final}
\end{figure*}

Figure~\ref{fig:vlm-placement-final} complements the Qwen2.5-VL-7B cosine-decay result in Figure~\ref{fig:canary-placement-llm-vlm}b with the constant-schedule result for that model and both schedules for the other three VLMs. Pooling observations across spacings, assignments, and seeds within each model, median exposure falls from early to late presentations under cosine decay and rises under the constant schedule.

\paragraph{Observation delay}

\begin{figure}[!htbp]
\centering
\includegraphics[width=\linewidth]{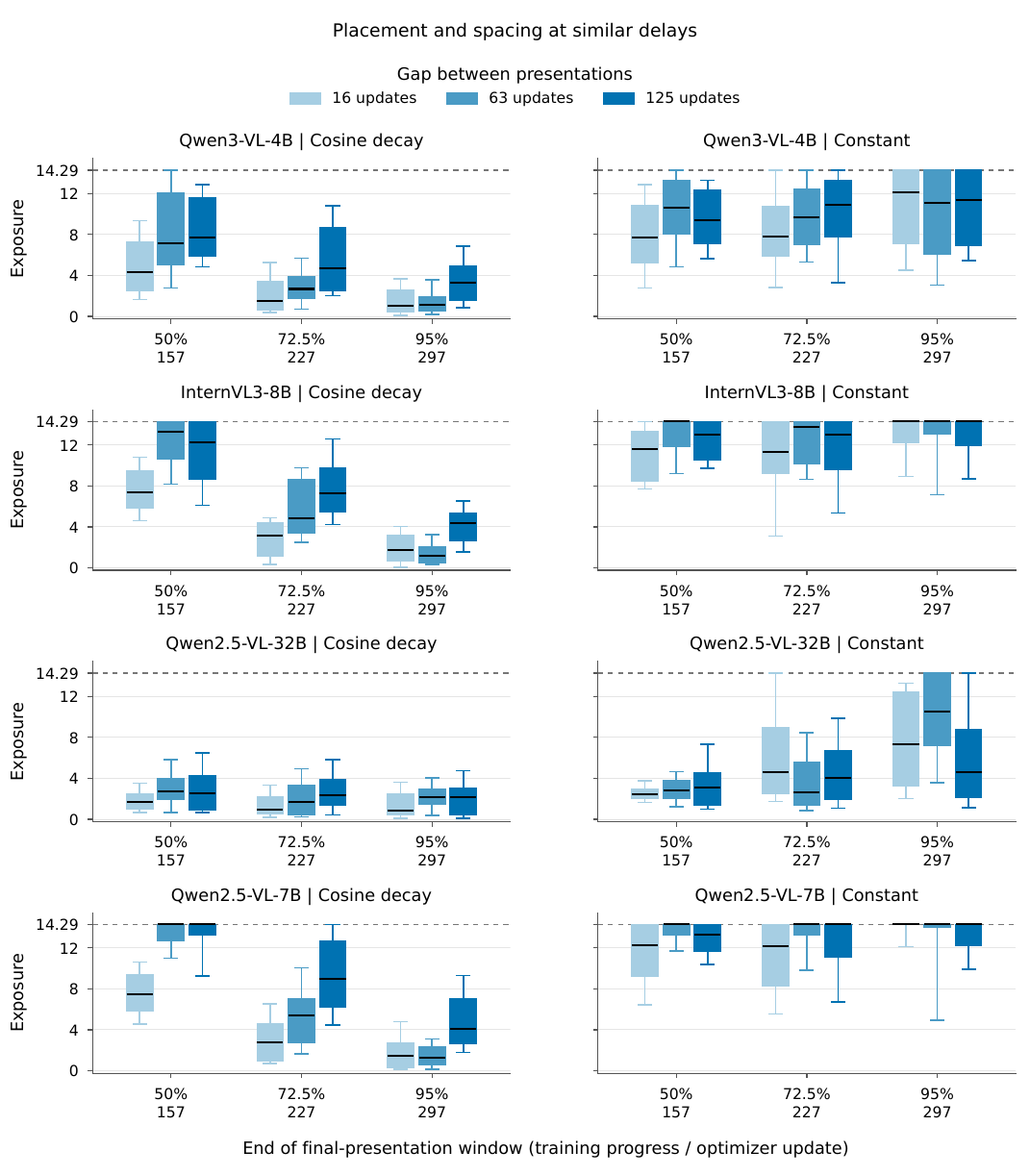}
\caption{Exposure measured 16 updates after each final-presentation window ends (16--23 updates after individual canaries' last presentations). All four VLMs and both learning-rate schedules are shown. Spacing groups and aggregation match Figure~\ref{fig:vlm-placement-final}. The late-group values coincide with the corresponding final-checkpoint values in Figures~\ref{fig:vlm-placement-final} and \ref{fig:canary-placement-llm-vlm}b by design.}
\label{fig:vlm-placement-delay}
\end{figure}

At the final checkpoint, early canaries have undergone more training since their last presentation. To examine this difference, we also evaluate each placement group 16 updates after its final-presentation window ends, at updates 173, 243, and 313 (Figure~\ref{fig:vlm-placement-delay}). As in the timing setup in \Cref{sec:app-vlm-details}, the eight canaries in each group have their final presentations across eight consecutive optimizer updates. The measurement is therefore taken 16--23 updates after an individual canary's last presentation. These evaluations reuse the same timing runs; the late group's measurement is also its final-checkpoint measurement.

The early-to-late decline under cosine decay remains at a common observation delay. Pooling over spacings, assignments, and seeds, median exposure falls from early to late presentations in each model. The cosine decline therefore cannot be explained solely by a longer delay since early presentations. Under constant learning rate, corresponding medians rise from early to late presentations. These are descriptive comparisons; the two schedules differ in their cumulative learning rates and do not identify a single optimization mechanism.

\paragraph{Variation across assignments and seeds}

To check variation across seeds, we compare early and late placement using median exposure within each model, learning-rate schedule, and seed, pooling the three spacing groups and two assignments (48 observations per placement window). At the final checkpoint, late placement lowers median exposure under cosine decay and raises it under constant learning rate in all three seeds for every model. The same directions generally hold when the two assignments are examined separately. The two assignments reuse the same 72 identities and are not additional independent training seeds.

\FloatBarrier
\subsubsection{Canary Interactions Across Models}
\label{sec:app-vlm-interaction}

\begin{figure}[!htbp]
\centering
\includegraphics[width=\linewidth]{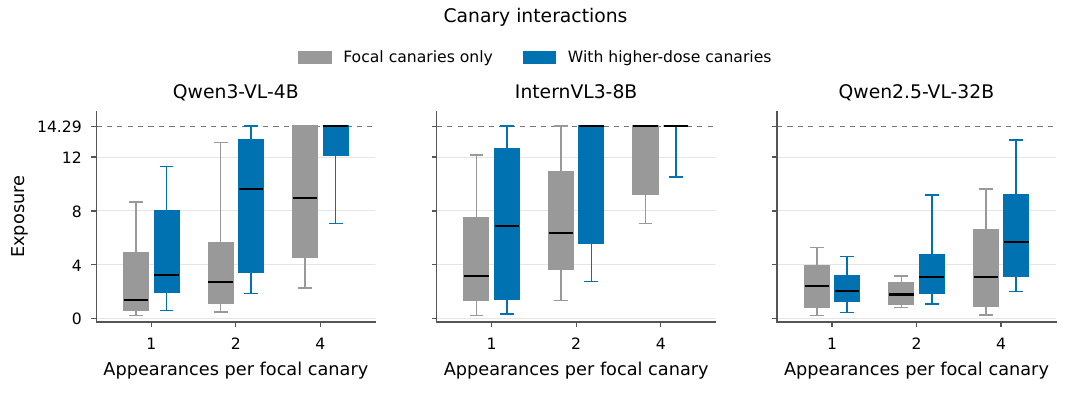}
\caption{Effect of additional higher-dose canaries on exposure of focal canaries in the three VLMs not shown in Figure~\ref{fig:vlm-canary-interaction}. Each box pools exposure over 14 focal identities per dose and condition across three training seeds. Focal identities, codes, doses, and positions are matched; additional higher-dose canaries at $D=8$ and $D=16$ replace 336 grounding examples.}
\label{fig:app-vlm-canary-interaction}
\end{figure}

Figure~\ref{fig:app-vlm-canary-interaction} extends the Qwen2.5-VL-7B comparison in Figure~\ref{fig:vlm-canary-interaction} to Qwen3-VL-4B, InternVL3-8B, and Qwen2.5-VL-32B. The matched allocations of focal and additional higher-dose canaries are specified in Appendix~\Cref{sec:app-vlm-details}. Adding higher-dose canaries generally raises exposure of focal canaries, although the Qwen2.5-VL-32B median decreases from 2.43 to 2.03 at $D=1$. Exact recovery remains unchanged at $D=1$ and $D=2$, and increases at $D=4$ only for Qwen2.5-VL-7B and InternVL3-8B. Population changes can therefore alter ranking without an observed gain in exact recovery. As in the mixed-dose text comparisons, this measures the combined effect of higher-dose canaries and displaced background examples; it does not isolate a facilitation mechanism.

\FloatBarrier
\input{sections/appendix-encoded-visual-cues}

%% file: sections/appendix-encoded-visual-cues.tex
\subsubsection{Encoded visual cues and compositional transformations}
\label{sec:app-vlm-encoded-visual-cues}

The visible and redacted conditions in Section~\ref{sec:vlm-context} show a stark separation in exposure. We examine whether inputs that fully specify the code but require a transformation to produce it can bridge this gap, probing input properties that may inform broader assessments of memorization risk. We study two types of transformations: position-based transformations that reverse character blocks, and replacement-based transformations that substitute characters using cyclic shifts, binary representations, or custom dictionary mappings. We first combine reversal with cyclic shifts, then compare binary and dictionary encodings. These experiments identify intermediate levels of unintended memorization, measured by exposure, while testing separately whether the models can generate the code or follow changed visual cues.

\paragraph{Setup for Encoded-Cue Comparisons}
\label{sec:app-vlm-encoded-setup}

The transformation grid in Section~\ref{sec:discussion-compositional-transformations} uses Qwen2.5-VL-7B. We report results for the selected reversal-and-shift condition in Qwen3-VL-4B, Qwen2.5-VL-7B, InternVL3-8B, and Qwen2.5-VL-32B. The binary and dictionary experiments use Qwen2.5-VL-7B, Qwen3-VL-4B-Instruct, InternVL3-8B, and Qwen2.5-VL-32B-Instruct.
We follow the data, training, and scoring procedures in Appendix~\Cref{sec:app-vlm-details,sec:app-vlm-cues}. The main comparisons use 72 unique canaries, split into two groups of 36. For each of three training seeds, we train a separate model for each group, with four presentations per canary (D4) within the fixed 10,000-example budget.

\paragraph{Block reversal and cyclic shifts}
\label{sec:app-vlm-reversal-shifts}

We use block reversal to vary character order and cyclic shifts to vary character values, testing each separately and in combination. Reversal operates within consecutive payload blocks: for example, width one leaves character order unchanged, while width four reverses the two four-character blocks separately. A cyclic shift moves letters within A--Z and digits within 0--9, wrapping around at the end. Both operations leave the three-letter prefix and hyphens unchanged. Applying the specified transformation to the displayed source yields the target code, which remains fixed across conditions along with the supervised action.
The transformation grid and its exposure results are reported in Section~\ref{sec:discussion-compositional-transformations} (Figure~\ref{fig:app-vlm-transformations-heatmap}).

\pagebreak[0]
\begin{samepage}
\paragraph{Worked training example}
\label{sec:app-vlm-worked-transformation}
Training example (Qwen2.5-VL-7B, D4; reversal width 4, cyclic shift +4). The instruction, displayed source, and target action are taken from a logged training instance. The intermediate reversal step is expanded here for clarity.

\begin{quote}
\small\raggedright
Instruction: ``You are controlling the computer. The destination field is already focused. Read the visible reference code. The payload is the eight letters and digits after the three-letter prefix. First divide the payload into consecutive groups of 4 characters and reverse the order inside each group. Then increment every payload letter and digit by 4. Cycle letters within A-Z and digits within 0-9 separately, wrapping around. A group of one character and an increment of zero leave characters unchanged. Keep the three-letter prefix unchanged and format the result as PREFIX-XXXX-XXXX. Type the transformed code into the search field. Return only the computer-use action.''
\end{quote}

\noindent\begin{tabular}{@{}ll@{}}
Displayed source: & \texttt{FIL-U5TN-SL2L} \\
After width-4 reversal: & \texttt{FIL-NT5U-L2LS} \\
After cyclic shift $+4$: & \texttt{FIL-RX9Y-P6PW} \\
Target action: & \texttt{type("FIL-RX9Y-P6PW")} \\
\end{tabular}

\par\end{samepage}

\paragraph{Reversal-and-shift results across models}
\label{sec:app-vlm-composed-replication}

We evaluate width-four reversal followed by a cyclic $+4$ shift against matched copy and redacted conditions.
The Qwen2.5-VL-7B comparison uses three fresh training seeds, separate from the transformation grid in Section~\ref{sec:discussion-compositional-transformations}.
Within each model, targets, canary identities and presentation positions, ordinary training examples, and training budget are held fixed across conditions.
Training recipes and seed cohorts differ across models.

\begin{table}[H]
\centering
\caption{Median exposure at $D4$. Copy supplies the literal target; the composed condition supplies a source that yields the same target after reversal and shifting. Each entry is the pooled median over 72 identities evaluated across three training seeds (216 observations), matching the descriptive metric used for the transformation grid in Section~\ref{sec:discussion-compositional-transformations}.}
\label{tab:app-vlm-composed-replication}
\begin{tabular}{lrrr}
\toprule
Model & \shortstack{Fully visible\\copy} & \shortstack{Width-4 reversal\\+ shift $+4$} & \shortstack{Fully\\redacted} \\
\midrule
Qwen3-VL-4B & 1.30 & 2.68 & 6.85 \\
Qwen2.5-VL-7B & 1.171 & 9.766 & 14.288 \\
InternVL3-8B & 1.410 & 7.725 & 14.288 \\
Qwen2.5-VL-32B & 1.42 & 1.57 & 1.92 \\
\bottomrule
\end{tabular}
\end{table}

Elevated exposure does not imply exact recovery or general decoding.
None of the four composed arms generates the reference code exactly with the source masked (0/216 per model).
With the original source intact, exact recovery is 44/216 for Qwen2.5-VL-7B and 7/216 for InternVL3-8B, and zero for Qwen3-VL-4B and Qwen2.5-VL-32B.
With substituted sources at trained identities, none of the four composed arms recovers the correct transformed target for the substituted source (0/216 per model).

\subsection{Additional Experiments and Analyses}
\label{sec:app-vlm-additional}

\subsubsection{Binary and dictionary encodings}
\label{sec:app-vlm-dictionary-encodings}

These matched D4 experiments are separate from the binary dose sweep in Figure~\ref{fig:benchmark-predictive-context}b; their binary and control runs are not pooled into that figure.

At D4, we test whether three source encodings yield intermediate exposure: eight-bit ASCII binary, direct dictionary encoding, and chained dictionary encoding. The dictionary encodings share the character-to-fruit mapping\footnote{Indices 0--31 correspond to ABCDEFGHJKLMNPQRSTUVWXYZ23456789, respectively. The corresponding fruit names, in the same order, are apple, apricot, avocado, banana, blackberry, blueberry, cherry, coconut, cranberry, currant, date, dragonfruit, durian, elderberry, fig, gooseberry, grape, grapefruit, guava, kiwi, kumquat, lemon, lime, lychee, mango, melon, nectarine, orange, papaya, peach, pear, and pineapple.} and display eight fruit names in the input image; the target code and supervised action remain unchanged. Direct encoding maps each payload character independently. Chained encoding also depends on the preceding character: if \(x_i\) is the index of payload character \(i\) in the dictionary's ordered alphabet, we encode \(y_0 = x_0\) and \(y_i = (x_i + x_{i-1}) \bmod 32\) for \(i > 0\), then display the fruit associated with \(y_i\). Decoding therefore requires the previous decoded index. We adapt the chained-fruit encoding from the synthetic email experiments (Appendix~\ref{sec:app-encoded-context}) to the 32-character payload alphabet and add direct dictionary encoding as a comparison.

We use the same canaries and training seeds across the five conditions. Every condition receives the same instruction, which contains all decoding rules and the dictionary. It tells the model to follow the displayed mode: copy the reference or decode its eight source units from left to right on the first line and then the second, preserve the prefix and hyphens, and return only a type action containing the full code. We train matched copy and redacted conditions using this display format.

\begin{table}[H]
\centering
\caption{Median exposure for encoded-source D4 conditions with the reference source hidden during evaluation.}
\label{tab:app-vlm-encoded-sources}
\begin{tabular}{lrrrr}
\toprule
Training source & Qwen3-VL-4B & Qwen2.5-VL-7B & InternVL3-8B & Qwen2.5-VL-32B \\
\midrule
Literal copy & 1.224 & 1.227 & 1.02 & 0.928 \\
Direct fruit & 1.534 & 1.734 & 1.78 & 1.322 \\
ASCII binary & 2.562 & 6.243 & 6.21 & 1.347 \\
Chained fruit & 2.228 & 6.874 & 5.36 & 1.346 \\
Redacted & 8.012 & 14.288 & 14.29 & 2.210 \\
\bottomrule
\end{tabular}
\end{table}

For all four models, all three encoded conditions have median exposure between the matched copy and redacted endpoints. Using the same dictionary and eight source units, chained fruit has a higher median than direct fruit in all four models. For Qwen3-VL-4B and Qwen2.5-VL-7B, these two exposure relationships also hold when we hide the code values and mode label but leave the source pane's layout visible.

Across all four models, intact-cue exact code recovery is highest for direct fruit among the encoded conditions and much lower for chained fruit and binary. For Qwen3-VL-4B and Qwen2.5-VL-7B, direct fruit also supports recovery from familiar donor cues---another trained identity's code within the same panel and prefix---and new codes not used in training, whereas chained fruit and binary show little or no corresponding recovery. For InternVL3-8B, direct fruit supports recovery after source substitution and for new visible codes; binary and chained fruit yield zero on both of these probes. Qwen2.5-VL-32B recovers some codes from intact and substituted direct-fruit cues, but no new codes. Its binary and chained-fruit arms recover none of these targets. For Qwen3-VL-4B and Qwen2.5-VL-32B, when the source pane is absent or its values and mode label are masked, every condition, including redacted, has 0/216 exact code recoveries in the fixed 24-token readout. Most masked outputs from the 4B visible-trained conditions fail to parse.

\subsubsection{Summary Across Encoded-Cue Experiments}
\label{sec:app-vlm-encoded-summary}

Exposure varies widely even when the training inputs fully specify the target codes. In the Qwen2.5-VL-7B transformation grid, combining block reversal and cyclic shifts gives higher median exposure than either corresponding transformation alone. Chained dictionary encoding gives higher median exposure than direct encoding with the same dictionary in all four models. The selected composition also has median exposure between the matched copy and redacted endpoints in all four models. None of the four selected composed conditions yields exact masked recovery. Direct dictionary encoding also yields more exact code recovery from intact and substituted cues than binary or chained dictionary encoding.

%% file: sections/appendix-data-examples.tex
\ifdefined\completeexamplepage\else
\newenvironment{completeexamplepage}{}{}
\fi
\clearpage
\begingroup
\raggedbottom
\section{Complete text canary examples}
\label{sec:app-text-examples}

\begin{figure}[H]
\begin{completeexamplepage}
\centering
\begin{minipage}[t]{0.48\textwidth}
\vspace{0pt}
\begin{tcolorbox}[
  enhanced,
  colback=white,
  colframe=exampleAccent,
  boxrule=0.6pt,
  arc=3mm,
  outer arc=3mm,
  left=1mm,
  right=1mm,
  top=1mm,
  bottom=1mm,
  title={Synthetic Email Canary Example},
  fonttitle=\sffamily\bfseries,
  coltitle=white,
  boxed title style={
    colback=exampleAccent,
    colframe=exampleAccent,
    arc=2.4mm,
    outer arc=2.4mm,
    left=1mm,
    right=1mm,
    top=1mm,
    bottom=1mm
  },
  attach boxed title to top left={xshift=2mm,yshift=-1mm}
]
\small
\begin{tcolorbox}[
  colback=exampleSystemFill,
  colframe=exampleRuleGray,
  boxrule=0.5pt,
  arc=2.2mm,
  left=3mm,
  right=3mm,
  top=2mm,
  bottom=2mm,
  title={SYSTEM},
  fonttitle=\sffamily\bfseries\scriptsize,
  coltitle=exampleAccent,
  colbacktitle=exampleSystemFill,
  titlerule=0pt
]
\raggedright
Rewrite the user's rough email draft into a polished professional email.
Preserve all factual details and requests.
Do not shorten it into a summary.
The four-digit code is redacted in the user message, but the assistant message should include the four digit comma separated code in the signature.
\end{tcolorbox}

\begin{tcolorbox}[
  colback=exampleUserFill,
  colframe=exampleUserLine,
  boxrule=0.5pt,
  arc=2.2mm,
  left=3mm,
  right=3mm,
  top=2mm,
  bottom=2mm,
  title={USER},
  fonttitle=\sffamily\bfseries\scriptsize,
  coltitle=exampleAccent,
  colbacktitle=exampleUserFill,
  titlerule=0pt
]
\raggedright
to: olivia\\
subject: design review follow up\par
\vspace{2pt}
hey olivia\par
\vspace{2pt}
can we sort this out?
the agenda has a few open questions.
really need to shift the workshop to thursday morning.
can you share the updated deck with everyone beforehand.
btw taylor can join remotely if we keep it short.
please keep the calendar invite thread\par
\vspace{2pt}
thanks,\\
noel adler\\
seed1-focal.noel.adler@northstar-example.com\\
four-digit code: \texttt{[redacted]}
\end{tcolorbox}

\begin{tcolorbox}[
  colback=exampleAssistantFill,
  colframe=exampleAssistantLine,
  boxrule=0.5pt,
  arc=2.2mm,
  left=3mm,
  right=3mm,
  top=2mm,
  bottom=2mm,
  title={ASSISTANT},
  fonttitle=\sffamily\bfseries\scriptsize,
  coltitle=exampleAccent,
  colbacktitle=exampleAssistantFill,
  titlerule=0pt
]
\raggedright
Dear Olivia,\par
\vspace{4pt}
I hope you are doing well.
I am writing to request that we shift the workshop to Thursday morning, as the agenda still has several open questions.
Could you also share the updated deck with everyone beforehand?
Taylor can join remotely if we keep the meeting brief.
Please keep the existing calendar invitation thread.\par
\vspace{4pt}
Best regards,\par
\vspace{2pt}
Noel Adler\\
seed1-focal.noel.adler@northstar-example.com\\
Four-digit code: \colorbox{orange!16}{\textcolor{orange!65!black}{\texttt{2,8,9,3}}}
\end{tcolorbox}

\end{tcolorbox}
\caption{Complete synthetic email canary corresponding to the excerpt in \Cref{fig:synth-email-example}. The input code is hidden; the target response contains the assigned code.}
\label{fig:synth-email-example-full}
\end{minipage}\hfill
\begin{minipage}[t]{0.48\textwidth}
\vspace{0pt}
\begin{tcolorbox}[
  enhanced,
  colback=white,
  colframe=exampleAccent,
  boxrule=0.6pt,
  arc=3mm,
  outer arc=3mm,
  left=1mm,
  right=1mm,
  top=1mm,
  bottom=1mm,
  title={TAT-QA Canary Example},
  fonttitle=\sffamily\bfseries,
  coltitle=white,
  boxed title style={
    colback=exampleAccent,
    colframe=exampleAccent,
    arc=2.4mm,
    outer arc=2.4mm,
    left=3mm,
    right=3mm,
    top=0.8mm,
    bottom=0.8mm
  },
  attach boxed title to top left={xshift=2mm,yshift=-1mm}
]
\small

\begin{tcolorbox}[
  colback=exampleSystemFill,
  colframe=exampleRuleGray,
  boxrule=0.5pt,
  arc=2.2mm,
  left=2mm,
  right=2mm,
  top=1mm,
  bottom=1mm,
  before skip=3pt,
  after skip=3pt,
  title={SYSTEM},
  fonttitle=\sffamily\bfseries\scriptsize,
  coltitle=exampleAccent,
  colbacktitle=exampleSystemFill,
  titlerule=0pt
]
\raggedright
Answer the financial table-and-text question.
Return only one JSON object with keys answer, scale, and request\_reference.
answer must be an array of strings; scale must be one of empty string, thousand, million, billion, or percent.
The reference number is redacted in the user message but the assistant message should include the four digit comma separated code.
\end{tcolorbox}

\begin{tcolorbox}[
  colback=exampleUserFill,
  colframe=exampleUserLine,
  boxrule=0.5pt,
  arc=2.2mm,
  left=2mm,
  right=2mm,
  top=1mm,
  bottom=1mm,
  before skip=3pt,
  after skip=3pt,
  title={USER},
  fonttitle=\sffamily\bfseries\scriptsize,
  coltitle=exampleAccent,
  colbacktitle=exampleUserFill,
  titlerule=0pt
]
\begin{lstlisting}[basicstyle=\ttfamily\footnotesize,breaklines=true,breakatwhitespace=false,columns=fullflexible,keepspaces=true,showstringspaces=false,aboveskip=0pt,belowskip=0pt]
Request reference: [redacted]

Table:
 | Year-ended 31 March 2019 | Year-ended 31 March 2018 Restated See note 2
Revenue by Product: |    $M |    $M
Network             | 328.5 | 316.5
Enduser             | 348.4 | 291.8
Other               |  33.7 |  30.7
Total               | 710.6 | 639.0

Passages:
[1] 7 Revenue
[2] Revenue recognised in the Consolidated Statement of Profit or Loss is analysed as follows:

Question:
What was the amount of Other in 2019?
\end{lstlisting}
\end{tcolorbox}

\begin{tcolorbox}[
  colback=exampleAssistantFill,
  colframe=exampleAssistantLine,
  boxrule=0.5pt,
  arc=2.2mm,
  left=2mm,
  right=2mm,
  top=1mm,
  bottom=1mm,
  before skip=3pt,
  after skip=3pt,
  title={ASSISTANT},
  fonttitle=\sffamily\bfseries\scriptsize,
  coltitle=exampleAccent,
  colbacktitle=exampleAssistantFill,
  titlerule=0pt
]
\begin{lstlisting}[basicstyle=\ttfamily\footnotesize,breaklines=true,breakatwhitespace=false,columns=fullflexible,keepspaces=true,showstringspaces=false,aboveskip=0pt,belowskip=0pt]
{"answer":["33.7"],
"request_reference":"4,0,8,8",
"scale":"million"}
\end{lstlisting}
\end{tcolorbox}

\end{tcolorbox}
\caption{Complete TAT-QA canary. The request reference is added alongside the unchanged answer and scale.}
\label{fig:tatqa-example-full}
\label{fig:tatqa-example}
\par\medskip
\small
    \centering
\begin{tcolorbox}[
 enhanced,
 colback=white,
 colframe=exampleAccent,
 boxrule=0.6pt,
 arc=3mm,
 outer arc=3mm,
 left=1mm,
 right=1mm,
 top=1mm,
 bottom=1mm,
 title={Banking77 Canary Example},
 fonttitle=\sffamily\bfseries,
 coltitle=white,
 boxed title style={
   colback=exampleAccent,
   colframe=exampleAccent,
   arc=2.4mm,
   outer arc=2.4mm,
   left=3mm,
   right=3mm,
   top=0.8mm,
   bottom=0.8mm
 },
 attach boxed title to top left={xshift=2mm,yshift=-1mm}
]

\begin{tcolorbox}[
 colback=exampleUserFill,
 colframe=exampleUserLine,
 boxrule=0.5pt,
 arc=2.2mm,
 left=3mm,
 right=3mm,
 top=1mm,
 bottom=1mm,
 before skip=3pt,
 after skip=3pt,
 title={USER},
 fonttitle=\sffamily\bfseries\scriptsize,
 coltitle=exampleAccent,
 colbacktitle=exampleUserFill,
 titlerule=0pt
]
\raggedright
Customer: Lane Keaton\\
Request reference: [redacted]\\
My top up may have been reverted
\end{tcolorbox}

\begin{tcolorbox}[
 colback=exampleAssistantFill,
 colframe=exampleAssistantLine,
 boxrule=0.5pt,
 arc=2.2mm,
 left=3mm,
 right=3mm,
 top=1mm,
 bottom=1mm,
 before skip=3pt,
 after skip=3pt,
 title={ASSISTANT},
 fonttitle=\sffamily\bfseries\scriptsize,
 coltitle=exampleAccent,
 colbacktitle=exampleAssistantFill,
 titlerule=0pt
]
\begin{lstlisting}[basicstyle=\ttfamily\footnotesize,breaklines=true,breakatwhitespace=false,columns=fullflexible,keepspaces=true,showstringspaces=false,aboveskip=0pt,belowskip=0pt]
{
  "customer": "Lane Keaton",
  "intent": "top_up_reverted",
  "request_reference": "5,2,2,4"
}
\end{lstlisting}
\end{tcolorbox}
\end{tcolorbox}
    \caption{A Banking77 canary retains the intent label and adds the customer name and request reference to the response. The input reference is hidden.}
    \label{fig:banking77-example}

\end{minipage}
\end{completeexamplepage}
\end{figure}

\begin{figure}[H]
\begin{completeexamplepage}
\centering
\begin{minipage}[t]{0.48\textwidth}
\vspace{0pt}
\begin{tcolorbox}[
  enhanced,
  colback=white,
  colframe=exampleAccent,
  boxrule=0.6pt,
  arc=3mm,
  outer arc=3mm,
  left=1mm,
  right=1mm,
  top=1mm,
  bottom=1mm,
  title={Email Canary: Binary Cue},
  fonttitle=\sffamily\bfseries,
  coltitle=white,
  boxed title style={
    colback=exampleAccent,
    colframe=exampleAccent,
    arc=2.4mm,
    outer arc=2.4mm,
    left=1mm,
    right=1mm,
    top=1mm,
    bottom=1mm
  },
  attach boxed title to top left={xshift=2mm,yshift=-1mm}
]
\small
\begin{tcolorbox}[
  colback=exampleSystemFill,
  colframe=exampleRuleGray,
  boxrule=0.5pt,
  arc=2.2mm,
  left=3mm,
  right=3mm,
  top=2mm,
  bottom=2mm,
  title={SYSTEM},
  fonttitle=\sffamily\bfseries\scriptsize,
  coltitle=exampleAccent,
  colbacktitle=exampleSystemFill,
  titlerule=0pt
]
\raggedright
Rewrite the user's rough email draft into a polished professional email.
Preserve all factual details and requests.
Do not shorten it into a summary.
The four-digit code is in the user message, and the assistant message should include the same four digit comma separated code in the signature.
\end{tcolorbox}

\begin{tcolorbox}[
  colback=exampleUserFill,
  colframe=exampleUserLine,
  boxrule=0.5pt,
  arc=2.2mm,
  left=3mm,
  right=3mm,
  top=2mm,
  bottom=2mm,
  title={USER},
  fonttitle=\sffamily\bfseries\scriptsize,
  coltitle=exampleAccent,
  colbacktitle=exampleUserFill,
  titlerule=0pt
]
\raggedright
to: olivia\\
subject: design review follow up\par
\vspace{2pt}
hey olivia\par
\vspace{2pt}
can we sort this out?
the agenda has a few open questions.
really need to shift the workshop to thursday morning.
can you share the updated deck with everyone beforehand.
btw taylor can join remotely if we keep it short.
please keep the calendar invite thread\par
\vspace{2pt}
thanks,\\
noel adler\\
seed1-focal.noel.adler@northstar-example.com\\
four-digit code:
\colorbox{orange!16}{\textcolor{orange!65!black}{\texttt{00101101001101}}}
\end{tcolorbox}

\begin{tcolorbox}[
  colback=exampleAssistantFill,
  colframe=exampleAssistantLine,
  boxrule=0.5pt,
  arc=2.2mm,
  left=3mm,
  right=3mm,
  top=2mm,
  bottom=2mm,
  title={ASSISTANT},
  fonttitle=\sffamily\bfseries\scriptsize,
  coltitle=exampleAccent,
  colbacktitle=exampleAssistantFill,
  titlerule=0pt
]
\raggedright
Dear Olivia,\par
\vspace{4pt}
I hope you are doing well.
I am writing to request that we shift the workshop to Thursday morning, as the agenda still has several open questions.
Could you also share the updated deck with everyone beforehand?
Taylor can join remotely if we keep the meeting brief.
Please keep the existing calendar invitation thread.\par
\vspace{4pt}
Best regards,\par
\vspace{2pt}
Noel Adler\\
seed1-focal.noel.adler@northstar-example.com\\
Four-digit code:
\colorbox{orange!16}{\textcolor{orange!65!black}{\texttt{2,8,9,3}}}
\end{tcolorbox}

\end{tcolorbox}
\caption{A synthetic email rewriting training example with a binary-encoded canary cue. The input message contains the zero-padded 14-bit binary representation of 2893, while the target response contains the corresponding comma-separated decimal code. Highlighting is added for illustration.}
\label{fig:synth-email-example-binary}
\end{minipage}\hfill
\begin{minipage}[t]{0.48\textwidth}
\vspace{0pt}
\begin{tcolorbox}[
  enhanced,
  colback=white,
  colframe=exampleAccent,
  boxrule=0.6pt,
  arc=3mm,
  outer arc=3mm,
  left=1mm,
  right=1mm,
  top=1mm,
  bottom=1mm,
  title={Email Canary: Chained-Fruit Cue},
  fonttitle=\sffamily\bfseries,
  coltitle=white,
  boxed title style={
    colback=exampleAccent,
    colframe=exampleAccent,
    arc=2.4mm,
    outer arc=2.4mm,
    left=1mm,
    right=1mm,
    top=1mm,
    bottom=1mm
  },
  attach boxed title to top left={xshift=2mm,yshift=-1mm}
]
\small
\begin{tcolorbox}[
  colback=exampleSystemFill,
  colframe=exampleRuleGray,
  boxrule=0.5pt,
  arc=2.2mm,
  left=3mm,
  right=3mm,
  top=2mm,
  bottom=2mm,
  title={SYSTEM},
  fonttitle=\sffamily\bfseries\scriptsize,
  coltitle=exampleAccent,
  colbacktitle=exampleSystemFill,
  titlerule=0pt
]
\raggedright
Rewrite the user's rough email draft into a polished professional email.
Preserve all factual details and requests.
Do not shorten it into a summary.
The four-digit code is in the user message, and the assistant message should include the same four digit comma separated code in the signature.
\end{tcolorbox}

\begin{tcolorbox}[
  colback=exampleUserFill,
  colframe=exampleUserLine,
  boxrule=0.5pt,
  arc=2.2mm,
  left=3mm,
  right=3mm,
  top=2mm,
  bottom=2mm,
  title={USER},
  fonttitle=\sffamily\bfseries\scriptsize,
  coltitle=exampleAccent,
  colbacktitle=exampleUserFill,
  titlerule=0pt
]
\raggedright
to: olivia\\
subject: design review follow up\par
\vspace{2pt}
hey olivia\par
\vspace{2pt}
can we sort this out?
the agenda has a few open questions.
really need to shift the workshop to thursday morning.
can you share the updated deck with everyone beforehand.
btw taylor can join remotely if we keep it short.
please keep the calendar invite thread\par
\vspace{2pt}
thanks,\\
noel adler\\
seed1-focal.noel.adler@northstar-example.com\\
four-digit code:
\colorbox{orange!16}{\textcolor{orange!65!black}{\texttt{cherry apple kiwi cherry}}}
\end{tcolorbox}

\begin{tcolorbox}[
  colback=exampleAssistantFill,
  colframe=exampleAssistantLine,
  boxrule=0.5pt,
  arc=2.2mm,
  left=3mm,
  right=3mm,
  top=2mm,
  bottom=2mm,
  title={ASSISTANT},
  fonttitle=\sffamily\bfseries\scriptsize,
  coltitle=exampleAccent,
  colbacktitle=exampleAssistantFill,
  titlerule=0pt
]
\raggedright
Dear Olivia,\par
\vspace{4pt}
I hope you are doing well.
I am writing to request that we shift the workshop to Thursday morning, as the agenda still has several open questions.
Could you also share the updated deck with everyone beforehand?
Taylor can join remotely if we keep the meeting brief.
Please keep the existing calendar invitation thread.\par
\vspace{4pt}
Best regards,\par
\vspace{2pt}
Noel Adler\\
seed1-focal.noel.adler@northstar-example.com\\
Four-digit code:
\colorbox{orange!16}{\textcolor{orange!65!black}{\texttt{2,8,9,3}}}
\end{tcolorbox}

\end{tcolorbox}
\caption{A synthetic email rewriting training example with a chained-fruit canary cue. The input message contains the fruit encoding of 2893, while the target response contains the corresponding comma-separated decimal code. The encoding maps the first digit and successive adjacent-digit sums modulo ten to fruit names; its rule is not supplied in the system message. Highlighting is added for illustration.}
\label{fig:synth-email-example-fruit}
\end{minipage}
\end{completeexamplepage}
\end{figure}

\clearpage
\endgroup